\documentclass[useAMS,usegraphicx,usenatbib]{mn2e}
\usepackage{times}

\newcommand{\Msun}{\ensuremath{\,{\rm M}_\odot}}            % Solar mass symbol
\newcommand{\Rsun}{\ensuremath{\,{\rm R}_\odot}}            % Solar radius symbol
\newcommand{\psun}{\ensuremath{\,\rho_\odot}}               % Solar density symbol
\newcommand{\Mjup}{\ensuremath{\,{\rm M}_{\rm Jup}}}        % Jupiter mass symbol
\newcommand{\Rjup}{\ensuremath{\,{\rm R}_{\rm Jup}}}        % Jupiter radius symbol
\newcommand{\pjup}{\ensuremath{\,\rho_{\rm Jup}}}           % Jupiter density symbol
\newcommand{\Teff}{\ensuremath{T_{\rm eff}}}                % Effective temperature symbol
\newcommand{\Teq}{\ensuremath{T_{\rm eq}^{\,\prime}}}                  % Equilibrium temperature symbol
\newcommand{\safronov}{\ensuremath{\Theta}}                 % Safronov number symbol
\newcommand{\logg}{\ensuremath{\log g}}                     % log(g) symbol
\newcommand{\FeH}{\ensuremath{\left[\frac{\rm Fe}{\rm H}\right]}}   % [Fe/H] symbol
\newcommand{\MoH}{\ensuremath{\left[\frac{\rm M}{\rm H}\right]}}    % [M/H] symbol
\newcommand{\Mlt}{\ensuremath{\alpha_{\rm MLT}}}            % Mixing length parameter symbol
\newcommand{\Porb}{\ensuremath{P_{\rm orb}}}                % Orbital period symbol

\newcommand{\Vsini}{\ensuremath{V_{\sin\!i}}}               % Stellar projected rotational velocity symbol
\newcommand{\kms}{\,km\,s$^{-1}$}                           % km/s symbol
\newcommand{\ms}{\,m\,s$^{-1}$}                             % m/s^2 symbol
\newcommand{\mss}{\,m\,s$^{-2}$}                            % m/s^2 symbol
\newcommand{\as}{\ensuremath{^{\prime\prime}}}              % Arcsecond symbol
\newcommand{\chir}{\ensuremath{\chi_\nu^{\,2}}}             % Reduced chi-squared symbol

\newcommand{\mc}[1]{\multicolumn{2}{c}{#1}}
\newcommand{\mcc}[1]{\multicolumn{3}{c}{#1}}
\newcommand{\er}[3]{\ensuremath{#1^{+#2}_{-#3}}}
\newcommand{\erc}[3]{\mc{\ensuremath{#1^{+#2}_{-#3}}}}

\newcommand{\ercc}[3]{\mcc{\ensuremath{#1^{+#2}_{-#3}}}}
\newcommand{\ermcc}[5]{\mcc{\ensuremath{{#1\,^{+#2}_{-#3}}\,^{+#4}_{-#5}}}}
\newcommand{\reff}[1]{{#1}}                                  % makes corrections bold-face if wanted.

\title[Homogeneous studies of transiting extrasolar planets. III.]
      {Homogeneous studies of transiting extrasolar planets. III. Additional planets and stellar models}

\author[John Southworth]
       {John Southworth\thanks{E-mail: jkt@astro.keele.ac.uk} \\
        Astrophysics Group, Keele University, Staffordshire, ST5 5BG, UK}

\begin{document} \maketitle %%%%%%%%%%%%%%%%%%%%%%%%%%%%%%%%%%%%%%%%%%%%%%%%%%%%%%%%%%%%%%%%%%%%%%%%%%%%%%%%%%
%%%%%%%%%%%%%%%%%%%%%%%%%%%%%%%%%%%%%%%%%%%%%%%%%%%%%%%%%%%%%%%%%%%%%%%%%%%%%%%%%%%%%%%%%%%%%%%%%%%%%%%%%%%%%%

\begin{abstract}
I derive the physical properties of thirty transiting extrasolar planetary systems using a homogeneous analysis of published data. The light curves are modelled with the {\sc jktebop} code, with special attention paid to the treatment of limb darkening, orbital eccentricity, and error analysis. The light from some systems is contaminated by faint nearby stars, which if ignored will systematically bias the results. I show that it is not \reff{realistically} possible to account for this using only transit light curves: light curve solutions must be constrained by measurements of the amount of contaminating light. A contamination of 5\% is enough to make the measurement of a planetary radius 2\% too low.

The physical properties of the thirty transiting systems are obtained by interpolating in tabulated predictions from theoretical stellar models to find the best match to the light curve parameters and the measured stellar velocity amplitude, temperature and metal abundance. Statistical errors are propagated by a perturbation analysis which constructs complete error budgets for each output parameter. These error budgets are used to compile a list of systems which would benefit from additional photometric or spectroscopic measurements.

The systematic errors arising from the inclusion of stellar models are assessed by using five independent sets of theoretical predictions for low-mass stars. This model dependence sets a lower limit on the accuracy of measurements of the physical properties of the systems, ranging from 1\% for the stellar mass to 0.6\% for the mass of the planet and 0.3\% for other quantities. The stellar density and the planetary surface gravity and equilibrium temperature are not affected by this model dependence. An external test on these systematic errors is performed by comparing the two discovery papers of the WASP-11\,/\,HAT-P-10 system: these two studies differ in their assessment of the ratio of the radii of the components and the effective temperature of the star.

I find that the correlations of planetary surface gravity and mass with orbital period have significance levels of only \reff{3.1}$\sigma$ and \reff{2.3}$\sigma$, respectively. \reff{The significance of the latter has not increased} with the addition of new data since Paper\,II. The division of planets into two classes based on Safronov number is increasingly blurred. Most of the objects studied here would benefit from improved photometric and spectroscopic observations, as well as improvements in our understanding of low-mass stars and their effective temperature scale.
\end{abstract}

\begin{keywords}
stars: planetary systems --- stars: binaries: eclipsing --- stars: binaries: spectroscopic --- stars: fundamental parameters
\end{keywords}

%%%%%%%%%%%%%%%%%%%%%%%%%%%%%%%%%%%%%%%%%%%%%%%%%%%%%%%%%%%%%%%%%%%%%%%%%%%%%%%%%%%%%%%%%%%%%%%%%%%%%%%%%%%%%%%%%%%%%%%

\section{Introduction}                                                                                \label{sec:intro}

The study of extrasolar planets is scientifically and culturally important, and after a late start \citep{MayorQueloz95nat} the number of known planets is escalating rapidly\footnote{see the Extrasolar Planets Encyclop{\ae}dia, {\tt http://exoplanet.eu/}}. Transiting planets are the crown jewels of this population as, with the exception of our own Solar system, they are the only planets whose masses and radii are directly measurable. In addition to this, it is possible to put constraints on the properties of their atmospheres, in which much interesting physics occurs, through measurements of the depths of transits and occultations at different wavelengths.

Whilst transiting extrasolar planets (TEPs) offer unique scientific possibilities, their study involves several complications. The most significant is that it is not in general possible to measure the mass and radius of a planet through basic observations alone. Additional constraints are needed, and are usually provided by forcing the properties of the host stars to match theoretical expectations\footnote{In principle, astrometric observations could either replace radial velocity measurements, or augment them and thus provide the missing constraint, but this has not yet been achieved in practise.}. This introduces not only a model dependence (i.e.\ systematic error), but also the possibility of inconsistent results if different theoretical predictions are used for some TEPs. Systematic errors can blur any distinctions between planets, making it hard to pick out discrete groups of TEPs from the varied general population.

This systematic error cannot be abolished, but it can at least be standardised. In this series of papers I am analysing the known transiting systems using rigorously homogeneous methods, with the aim of removing the systematic differences in measurements of the physical properties of TEPs. The resulting physical properties are therefore statistically compatible, and any structure in \reff{distributions of parameters} is maximised.

In Paper\,I \citep{Me08mn} I analysed the light curves of the fourteen transiting systems for which high-precision photometry was then available, paying careful attention to the role of limb darkening and to the estimation of comprehensive errorbars. Paper\,II \citep{Me09mn} used these results plus the predictions of three different theoretical stellar models to measure the physical properties of the fourteen TEPs. In this work I broaden the analysis to thirty TEPs and five sets of theoretical stellar models, resulting in improved statistics and better systematic error estimates.

There are a few homogeneous analyses of transiting systems available in the literature. A good analysis of 23 systems was presented by \citet[][hereafter TWH08]{Torres++08apj}, but these authors tried only two different theoretical model sets and did not assign systematic errors to their results. Analogously, such work would also benefit from homogeneous analysis of the spectra of the host stars in order to put their effective temperature and chemical abundance measurements on a consistent scale. Steps towards this goal were pioneered by \citet{ValentiFischer05apjs} and are being continued by \citet{Ammler+09aa} and \citet{Ghezzi+10iaus}, but a homogeneous study of the host stars of all known TEPs is not currently available.

In Sect.\,\ref{sec:lc} I present the methods used to analyse the light curves of the thirty TEPs included in this work. Sect.\,\ref{sec:model} discusses the five theoretical stellar model sets and their application to determining the physical properties of the TEPs. Sect.\,\ref{sec:teps} presents the new results for these objects, Sect.\,\ref{sec:syserr} discusses the influence of systematic errors due to the use of theoretical models, and in Sect.\,\ref{sec:absdim} I summarise the physical properties of the known TEPs and explore correlations between various parameters. Those readers interested in the general properties of TEPs rather than specific systems can skip Sect.\,\ref{sec:teps} without problem.

%%%%%%%%%%%%%%%%%%%%%%%%%%%%%%%%%%%%%%%%%%%%%%%%%%%%%%%%%%%%%%%%%%%%%%%%%%%%%%%%%%%%%%%%%%%%%%%%%%%%%%%%%%%%%%%%%%%%%%%

\section{Light curve analysis: JKTEBOP}                                                                  \label{sec:lc}

I have modelled the light curves of each TEP using the {\sc jktebop}%
%-------------------------------------------footnote---------------------------------------------
\footnote{{\sc jktebop} is written in {\sc fortran77} and the source code is available at {\tt http://www.astro.keele.ac.uk/$\sim$jkt/codes/jktebop.html}}
%-------------------------------------------footnote---------------------------------------------
code \citep{Me++04mn,Me++04mn2}. {\sc jktebop} grew out of the original {\sc ebop} program written for eclipsing binary star systems \citep{PopperEtzel81aj,Etzel81conf} and implementing the NDE model \citep{NelsonDavis72apj}. {\sc jktebop} uses biaxial spheroids to model the component stars (or star and planet) so allows for departures from sphericity. The shapes of the components are governed by the mass ratio, $q$, although the results in this work are all extremely insensitive to the value of this parameter.

The main parameters of a {\sc jktebop} fit are the orbital inclination, $i$, and the fractional radii of the two stars%
%-------------------------------------------footnote---------------------------------------------
\footnote{Throughout this work stellar parameters are indicated by a subscripted `A' and planet parameters by a subscripted 'b', to conform to IAU nomenclature.}%
%-------------------------------------------footnote---------------------------------------------
, $r_{\rm A}$ and $r_{\rm b}$. The fractional radii are defined as
\begin{equation}
r_{\rm A} = \frac{R_{\rm A}}{a} \qquad \qquad r_{\rm b} = \frac{R_{\rm b}}{a}
\end{equation}
where $R_{\rm A}$ and $R_{\rm b}$ are the stellar and planetary radii and $a$ is the orbital semimajor axis. $r_{\rm A}$ and $r_{\rm b}$ correspond to radii of spheres of the same volume as the biaxial spheroids. In {\sc jktebop} the fractional radii are re-parameterised as their sum and ratio:
\begin{equation}
r_{\rm A} + r_{\rm b} \qquad \qquad k = \frac{r_{\rm b}}{r_{\rm A}} = \frac{R_{\rm b}}{R_{\rm A}}
\end{equation}
because these are only weakly correlated with each other. In general the orbital period, $\Porb$, is taken from the literature and the time of transit midpoint, $T_0$, is included as a fitted parameter in each {\sc jktebop} run.

\subsection{Treatment of limb darkening}

The limb darkening (LD) of the star is an important `nuisance parameter' affecting transit light curves which can be parametrised using any of five LD laws in {\sc jktebop}. Wherever possible the LD coefficients are included as fitted parameters, but when there is insufficient information for this the coefficients are fixed at theoretical values. For each light curve I have obtained solutions with each of the five LD laws (see Paper\,I for their definition) and with both LD coefficients fixed, with the linear coefficient fitted and the nonlinear coefficient fixed (hereafter referred to as `LD fit/fix'), and with both coefficients fitted.

Theoretical LD coefficients have been taken from \citet{Vanhamme93aj}, \citet{Claret00aa,Claret04aa2} and \citet{ClaretHauschildt03aa}. The tabulated values have been bilinearly interpolated, using the {\sc jktld} code%
%-------------------------------------------footnote---------------------------------------------
\footnote{{\sc jktld} is written in {\sc fortran77} and the source code is available at {\tt http://www.astro.keele.ac.uk/$\sim$jkt/codes/jktld.html}}%
%-------------------------------------------footnote---------------------------------------------
, to the known effective temperature (\Teff) and surface gravity (\logg) of the star. I find that there is usually a spread of 0.1--0.2 in the theoretical LD coefficients for cool stars, so when the nonlinear LD coefficient is not included as a fitted parameter it is perturbed in the Monte Carlo simulations by $\pm$0.1 on a flat distribution to account for this. The dependence on theoretical calculations in this case is still exceptionally small, because the linear and nonlinear coefficients of the LD laws are highly correlated with each other \citep{Me++07aa}.

Once solutions have been obtained for the five LD laws, the final result is calculated by taking the \reff{weighted} means of the parameter values for the four two-parameter LD laws (i.e.\ the linear law is not used). The parameter errorbars are taken to be the largest of the individual errorbars (see below) plus a contribution to account for scatter of the parameter values from different LD law solutions.

\subsection{Error analysis}

For each solution I run 1000 Monte Carlo simulations \citep{Me+04mn3,Me+05mn} to provide robust estimates of the 1$\sigma$ statistical errorbars. The starting parameter values are perturbed for each simulation to avoid sticking artificially close to the original best fit. If the reduced $\chi^2$ of the fit, \chir, is greater than unity, the Monte Carlo errorbars are multiplied by $\sqrt{\chir}$ to account for this.

Monte Carlo simulations do not fully account for the presence of correlated (`red') noise, which is an unavoidable reality in high-precision light curves of bright stars. I therefore also run a residual permutation (or ``prayer bead'') algorithm \citep{Jenkins++02apj} with the quadratic LD law. If there is significant correlated noise the residual-permutation errorbars will exceed the Monte-Carlo errorbars. I then take the larger of the two error estimates to represent the final errorbars of the photometric parameters.

\subsection{Orbital eccentricity}                                                                \label{sec:lc:ecc}

Some TEPs have a non-circular orbit which must be accounted for in the light curve analysis. Orbital eccentricity is very difficult to detect from the shape of a transit light curve \citep{Kipping08mn} but can have a significant effect on the resulting parameters (for an example see Sect.\,\ref{sec:teps:xo3}). Non-circular orbits normally become apparent from radial velocity (RV) measurements of the parent stars. These RVs can then be used to determine the eccentricity, $e$, and the longitude of periastron, $\omega$, of the binary orbit,

{\sc jktebop} has been modified to account for orbital eccentricity by including the possibility of specifying values for either $e$ and $\omega$ or the combinations $e\cos\omega$ and $e\sin\omega$. These values and their uncertainties are then simply treated as extra observations, and $e$ and $\omega$ (or their combinations) are included as fitted parameters. In this way the uncertainties in $e$ and $\omega$ are correctly propagated into the errorbars in the other photometric parameters. I prefer to work with $e\cos\omega$ and $e\sin\omega$ rather than $e$ and $\omega$, because the latter two quantities are strongly correlated with each other \citep[e.g.][]{Bruntt+06aa}.

\subsection{Contaminating light}                                                                      \label{sec:lc:l3}

It is possible for additional light to contaminate photometric observations of transiting planets. Any extra light from nearby faint stars will dilute the transit depth, causing a systematic error in the light curve parameters. This idea is becoming more important for several reasons. Firstly, the CoRoT satellite has a large point spread function (PSF) which usually contains a number of stars aside from the one hosting a transiting planet. Secondly, observations using telescope defocussing are potentially more susceptible to contaminating light. Thirdly, \citet{Daemgen+09aa} have detected faint companions to three transiting systems (TrES-2, TrES-4 and WASP-2) from ground-based high-resolution observations of fourteen TEPs obtained with a lucky-imaging camera. These companions could be bound to their respective transiting systems, or may just be asterisms.

\reff{Temporarily ignoring the orbital ephemeris ($\Porb$ and $T_0$),} there are three main observables in a transit shape: its depth, duration, and the duration of totality (Paper\,I). From transit light curves we measure three quantities, which in the case of {\sc jktebop} are $r_{\rm A}$, $r_{\rm b}$ and $i$. It is therefore expected to be impossible to fit directly for contaminating light, as this would require measuring four independent parameters using only three observables. This expectation will now be verified.

\subsubsection{The effect of third light}

\begin{figure} \includegraphics[width=\columnwidth,angle=0]{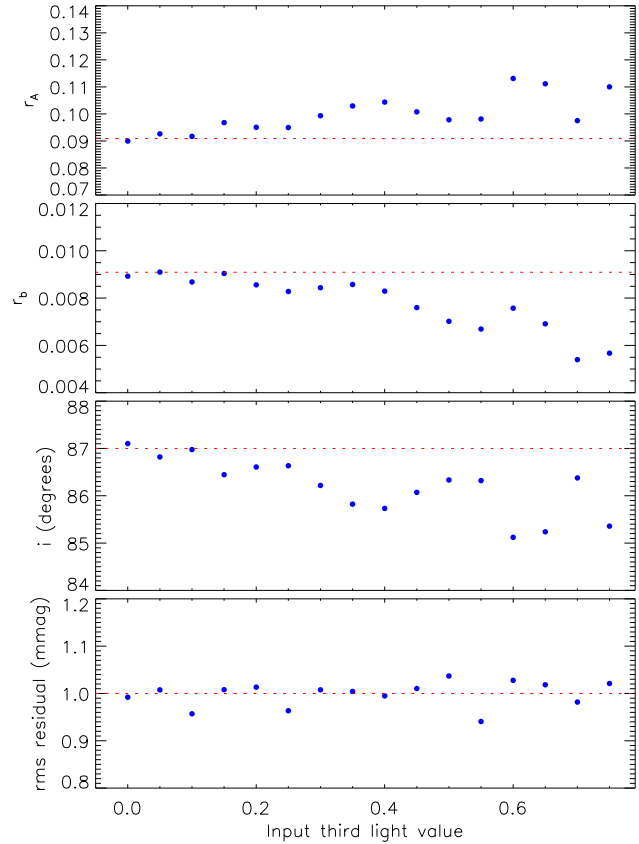}
\caption{\label{fig:lc:l3:zero} Plot of the variation of parameters fitted
to a set of synthetic transit light curves with 1\,mmag of Gaussian noise
added. The synthetic datasets were generated for a range of third light
values but fitted with the assumption of $L_3 = 0$. Dotted lines show the
input parameter values for the synthetic light curve calculations.} \end{figure}

I have explored the possibility of measuring contaminating light by simulating a set of light curves with reasonable parameters ($r_{\rm A}+r_{\rm b}=0.1$, $k=0.1$, $i=87^\circ$). I added contaminating light (by convention referred to as `third light' and expressed as a fraction of the total system light) by amounts ranging from $L_3 = 0$ to 0.75 in steps of 0.05. These were transformed into typical good ground-based light curves by retaining \reff{approximately} 400 points within each transit and adding a Gaussian scatter with standard deviation 1\,mmag. These synthetic light curves were then fitted with {\sc jktebop} under the assumption that $L_3 = 0$. The results (Fig.\,\ref{fig:lc:l3:zero}) show that the presence of $L_3$ results in systematic overestimates of $r_{\rm A}$ and underestimates of $r_{\rm b}$ and $i$. The bottom panel of Fig.\,\ref{fig:lc:l3:zero} shows that the quality of the fit does {\em not} get worse as $L_3$ increases. Unaccounted third light therefore biases the resulting parameters without being detectable through its impact on the quality of the fit.

\begin{figure} \includegraphics[width=\columnwidth,angle=0]{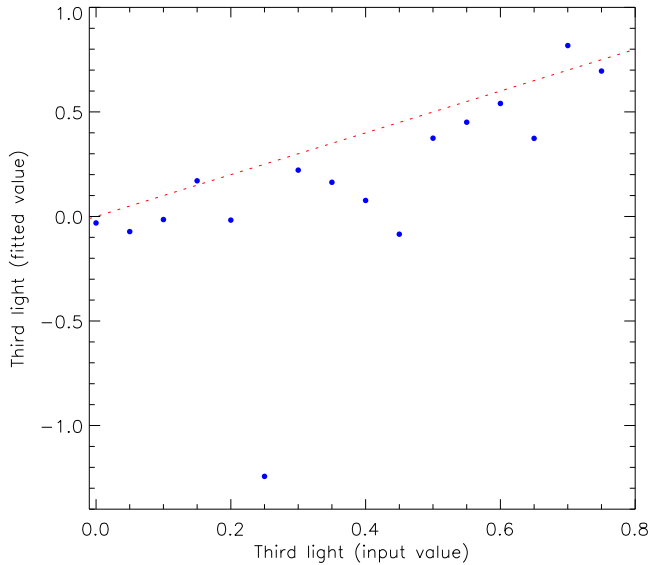}
\caption{\label{fig:lc:l3:comp} Plot of fitted versus input values of third
light for the same light curves as in Fig.\,\ref{fig:lc:l3:zero}, but with
$L_3$ included as a fitted parameter. The dotted line shows parity. Note
the large scale on the $y$-axis.} \end{figure}

\begin{figure} \includegraphics[width=\columnwidth,angle=0]{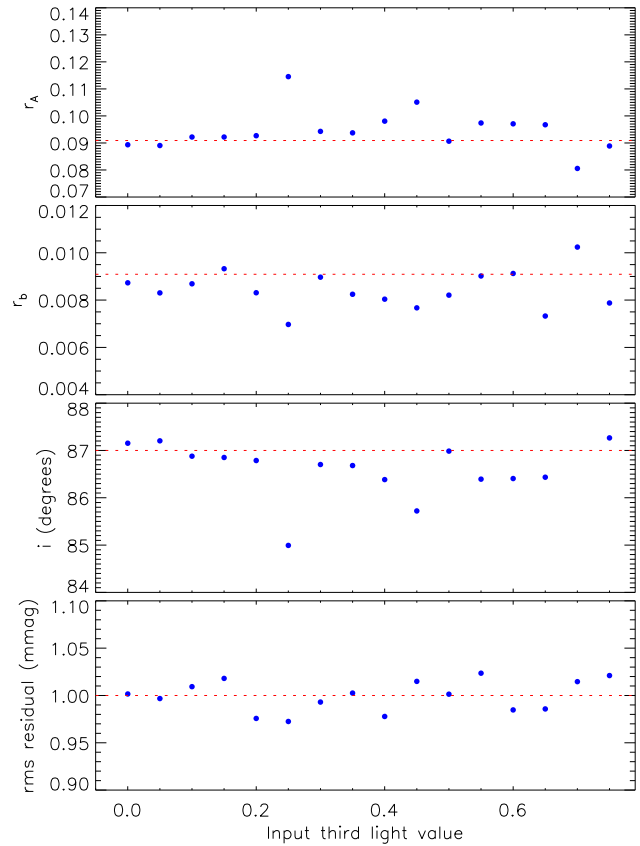}
\caption{\label{fig:lc:l3:nonzero} \reff{Plot of the variation of parameters fitted
to the same light curves as in Fig.\,\ref{fig:lc:l3:zero}, but in this case with
$L_3$ included as a fitted parameter. The dotted lines show parity.}} \end{figure}

As a second test I modelled the same synthetic light curves again, this time fitting for third light. The resulting values of $L_3$ are shown in Fig.\,\ref{fig:lc:l3:comp} and are extremely scattered as well as biased to smaller values. {\sc jktebop} deliberately does not restrict photometric parameters to physically realistic values (e.g.\ $L_3 \geqslant 0$), to avoid statistical biases in the uncertainties arising from Monte Carlo simulations. Fig.\,\ref{fig:lc:l3:comp} demonstrates that there {\em is} a very small amount of information on $L_3$ in a good ground-based light curve, but that this information is far too sparse to be useful. \reff{Fig.\,\ref{fig:lc:l3:nonzero} shows the resulting values of the other main parameters: $r_{\rm b}$ and $i$ are biased towards lower values and there is no trend visible in the sizes of the residuals.}

In reality a large value of $L_3$ is not expected because such a bright star would show up in the spectroscopic observations of a transiting system. Fainter stars can be found via high-resolution imaging if they are slightly away from the planet host star \citep{Daemgen+09aa}. But it may never be possible to rule out the presence of a much fainter star ($L_3 < 5$\% depending on the quality of the spectroscopic observations) which almost exactly coincide with the planet host. As a guide, 5\% of third light can be compensated for by increasing $r_{\rm A}$ by 1\%, and decreasing $r_{\rm b}$ by 2\% and $i$ by 0.1$^\circ$. It can therefore change the derived radius of the planet by several percent.

\subsubsection{Accounting for third light}

The observations of \citet{Daemgen+09aa} make it possible to account for third light when analysing transit light curves. They measured magnitude differences (light ratios) in the SDSS $i$ and $z$ passbands. When necessary I have propagated the light ratios to other passbands by convolving synthetic spectra from {\sc atlas9} model atmospheres \citep{Kurucz79apjs,Kurucz93} with passband response functions made available by the Isaac Newton Group\footnote{\tt http://catserver.ing.iac.es/filter/}.

Armed with $L_3$ values for the correct passbands, I have included these in the same way as $e$ and $\omega$. {\sc jktebop} was modified to accept measured $L_3$ values as observations, and $L_3$ was included as a fitted parameter. Note that several published studies have instead simply subtracted $L_3$ from a light curve before analysis, which is statistically incorrect as it neglects the uncertainty in $L_3$.

\citet{Daemgen+09aa} surveyed fourteen transiting systems and detected faint companions to three of them. The companions are within 0.7--1.6\as\ of the transit host stars, and are fainter by generally 4\,mag in the $i$ band. Daemgen et al.\ found that their presence changed the physical properties of the TEPs by roughly 1$\sigma$. Of the three affected objects, TrES-2 and TrES-4 are analysed in the current work and WASP-2 is the subject of a separate publication \citep{Me+10}.

%%%%%%%%%%%%%%%%%%%%%%%%%%%%%%%%%%%%%%%%%%%%%%%%%%%%%%%%%%%%%%%%%%%%%%%%%%%%%%%%%%%%%%%%%%%%%%%%%%%%%%%%%%%%%%%%%%%%%%%

\section{Incorporating stellar models: ABSDIM}                                                        \label{sec:model}

\begin{table*}
\caption{\label{table:models} Physical ingredients and coverage of the stellar
models used in this work. $Y_{\rm ini}$ is the primordial helium abundance,
$\Delta Y/\Delta Z$ is the helium-to-metals enrichment ratio, $Z_{\sun}$ is the
solar metal abundance (fraction by mass) and \Mlt\ is the mixing length parameter.}
\setlength{\tabcolsep}{5pt}
\begin{tabular}{lllllllll} \hline \hline
Model set     & Reference                                         & Range in     & Range in metal    & $Y_{\rm ini}$ & \underline{$\Delta Y$} & $Z_{\sun}$ & \Mlt & Notes \\
\             &                                                   & mass (\Msun) & abundance ($Z$)   &               &            $\Delta Z$  &            &      &       \\
\hline
{\it Claret}  & \citet{Claret04aa,Claret05aa,Claret06aa2,Claret07aa2} & 0.2 to 1.5   & 0.01 to 0.05      & 0.24  &   2.0   & 0.02   & 1.68  & Calculated on request       \\
{\it Y$^2$}   & \citet{Demarque+04apjs}                               & 0.4 to 5.2   & 0.00001 to 0.08   & 0.23  &   2.0   & 0.02   & 1.743 &                             \\
% {\it Padova}  & \citet{Bertelli+08aa}                               & 0.15 to 2.5  & 0.0001 to 0.07    &   \mc{choice}   & 0.017  & 1.68  & One $Y$ chosen for each $Z$ \\
{\it Teramo}  & \citet{Pietrinferni+04apj}                            & 0.5 to 10.0  & 0.0001 to 0.04    & 0.245 &$\sim$1.4& 0.0198 & 1.913 &                             \\
{\it VRSS}    & \citet{Vandenberg++06apjs}                            & 0.4 to 4.0   & 0.005 to 0.050    &0.23544&   2.2   & 0.188  & 1.90  &                             \\
{\it DSEP}    & \citet{Dotter+08apjs}                                 & 0.1 to 5.0   & 0.000041 to 0.0404& 0.245 &   1.6   & 0.0189 & 1.938 &                             \\
% {\it Padova}          & \citet{Girardi+00aas}                       & 0.15 to 7.0  & 0.0004 to 0.03    & 0.23  &   2.25  & 0.019  & 1.68  &                             \\
% {\it Siess}           & \citet{Siess++00aa}                         & 0.1 to 7.0   & 0.01 to 0.04      & 0.235 &   2.1   & 0.02   & 1.605 & Includes pre-MS phase       \\
% {\it Cambridge\,2000} & \citet{Pols+98mn}                           & 0.5 to 50    & 10$^{-4}$ to 0.03 & 0.24  &   2.0   & 0.0188 & 2.0   & Models with overshooting    \\
% {\it Cambridge\,2007} & \citet{EldridgeTout04mn2}                   & 0.5 to 2.0   & 0.01 to 0.05      & 0.24  &   2.0   & 0.0188 & 2.0   & Calculated for this work    \\
\hline \hline \end{tabular} \end{table*}

Analysis of a transit light curve gives the quantities $r_{\rm A}$, $r_{\rm b}$ and $i$\,\footnote{\reff{From transit light curves we also get \Porb\ and $T_0$. The uncertainty in \Porb\ is generally negligible, and $T_0$ does not enter the {\sc absdim} analysis.}}. From RV measurements of the parent star it is possible to obtain $e$, $\omega$, and the velocity amplitude of the star, $K_{\rm A}$. With these observables we remain unfortunately one piece of information short of being able to calculate the full physical properties of the system. An additional constraint is needed, and this is generally supplied by forcing the properties of the star to match the predictions of theoretical stellar evolutionary models. To guide this process we can use the spectroscopically measured \Teff\ and metal abundance, \FeH, of the star.

In the current work I adopt the method outlined in Paper\,II, in which the variable governing the solution process is taken to be $K_{\rm b}$, the orbital velocity amplitude of the {\em planet}. An initial value of $K_{\rm b}$ is defined, usually in the region of 150\kms, and the full physical properties of the system are calculated using standard formulae \citep[e.g.][]{Hilditch01book}. Armed with the resulting stellar mass, $M_{\rm A}$, and \FeH, I linearly interpolate within tabulated theoretical model results to find the predicted radius and \Teff\ of the star. This process is iteratively repeated whilst varying $K_{\rm b}$ in order to minimise the figure of merit
\begin{equation}
{\rm fom} = \left(\frac{r_{\rm A}^{\rm (obs)}-(R_{\rm A}^{\rm (pred)}/a)}{\sigma{\rm (r_{\rm A}^{\rm (obs)})}}\right)^2 +
            \left(\frac{\Teff^{\rm (obs)}-\Teff^{\rm (pred)}}{\sigma(\Teff^{\rm (obs)})}\right)^2
\end{equation}
which results in the best-fitting system properties. In principle it is possible to also solve for the age of the system, but in practise the wide variety of evolutionary timescales of stars make this difficult. I therefore step through the possible ages of the star in 0.1\,Gyr increments, starting at zero age and finishing when the star leaves the main sequence, in order to find the best overall solution. I do not make any attempt to match spectroscopically measured \logg\ values as they are usually much less reliable than the surface gravity of the star calculated from the $M_{\rm A}$ and $R_{\rm A}$ obtained above. The above procedure implicitly applies the strong constraint on stellar density obtained from the light curve analysis \citep{SeagerMallen03apj}.

The uncertainties in the system properties are calculated by a perturbation analysis, in which each input parameter is modified by its 1$\sigma$ uncertainty and new solutions specified. The uncertainty for each output parameter is then calculated by adding the uncertainties due to each input parameter in quadrature. This perturbation analysis has the advantage of yielding detailed error budgets, where the effect of the uncertainty of every input parameter on every output parameter is specified. These error budgets indicate what additional observations are the best for improving our understanding of a specific TEP.

\subsection{Which stellar models to use?}                                                       \label{sec:model:which}

As outlined above, the physical properties of TEPs are calculated by forcing the properties of the parent star to match theoretical expectations. This dependence on theoretical predictions is a concern and will cause a systematic error. It is well known that whilst theoretical models are pretty good at reproducing the actual properties of stars, the various model sets are not flawless and do not agree perfectly with each other.

The existence of different theoretical model sets for low-mass stars opens the possibility of using several of them and explicitly deducing the systematic errors in TEP properties caused by their use. In Paper\,II six different sets of theoretical models were investigated and three adopted for calculating the planet properties. The {\it Siess} and {\it Cambridge-2007} models were in relatively poor agreement with other models, and the {\it Cambridge-2000} models had a lower coverage of the relevant parameter space than other models. I was therefore left with only three different model sets, which was insufficient to define high-quality systematic error estimates. On top of this, the {\it Padova} models included heavy element abundances only up to $Z = 0.03$ so did not cover quite a few transiting systems.

In the current paper I have therefore adopted the same solution procedure as introduced in Paper\,II, but with a significantly revised database of theoretical model predictions and with one more change. Instead of using the {\it Claret} models to define my baseline solutions and two other model sets to obtain systematic error estimates, I have used the unweighted mean and standard deviation of the results from all five model sets to describe the baseline solutions and systematic errors. The dependence of the final results on a single model set is therefore broken: all five model sets are treated equally and the choice of which sets of models to use becomes less important.

Of the sets of theoretical models included in Paper\,II, only the {\sc Y$^2$} models survive unchanged here (see Table\,\ref{table:models}). The {\it Claret} models have been supplemented by additional calculations for higher metal abundances of $Z = 0.06$ and 0.07. The third model set used here is {\it Teramo}\footnote{Obtained from the BaSTI database on 17/11/2009: \\{\tt http://albione.oa-teramo.inaf.it/}} \citep{Pietrinferni+04apj}, and I selected the ones with moderate convective core overshooting (for masses $>$1.1\Msun), the standard mass loss law ($\eta = 0.4$) and normal elemental abundances (scaled-solar, i.e.\ no enhancement of the $\alpha$-elements). For the fourth model set I acquired the Victoria-Regina ({\it VRSS}) models\footnote{Obtained on 18/11/2009 from {\tt http://www.cadc-ccda.hia-iha. nrc-cnrc.gc.ca/cvo/community/VictoriaReginaModels/}.} \citep{Vandenberg++06apjs} with scaled-solar elemental abundances. In these models the convective core overshooting parameter depends on mass and is empirically calibrated. The fifth and final model set ({\it DSEP}) comes from Dartmouth Stellar Evolution Database\footnote{Obtained on 18/11/2009 from {\tt http://stellar.dartmouth.edu/ $\sim$models/index.html}.} and again comprises the calculations with scaled-solar elemental abundances. I selected those models which follow the standard helium-to-metal enrichment law. The {\it DSEP} models include a contraction to the zero-age main sequence which can take tens of Myr.

\begin{figure*} \includegraphics[width=\textwidth,angle=0]{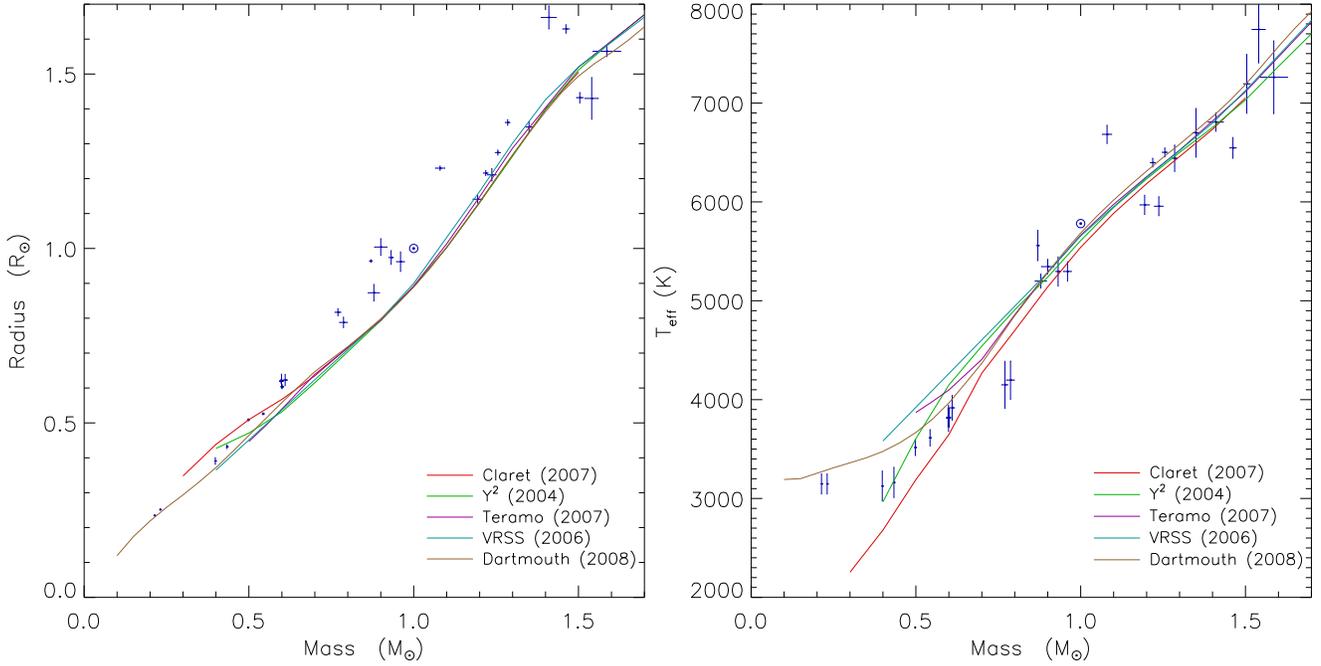}
\caption{\label{fig:model:mrt} Mass--radius (left) and mass--\Teff\ (right) plots
showing the predictions of the five sets of stellar models adopted in this work
(solid lines). The predictions are for an age of 0.5\,Gyr (to minimise the effects
of evolution to and from the ZAMS) and for the solar chemical composition (which
varies between models). The measured masses, radii and \Teff\ values of a sample of
detached eclipsing binaries (see Paper\,II) are shown for comparison, using blue
errorbars. The Sun is indicated by the usual $\odot$; note that the predictions of
the models do not pass through the solar values on these plots as the Sun is much
older than 0.5\,Gyr.} \end{figure*}

In Fig.\,\ref{fig:model:mrt} I compare the predictions in mass and radius of the five sets of stellar models, for an age of 0.5\,Gyr and for the adopted solar chemical composition (which differs between models). The models have a fairly good agreement in the mass--radius plane, particularly near 1.0\Msun\ as they are calibrated on the Sun, but a wider variety in the mass--\Teff\ plane. Also shown in Fig.\,\ref{fig:model:mrt} are the masses, radii and \Teff\ values of the sample of detached eclipsing binary star systems constructed in Paper\,II. It can be seen that the disagreement with the models is much larger than the errorbars. \reff{This conclusion holds for all chemical compositions for which the five sets of models are available. Similarly, adopting an age either before or beyond the main sequence can provide an agreement for individual eclipsing binaries, but not for all of them simultaneously.}

Fig.\,\ref{fig:model:mrt} illustrates what can be expected for the variation of systematic errors with mass. The models agree with each other very well in some regions, so systematics will be minimised, and less well at lower and higher masses, when systematics will be larger. The agreement between models is clearly much better than with the properties of well-studied eclipsing binaries. This means that using the five model sets will lead to only a {\em lower limit} on the systematic errors in the properties of TEPs. A probable upper limit to the systematic effects can be obtained by calculating solutions with an eclipsing binary mass--radius relation instead of a stellar model set; this is applied below and discussed further in Paper\,II.

\subsection{Calculating the physical properties of transiting planets}

Using the method and theoretical stellar models outlined above, the mass, radius, surface gravity and mean density of the star ($M_{\rm A}$, $R_{\rm A}$, $\log g_{\rm A}$, $\rho_{\rm A}$) and of the planet ($M_{\rm b}$, $R_{\rm b}$, $g_{\rm b}$, $\rho_{\rm b}$) can be calculated. For each TEP I have obtained results for each of the five stellar model sets and also using an empirical mass--radius relation defined by the eclipsing binaries. The final result for each parameter is the unweighted mean of the five stellar-model results. The statistical errorbar is taken to be the largest one from these five solutions and the systematic errorbar is taken to be the standard deviation of the parameter values from the five solutions. I include the final $K_{\rm b}$ value when possible to aid the comparison between different solutions for the same planet. $K_{\rm b}$ is the parameter through which all of the stellar model dependence enters.

The surface gravity of the planet, $g_{\rm b}$, can be calculated from purely geometrical observed quantities \citep{Me++07mn} so has no systematic error. Similarly, the stellar density, $\rho_{\rm A}$, is independent of the stellar models \citep{SeagerMallen03apj} if $M_{\rm A} \gg M_{\rm b}$ is assumed.

In addition to the above parameters, I have also calculated the equilibrium temperature ($T_{\rm eq}$) and \citet{Safronov72} number (\safronov) of the planet. $T_{\rm eq}$ is given by the equation
\begin{equation}
T_{\rm eq} = \Teff \left(\frac{1-A}{4F}\right)^{1/4} \left(\frac{R_{\rm A}}{2a}\right)^{1/2}
\end{equation}
where $A$ is the Bond albedo and $F$ is a heat redistribution factor. Because $A$ and $F$ are not known precisely I instead calculate a modified equilibrium temperature (\Teq) which equals $T_{\rm eq}$ if $A = 1 - 4F$:
\begin{equation} \label{eq:teq}
\Teq = \Teff \left(\frac{R_{\rm A}}{2a}\right)^{1/2} = \Teff \left(\frac{r_{\rm A}}{2}\right)^{1/2}
\end{equation}
The Safronov number is defined as the ratio of the escape velocity to the orbital velocity of the planet:
\begin{equation}
\Theta = \frac{1}{2} \left(\frac{V_{\rm esc}}{V_{\rm orb}}\right)^2
       = \left(\frac{a}{R_{\rm b}}\right) \left(\frac{M_{\rm b}}{M_{\rm A}}\right)
       = \frac{1}{r_{\rm b}} \frac{M_{\rm b}}{M_{\rm A}}
\end{equation}

From Eq.\,\ref{eq:teq} above it can be seen that \Teq\ depends only on the stellar \Teff\ and the fractional radius obtained from the light curve analysis. \Teq\ therefore turns out (like $g_{\rm b}$) to be independent of the stellar models, but does have some systematic error as it is dependent on the effective temperature scale of low-mass \reff{(F, G and K)} stars.

%%%%%%%%%%%%%%%%%%%%%%%%%%%%%%%%%%%%%%%%%%%%%%%%%%%%%%%%%%%%%%%%%%%%%%%%%%%%%%%%%%%%%%%%%%%%%%%%%%%%%%%%%%%%%%%%%%%%%%%

\section{Results for individual systems}                                                               \label{sec:teps}

\begin{table*} \caption{\label{tab:teps:lcpar} Parameters from the light curve
analyses presented here and in previous works, and used here to determine the
physical properties of the TEPs. The orbital periods are taken from the literature,
and the bracketed numbers represent the uncertainty in the preceding digits. Systems
for which orbital eccentricity was accounted for are indicated with a $\star$ in the
column marked ``$e$?''.}
\begin{tabular}{l l c r@{\,$\pm$\,}l r@{\,$\pm$\,}l r@{\,$\pm$\,}l l}
\hline \hline
System & Orbital period & $e$? & \mc{Orbital inclination,} & \mc{Fractional stellar}  & \mc{Fractional planetary} & Reference \\
       &     (days)     &      &    \mc{$i$ (degrees)}     & \mc{radius, $r_{\rm A}$} & \mc{radius, $r_{\rm b}$}  &           \\
% System & Orbital period (days) & \mc{Orbital inclination (degrees)} & \mc{Fractional stellar radius ($r_{\rm A}$)} & \mc{Fractional planetary radius ($r_{\rm b}$)} \\
\hline
GJ\,436     & 2.64389524 (76) &$\star$& 86.43 & 0.18       & 0.0731 & 0.0027             & 0.00605 & 0.00023              & Paper\,I \\
HAT-P-1     & 4.4652934 (93)  &       & 86.25 & 0.22       & 0.0935 & 0.0025             & 0.01051 & 0.00031              & This work \\
HAT-P-2     & 5.6334729 (61)  &$\star$& 85.9 & 1.5         & 0.1247 & 0.0106             & 0.00847 & 0.00082              & This work \\
HD\,149026  & 2.8758911 (25)  &       & 88.0 & 2.0         & \erc{0.140}{0.012}{0.006}   & \erc{0.0068}{0.0011}{0.0008}   & Paper\,I \\
HD\,189733  & 2.21857578 (80) &       & 85.78 & 0.25       & 0.1113 & 0.0031             & 0.0175 & 0.0005                & Paper\,I \\
HD\,209458  & 3.52474859 (38) &       & 86.590 & 0.046     & 0.11384 & 0.00041           & 0.01389 & 0.00006              & Paper\,I \\
OGLE-TR-10  & 3.101278 (4)    &       & 83.87 & 0.69       & 0.157 & 0.009               & 0.0182 & 0.0011                & Paper\,I \\
OGLE-TR-56  & 1.211909 (1)    &       & 79.8 & 2.4         & 0.245 & 0.026               & 0.0241 & 0.0034                & Paper\,I \\
OGLE-TR-111 & 4.0144479 (41)  &       & 88.11 & 0.66       & 0.0842 & 0.0038             & 0.01107 & 0.00067              & Paper\,I \\
OGLE-TR-113 & 1.4324757 (13)  &       & 87.7 & 1.4         & 0.1592 & 0.0043             & 0.02331 & 0.00089              & This work \\
OGLE-TR-132 & 1.689868 (3)    &       & 83.3 & 2.4         & 0.211 & 0.020               & 0.0198 & 0.0024                & Paper\,I \\
OGLE-TR-182 & 3.97910 (1)     &       & 84.3 & 1.2         & 0.137 & 0.014               & 0.0135 & 0.0013                & This work \\
OGLE-TR-211 & 3.67724 (3)     &       & 88.0 & 2.0         & \erc{0.1422}{0.0150}{0.0083}& \erc{0.01181}{0.00146}{0.00083}& This work \\
OGLE-TR-L9  & 2.485533 (7)    &       & 82.07 & 0.69       & 0.1731 & 0.0083             & 0.01910 & 0.00085              & This work \\
TrES-1      & 3.0300728 (6)   &       & 88.67 & 0.71       & 0.0964 & 0.0018             & 0.01331 & 0.00035              & Paper\,I \\
TrES-2      & 2.4706101 (18)  &       & 83.80 & 0.36       & 0.1282 & 0.0035             & 0.01658 & 0.00043              & This work \\
TrES-3      & 1.3061864 (5)   &       & 82.07 & 0.17       & \erc{0.1666}{0.0017}{0.0015}& \erc{0.02731}{0.00055}{0.00043}& This work \\
TrES-4      & 3.553945 (75)   &       & 81.53 & 0.60       & 0.1802 & 0.0083             & 0.0174 & 0.0012                & This work \\
WASP-1      & 2.519961 (18)   &       & 88.0 & 2.0         & \erc{0.1737}{0.0057}{0.0089}& \erc{0.0182}{0.0007}{0.0011}   & Paper\,I \\
WASP-2      & 2.15222144 (39) &       & 84.81 & 0.17       & 0.1238 & 0.0018             & 0.01643 & 0.00030              & \citet{Me+10} \\
WASP-3      & 1.846835 (2)    &       & 84.1 & 1.3         & 0.201& 0.010                & 0.0218 & 0.0011                & This work \\
WASP-4      & 1.33823150 (61) &       & 89.0 & 1.0         & \erc{0.1825}{0.0011}{0.0010}& \erc{0.02812}{0.00022}{0.00014}& \citet{Me+09mn2} \\
WASP-5      & 1.6284246 (13)  &       & 85.8 & 1.1         & 0.1847 & 0.0061             & 0.02050 & 0.00091              & \citet{Me+09mn}  \\
WASP-10     & 3.0927616 (10)  &$\star$& 88.81 & 0.40       & 0.0865 & 0.0041             & 0.01349 & 0.00065              & This work \\
WASP-18     & 0.94145181 (44) &$\star$& 85.0 & 2.1         & 0.2795 & 0.0084             & 0.0272 & 0.0012                & \citet{Me+09apj} \\
XO-1        & 3.9415128 (28)  &       & 89.06 & 0.84       & 0.0886 & 0.0019             & 0.01166 & 0.00035              & Paper\,I \\
XO-2        & 2.6158640 (16)  &       & 88.8 & 1.2         & \erc{0.1237}{0.0024}{0.0047}& \erc{0.01300}{0.00033}{0.00070}& This work \\
XO-3        & 3.1915289 (32)  &$\star$& 83.89 & 0.40       & 0.1447 & 0.0046             & 0.01317 & 0.00047              & This work \\
XO-4        & 4.12502 (2)     &       &\erc{89.9}{0.1}{3.9}& \erc{0.1300}{0.0283}{0.0051}& \erc{0.01124}{0.00334}{0.00054}& This work \\
XO-5        & 4.187757 (11)   &       & 87.04 & 0.65       & 0.1004 & 0.0049             & 0.01054 & 0.00073              & This work \\
\hline \hline \end{tabular} \end{table*}

\begin{table*} \caption{\label{tab:teps:synth} Measured quantities for the
parent stars which were adopted in the analysis presented in this work.}
\setlength{\tabcolsep}{2.5pt}
\begin{tabular}{l r@{\,$\pm$\,}l l r@{\,$\pm$\,}l l r@{\,$\pm$\,}l l}
\hline \hline
System & \multicolumn{3}{l}{Velocity amplitude (\ms)} & \mc{\Teff\ (K)} & Reference & \mc{\FeH} & Reference \\
\hline
GJ\,436     &  18.34& 0.52& \citet{Maness+07pasp}   & 3500 &100 & \citet{Bean++06apj}     &$-$0.03& 0.2  & \citet{Bonfils+05aa}    \\
HAT-P-1     &  59.3 & 1.4 & \citet{Johnson+08apj}   & 5975 & 50 & \citet{Bakos+07apj}     & 0.13  & 0.05 & \citet{Bakos+07apj}     \\
HAT-P-2     & 983.9 &17.2 & \citet{Pal+10mn}        & 6290 & 60 & \citet{Pal+10mn}        & 0.14  & 0.08 & \citet{Pal+10mn}        \\
HD\,149026  &  43.3 & 1.2 & \citet{Sato+05apj}      & 6147 & 50 & \citet{Sato+05apj}      & 0.36  & 0.05 & \citet{Sato+05apj}      \\
HD\,189733  & 200.56& 0.88& \citet{Boisse+09aa}     & 5050 & 50 & \citet{Bouchy+05aa}     &$-$0.03& 0.05 & \citet{Bouchy+05aa}     \\
HD\,209458  &  85.1 & 1.0 & \citet{Naef+04aa}       & 6117 & 50 & \citet{Santos++04aa}    & 0.02  & 0.05 & \citet{Santos++04aa}    \\
OGLE-TR-10  &  80   & 17  & \citet{Konacki+05apj}   & 6075 & 86 & \citet{Santos+06aa}     & 0.28  & 0.10 & \citet{Santos+06aa}     \\
OGLE-TR-56  & 212   & 22  & \citet{Bouchy+05aa2}    & 6119 & 62 & \citet{Santos+06aa}     & 0.25  & 0.08 & \citet{Santos+06aa}     \\
OGLE-TR-111 &  78   & 14  & \citet{Pont+04aa}       & 5044 & 83 & \citet{Santos+06aa}     & 0.19  & 0.07 & \citet{Santos+06aa}     \\
OGLE-TR-113 & 267   & 34  & TWH08                   & 4804 &106 & \citet{Santos+06aa}     & 0.15  & 0.10 & \citet{Santos+06aa}     \\
OGLE-TR-132 & 167   & 18  & \citet{Moutou+04aa}     & 6210 & 59 & \citet{Gillon+07aa3}    & 0.37  & 0.07 & \citet{Gillon+07aa3}    \\
OGLE-TR-182 & 120   & 17  & \citet{Pont+08aa}       & 5924 & 64 & \citet{Pont+08aa}       & 0.37  & 0.08 & \citet{Pont+08aa}       \\
OGLE-TR-211 &  82   & 16  & \citet{Udalski+08aa}    & 6325 & 91 & \citet{Udalski+08aa}    & 0.11  & 0.10 & \citet{Udalski+08aa}    \\
OGLE-TR-L9  & 510   & 170 & \citet{Snellen+09aa}    & 6933 & 58 & \citet{Snellen+09aa}    &$-$0.05& 0.20 & \citet{Snellen+09aa}    \\
TrES-1      & 115.2 & 6.2 & \citet{Alonso+04apj}    & 5226 & 50 & \citet{Santos+06aa}     & 0.06  & 0.05 & \citet{Santos+06aa}     \\
TrES-2      & 181.3 & 2.6 & \citet{Odonovan+06apj}  & 5795 & 73 & \citet{Ammler+09aa}     & 0.06  & 0.08 & \citet{Ammler+09aa}     \\
TrES-3      & 369   & 11  & \citet{Sozzetti+09apj}  & 5650 & 75 & \citet{Sozzetti+09apj}  &$-$0.19& 0.08 & \citet{Sozzetti+09apj}  \\
TrES-4      & 97.4  & 7.2 & \citet{Mandushev+07apj} & 6200 & 75 & \citet{Sozzetti+09apj}  & 0.14  & 0.09 & \citet{Sozzetti+09apj}  \\
WASP-1      & 111   & 9   & \citet{Wheatley+10}     & 6110 & 50 & \citet{Stempels+07mn}   & 0.23  & 0.08 & \citet{Stempels+07mn}   \\
WASP-2      & 153.6 & 3.0 & \citet{Triaud+10aa}     & 5150 & 80 & \citet{Triaud+10aa}     &$-$0.08& 0.08 & \citet{Triaud+10aa}     \\
WASP-3      & 290.5 & 9.5 & \citet{Tripathi+10}     & 6400 & 100&\citet{Pollacco+08mn}    & 0.00  & 0.20 & \citet{Pollacco+08mn}   \\
WASP-4      &\erc{242.1}{2.8}{3.1}&\citet{Triaud+10aa}&5500&100&\citet{Gillon+09aa}      &$-$0.03& 0.09 & \citet{Gillon+09aa}     \\
WASP-5      & 268.7 & 1.8 & \citet{Triaud+10aa}     & 5700 &100 & \citet{Gillon+09aa}     & 0.09  & 0.09 & \citet{Gillon+09aa}     \\
WASP-10     & 553.1 & 7.5 & \citet{Johnson+09apj}   & 4675 &100 & \citet{Christian+09mn}  & 0.03  & 0.20 & \citet{Christian+09mn}  \\
WASP-18     &1816.9 & 2.0 & \citet{Triaud+10aa}     & 6400 &100 & \citet{Hellier+09natur} & 0.00  & 0.09 & \citet{Hellier+09natur} \\
XO-1        & 116.0 & 9.0 & \citet{Mccullough+06apj}& 5750 & 50 & \citet{Mccullough+06apj}& 0.02  & 0.05 & \citet{Mccullough+06apj}\\
XO-2        &  85   & 8   & \citet{Burke+07apj}     & 5340 & 50 & \citet{Burke+07apj}     & 0.45  & 0.05 & \citet{Burke+07apj}     \\
XO-3        & 1488  & 10  & \citet{Winn+09apj}      & 6429 & 75 & \citet{Johnskrull+08apj}&$-$0.18& 0.05 & \citet{Johnskrull+08apj}\\
XO-4        & 163   & 16  & \citet{Mccullough+08xxx}& 6397 & 70 & \citet{Mccullough+08xxx}&$-$0.04& 0.05 & \citet{Mccullough+08xxx}\\
XO-5        & 144.9 & 2.0 & \citet{Pal+09apj}       & 5370 & 70 & \citet{Pal+09apj}       & 0.05  & 0.06 & \citet{Pal+09apj}       \\
\hline \hline \end{tabular} \end{table*}

In this section I present the photometric ({\sc jktebop}) and absolute-dimensions ({\sc absdim}) analyses of a set of thirty TEPs based on high-quality data. In many cases I adopt the {\sc jktebop} results from Paper\,I or later works \citep{Me+09mn,Me+09mn2,Me+09apj,Me+10}. The final {\sc jktebop} results of all TEPs are collected in Table\,\ref{tab:teps:lcpar}, which also includes the orbital periods and indicates for which systems a non-circular orbit was adopted. The mass ratio of each TEP is required as an input parameter for the light curve analysis, but in all cases its effect on the solution is negligible. Representative values have been taken from the literature but will not be discussed further.

The physical properties of all thirty TEPs are obtained or revised in the current work, using the new theoretical model sets discussed in Sect.\,\ref{sec:model}. This also requires \Teff, \FeH\ and $K_{\rm A}$ values for each system. These are summarised in Table\,\ref{tab:teps:synth}. The values are mostly unchanged for the fourteen TEPs studied in Paper\,II, but in a few cases improved values have become available and replace the previous entries. In Papers I and II the individual systems were tackled roughly in order of increasing complexity. The current work reverts to the more structured approach of attacking the TEPs in alphabetical order, beginning with those objects for which a light curve analysis is presented (Sects.\ \ref{sec:teps:hatp1} to \ref{sec:teps:xo5}, then moving on to those whose photometric parameters are adopted unchanged from Paper\,I (Sect.\,\ref{sec:teps:done}).

%%%%%%%%%%%%%%%%%%%%%%%%%%%%%%%%%%%%%%%%%%%%%%%%%%%%%%%%%%%%%%%%%%%%%%%%%%%%%%%%%%%%%%%%%%%%%%%%%%%%%%%%%%%%%%%%%%%%%%%

\subsection{HAT-P-1}                                                                             \label{sec:teps:hatp1}

\begin{figure} \includegraphics[width=\columnwidth,angle=0]{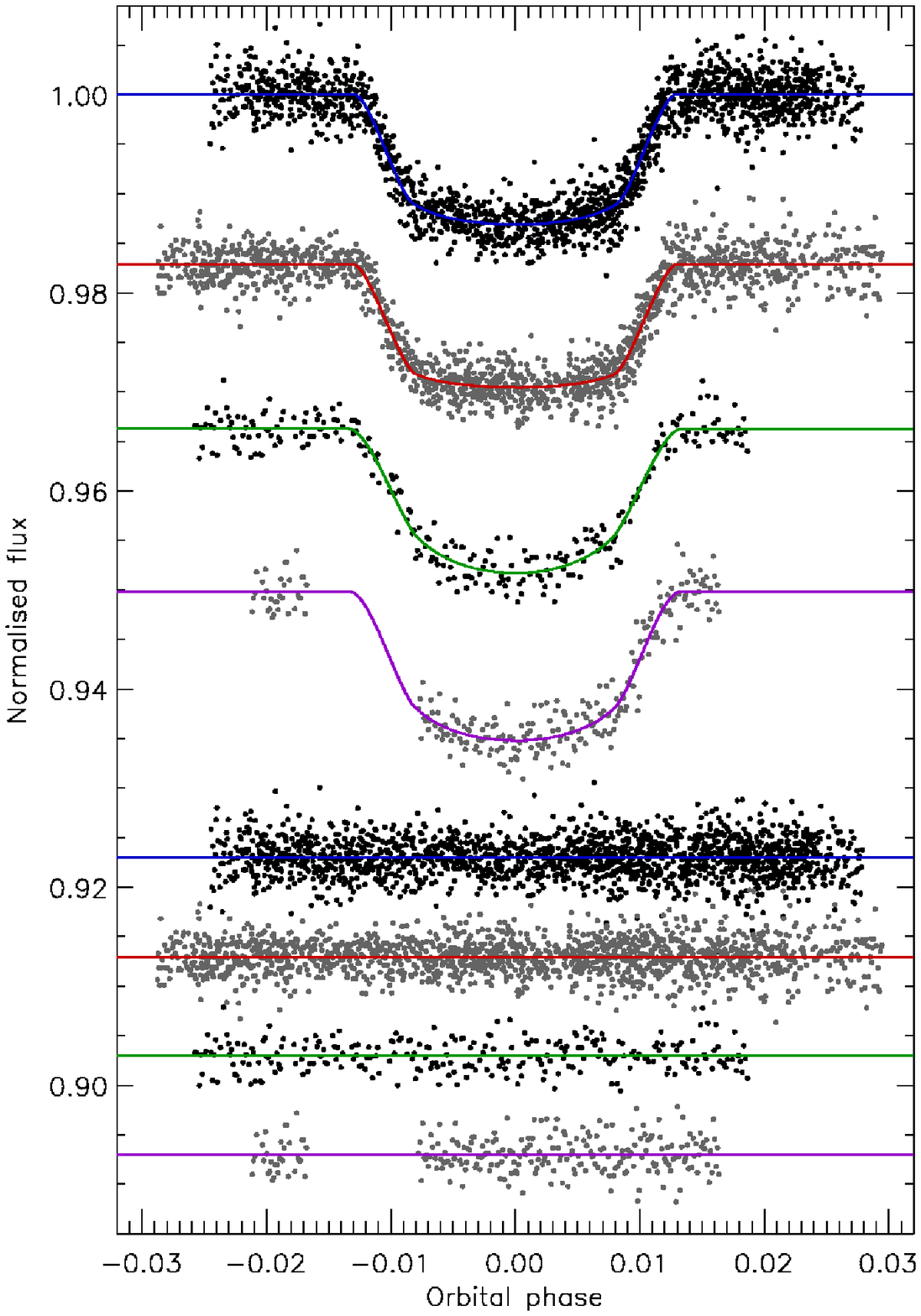}
\caption{\label{fig:hatp1:lc} Phased light curves of HAT-P-1 compared to the
best fits found using {\sc jktebop} and the quadratic LD law in Paper\,I and in
this work. The best fits and residuals are offset from unity and zero fluxes,
respectively, for display purposes. The light curves are, from top to bottom, FLWO
$z$-band \citep{Winn+07aj2}, Lick $Z$-band \citep{Winn+07aj2}, Magnum $V$-band
\citep{Johnson+08apj} and Nickel $Z$-band \citep{Johnson+08apj}.} \end{figure}

HAT-P-1 was found to be a TEP by \citet{Bakos+07apj} from data taken by the HAT survey \citep{Bakos+02pasp,Bakos+04pasp}. Its low mass (0.5\Mjup) and large radius (1.2\Rjup) make it one of the least dense exoplanets known. Excellent light curves from the FLWO 1.2\,m ($z$ band), Lick 1.0\,m Nickel ($Z$ band) and Wise 1.0\,m telescopes were presented by \citet{Winn+07aj2} and the first two of these were included in Paper\,I. Since then additional data from the Nickel ($Z$ band) and the Hawaiian 2\,m Magnum ($V$ band) telescopes have been obtained by \citet{Johnson+08apj}. In this work I have analysed the latter two datasets in order to refine the results from Paper\,I. In both cases I have adopted solutions with the linear LD coefficient fitted and the nonlinear coefficient fixed but perturbed in the Monte Carlo simulations (`LD fit/fix'). The residual permutation analyses indicate that correlated errors are important for both datasets.

The final light curve parameters are the weighted means of those for the four studied datasets. The results agree well with each other except for $k$, for which $\chir = 2.8$. The errorbar for $k$ has been multiplied by $\sqrt{2.8}$ to account for this. The light curve fits are plotted in Fig.\,\ref{fig:hatp1:lc} and summarised in Table\,A3. They are in good agreement with literature values.

The physical properties of HAT-P-1 have been calculated using the five different sets of stellar evolutionary models plus the empirical mass--radius relation from Paper\,II. The individual solutions are given in Table\,A4 and then compared with literature values, where a good agreement is found.

%%%%%%%%%%%%%%%%%%%%%%%%%%%%%%%%%%%%%%%%%%%%%%%%%%%%%%%%%%%%%%%%%%%%%%%%%%%%%%%%%%%%%%%%%%%%%%%%%%%%%%%%%%%%%%%%%%%%%%%

\subsection{HAT-P-2}                                                                             \label{sec:teps:hatp2}

\begin{figure} \includegraphics[width=\columnwidth,angle=0]{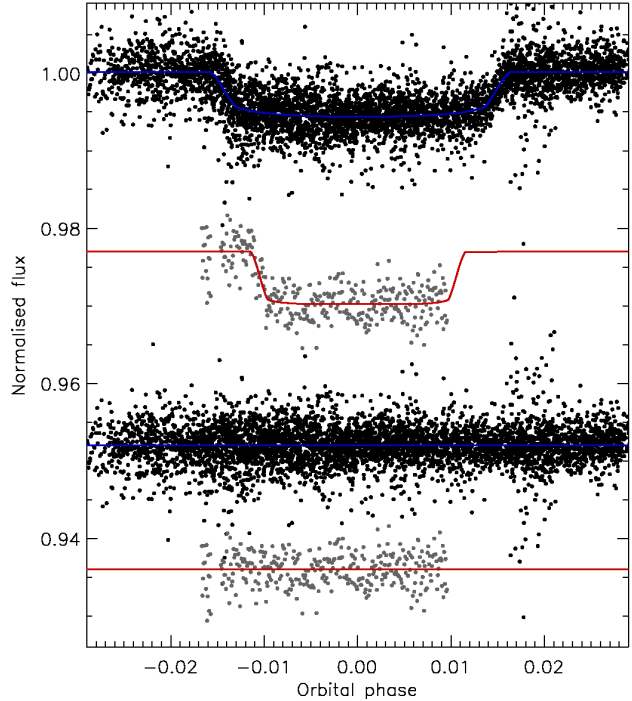}
\caption{\label{fig:hatp2:lc} Phased $z$-band light curves of HAT-P-2
compared to the best fits found using {\sc jktebop} and the quadratic LD
law. The best fits and residuals are offset for display purposes. The upper
light curve is from the FLWO 1.2\,m \citep{Bakos+07apj3,Pal+10mn} and the
lower is from the Perkins 1.8\,m \citep{Pal+10mn}.} \end{figure}

HAT-P-2 was discovered to be a TEP system by \citet{Bakos+07apj3}, under the name HD\,147506. It is a very bright system ($V=8.7$) with a massive planet ($M_{\rm b} = 8.74$\Mjup) in a highly eccentric orbit. It has been found not to exhibit a spin--orbit misalignment \citep{Winn+07apj,Loeillet+08aa}, in contrast to other massive TEPs on eccentric orbits \citep{Johnson+09pasp}.

Good $z$-band light curves of HAT-P-2 have been published by \citet{Bakos+07apj3}, covering one transit with the FLWO 1.2\,m, and by \citet{Pal+10mn}, covering another six transits with the FLWO 1.2\,m and the Perkins telescopes. Here I analyse the FLWO datasets together, omitting the small amount of data taken on the night of 2007/03/18, as well as the Perkins data. One complication is the orbital eccentricity: this was accounted for using the method discussed in Sect.\,\ref{sec:lc:ecc} and adopting the constraints $e\cos\omega = -0.5152 \pm 0.0036$ and $e\sin\omega = -0.0441 \pm 0.0084$ \citep{Pal+10mn}. In both cases correlated errors were not important and the LD fit/fix solutions were adopted. The best fits are shown in Fig.\,\ref{fig:hatp2:lc}.

The two light curve solutions unfortunately do not agree very well (9.3$\sigma$ discrepancy in $k$). I therefore adopt the FLWO 1.2\,m results, as these are the much more extensive of the two sets of data and have full coverage of the transit phases. The FLWO results agree well with those of \citet{Pal+10mn}, for which most of the data come from, but have a larger $r_{\rm A}$ and $r_{\rm b}$ than other literature values (Table\,A7).

As expected given the light curve results, my {\sc absdim} analysis returns system properties in good agreement with those of \citet{Pal+10mn} but not with other literature studies (Table\,A8). The prime mover in the most recent solutions is $r_{\rm A}$, which has a strong effect on the density of the star and thus the other physical properties. The radius of the planet is uncertain by 10\%, despite the existence of a high-quality light curve for HAT-P-2, because the transit depth is shallow (0.6\%). An improved photometric study is warranted.

%%%%%%%%%%%%%%%%%%%%%%%%%%%%%%%%%%%%%%%%%%%%%%%%%%%%%%%%%%%%%%%%%%%%%%%%%%%%%%%%%%%%%%%%%%%%%%%%%%%%%%%%%%%%%%%%%%%%%%%

\subsection{OGLE-TR-113}                                                                       \label{sec:teps:ogle113}

\begin{figure} \includegraphics[width=\columnwidth,angle=0]{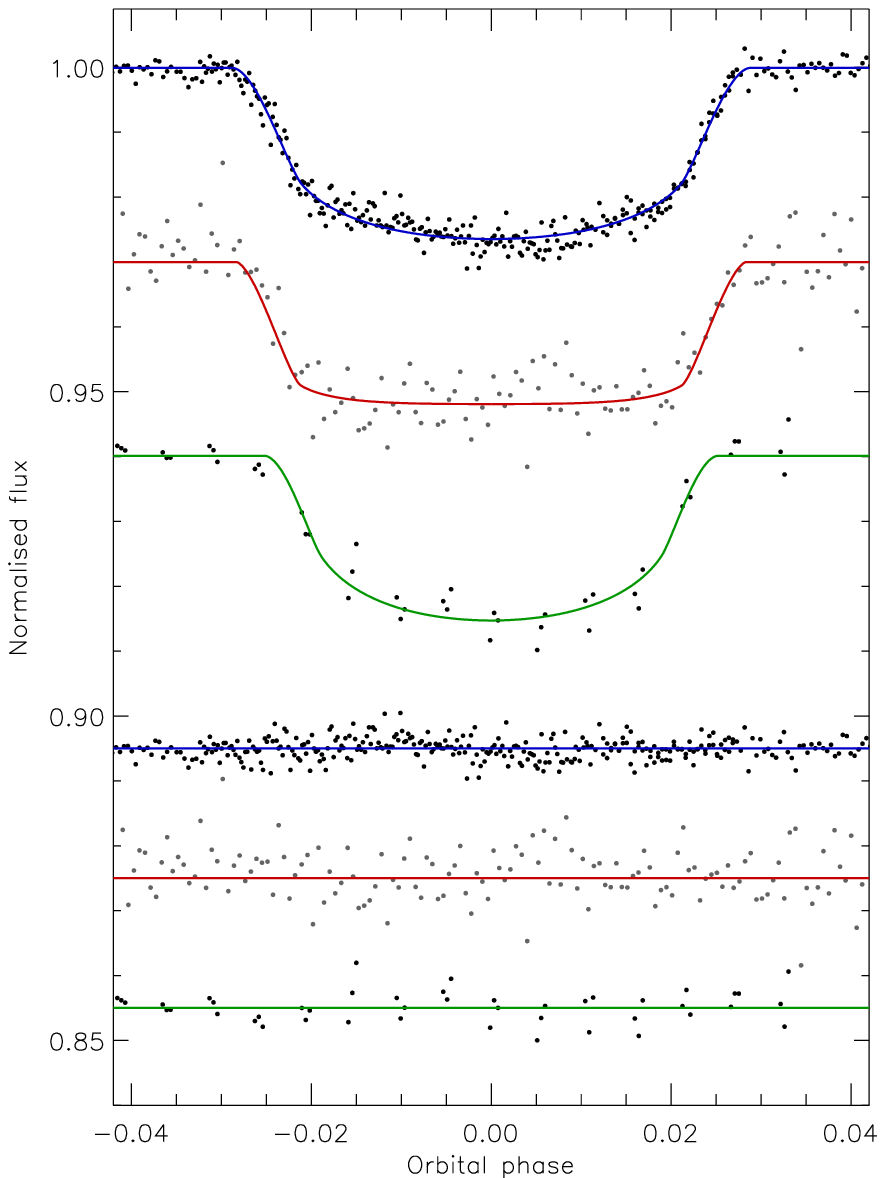}
\caption{\label{fig:ogle113:lc} Phased light curves of OGLE-TR-113 with the
best fits found using {\sc jktebop} and the quadratic LD law and residuals of
the fits. From top to bottom the datasets are \citet{Gillon+06aa} ($R$-band),
\citet{SnellenCovino07mn} ($K_s$-band, binned by $\times$5 for display purposes)
and \citet{Diaz+07apj} ($V$-band).} \end{figure}

Like OGLE-TR-132 (studied in Paper\,I), OGLE-TR-113 was identified as a possible planetary system by \citet{Udalski+02aca2} and its nature was confirmed by \citet{Bouchy+04aa} using the OGLE light curve and new RV measurements. It was independently confirmed as a TEP by \citet{Konacki+04apj}, also from the OGLE light curve and high-precision RVs, and an abundance analysis of the parent star has been presented by \citet{Santos+06aa}. Whilst OGLE-TR-113 exhibits a deep transit, its photometric tractability is hindered by the presence of a brighter star only 3\as\ away.

Apart from the OGLE discovery observations \citep{Udalski+02aca2}, three photometric studies of OGLE-TR-113 have been published. \citet{Gillon+06aa} used the ESO New Technology Telescope (NTT) and SUSI2 imager to observe two transits in the $R$ band, and obtained what is currently the best light curve of OGLE-TR-113. \citet{SnellenCovino07mn} observed a $K$-band transit and an occultation of the system using NTT/SOFI, and detected the occultation with a significance of 2.8$\sigma$. \citet{Diaz+07apj} obtained  $V$-band photometry of one transit using a Very Large Telescope (VLT) and the VIMOS instrument; additional data taken in the $I$ and $K_s$ bands are unavailable and of lower quality.

In this work I analyse the \citeauthor{Gillon+06aa} observations, the \citeauthor{SnellenCovino07mn} transit light curve, and the $V$-band data obtained by \citeauthor{Diaz+07apj} For the second of these three datasets I allowed for light from the planet with a surface brightness ratio of $0.07 \pm 0.02$. The surface brightness ratio is a parameter of the {\sc jktebop} model which is important for eclipsing binary systems but usually left at zero for transiting systems due to the faintness of the planet with respect to the star. The best fits are shown in Fig.\,\ref{fig:ogle113:lc} and given in Table\,A12. In all three cases correlated noise is not important. For the Snellen light curve I had to adopt solutions with both LD coefficients fixed, but for the other two I was able to use the LD fit/fix solutions. The final results for the Gillon and Snellen data are in good agreement. The solution of the D\'{\i}az data prefers a rather higher $i$ and lower $r_{\rm A}$ and $r_{\rm b}$. I therefore combined the Gillon and Snellen data results to obtain the final light curve parameters.

The resulting physical properties of OGLE-TR-113 are given in Table\,A13. The system age is constrained only to be more than a few Gyr, and in several cases is up against the edge of the stellar model grid at 20\,Gyr. Aside from that, the properties of the star and planet are rather well-determined but would benefit from an improved $K_{\rm A}$ value as well as a better light curve. The agreement with literature studies is good, although it seems that in some cases the published errorbars are smaller than one would expect.

%%%%%%%%%%%%%%%%%%%%%%%%%%%%%%%%%%%%%%%%%%%%%%%%%%%%%%%%%%%%%%%%%%%%%%%%%%%%%%%%%%%%%%%%%%%%%%%%%%%%%%%%%%%%%%%%%%%%%%%

\subsection{OGLE-TR-182}                                                                       \label{sec:teps:ogle182}

\begin{figure} \includegraphics[width=\columnwidth,angle=0]{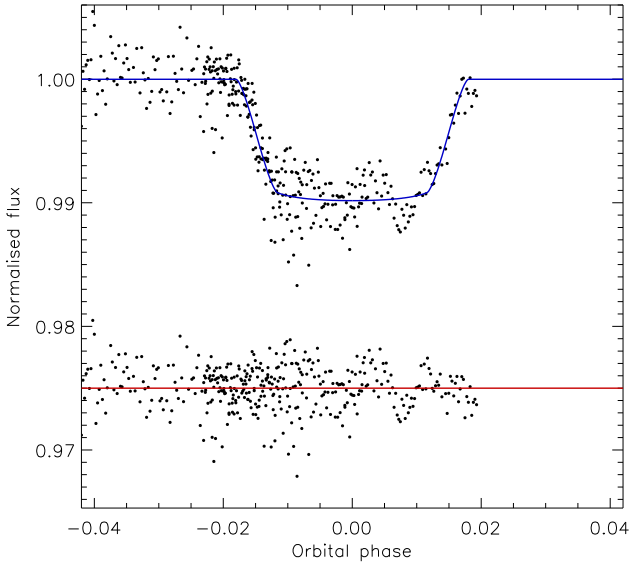}
\caption{\label{fig:ogle182:lc} Phased VLT light curve of OGLE-TR-182 from
\citet{Pont+08aa} compared to the best fit found using {\sc jktebop} and the
quadratic LD law. The residuals are offset from zero for display purposes.}
\end{figure}

OGLE-TR-182 is the sixth TEP discovered as a result of the OGLE search for light variability in selected fields in the Southern hemisphere. Its discovery and analysis was presented by \citet{Pont+08aa}, which remains the only study of this object to date. OGLE-TR-182 is difficult because of its faintness ($V = 16.8$ and $I = 15.9$) and crowded field. \citet{Pont+08aa} obtained 24 RV measurements using VLT/FLAMES/UVES, and a light curve with VLT/FORS1.

The VLT light curve is analysed here and is rather affected by correlated noise. Including the linear LD coefficients as a fitted parameter gives substantially better fits than with both LD coefficients fixed, but the data cannot support the determination of both LD coefficients. I therefore adopt the LD fit/fix solutions (see Fig.\,\ref{fig:ogle182:lc} and Table\,A15), which are not in good agreement with \citet{Pont+08aa}. Compared to these authors I find a solution with a lower $i$ and a correspondingly larger star and planet.

The physical properties of OGLE-TR-182 are summarised in Table\,A16 and point to a planet with a rather low density of 0.33\pjup. However, my results are rather different to those of \citet{Pont+08aa}, and are in poorer agreement with the measured spectroscopic \Teff\ and (rather uncertain) \logg\ measurement. My analysis procedure is more sensitive to the quality of the light curve than the more `global' approach taken by Pont et al., so is potentially more susceptible to correlated noise. This possibility should be investigated by acquiring a new light curve, under good seeing conditions to cope with the crowded field.

% Alternative: try fixing i with Teff like for OGLE-TR-211
% obs: 5924 ± 64 and 4.47 ± 0.17
% i     r1          Teff  logg
% 83    0.1520      5100  4.02
% 84    0.1400      5672  4.11
% 84.5  0.1342      5721  4.15
% 85    0.1286      5761  4.19
% 86    0.1186      5806  4.26
% 87    0.1103      5799  4.32
% 88    0.1037      5813  4.38
% 89    0.0997      5798  4.41
% 90    0.0983      5784  4.42
% nope - this does not work.

%%%%%%%%%%%%%%%%%%%%%%%%%%%%%%%%%%%%%%%%%%%%%%%%%%%%%%%%%%%%%%%%%%%%%%%%%%%%%%%%%%%%%%%%%%%%%%%%%%%%%%%%%%%%%%%%%%%%%%%

\subsection{OGLE-TR-211}                                                                       \label{sec:teps:ogle211}

\begin{figure} \includegraphics[width=\columnwidth,angle=0]{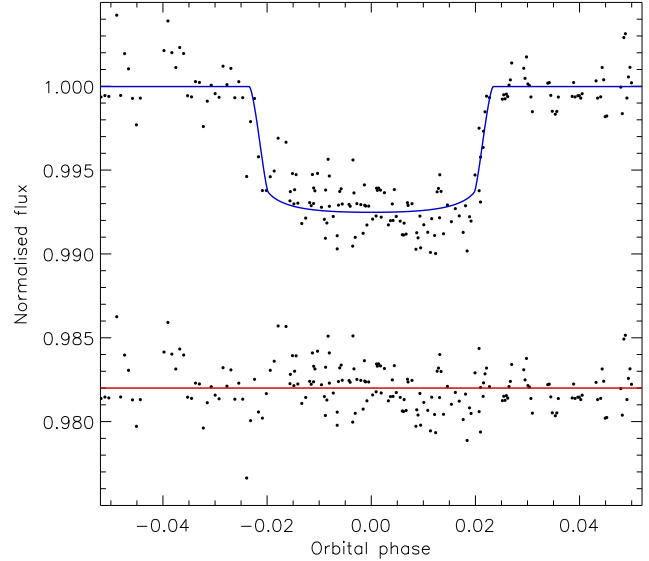}
\caption{\label{fig:ogle211:lc} Phased VLT light curve of OGLE-TR-211 from
\citet{Udalski+08aa} compared to the best fit found using {\sc jktebop} and the
quadratic LD law. The residuals are offset from zero for display purposes.}
\end{figure}

OGLE-TR-211 is the seventh TEP discovered using OGLE data \citep{Udalski+08aa}. Its relative faintness means that the available follow-up photometry and spectroscopy is not definitive. The parent star is more massive and also more evolved than the Sun, which results in the transit being rather shallow (Fig.\,\ref{fig:ogle211:lc}). Here I analyse the VLT light curve presented by \citet{Udalski+08aa}, ignoring the observational errors supplied with the data which are quite discretised (the only values are 0.001, 0.002 and 0.003) and contribute to instability in the light curve solution.

I am not able to get a determinate solution to the VLT data. Possible fits occupy a locus extending from a high $i$ with low $r_{\rm A}$ to a lower $i$ with a large $r_{\rm A}$. I have therefore calculated solutions for a range of $i$ values and retained only those which in the {\sc absdim} analysis result in a \Teff\ within a conservative 3$\sigma$ of the observed value. The observed stellar \logg\ did not provide a useful constraint. Allowable solutions extend from $i = 90^\circ$ down to a sharp cut-off around $i = 86.25\degr$ so I present solutions for $i = 86, 88$, and 90$^\circ$ in Table\,A17. For the final result I accept the LD fit/fix solutions for $i = 88^\circ$ but specify errors which account for both the Monte Carlo errorbars and the variation between the different solutions (Table\,A18). The correlated errors are again important, as can be seen in Fig.\,\ref{fig:ogle211:lc}.

% obs: 6325 ± 91 and 4.22 ± 0.17
% i     r1          Teff  logg
% 90    0.1367      6396  4.21
% 88    0.1424      6325  4.17
% 87    0.1476      6331  4.14
% 86.5  0.1509      6335  4.12
% 86.25 0.1527      6338  4.12
% 86.2  0.1531      5980  4.10
% 86.15 0.1535      5978  4.10
% 86.0  0.1546      5975  4.09
% 85    0.1631      5783  4.03

The physical properties of OGLE-TR-211 are shown in Table\,A19 and are in reasonable agreement with those of \citet{Udalski+08aa} except for the planetary mass. $M_{\rm b}$ depends mainly on the measured $K_{\rm A}$, for which both studies have used the same value, so it is not clear why such a discrepancy should arise. Table\,A19 includes the first determinations of the age and density of the star, planetary equilibrium temperature (which is quite high at $\Teq = \er{1686}{90}{55}$\,K) and Safronov number. The system age is relatively well determined ($2.6^{+0.6 \ +0.4}_{-0.7\ -0.3}$\,Gyr) because the star has evolved away from the zero-age main sequence. OGLE-TR-211 would certainly benefit from additional spectroscopic and photometric observations.

%%%%%%%%%%%%%%%%%%%%%%%%%%%%%%%%%%%%%%%%%%%%%%%%%%%%%%%%%%%%%%%%%%%%%%%%%%%%%%%%%%%%%%%%%%%%%%%%%%%%%%%%%%%%%%%%%%%%%%%

\subsection{OGLE-TR-L9}                                                                         \label{sec:teps:oglel9}

\begin{figure} \includegraphics[width=\columnwidth,angle=0]{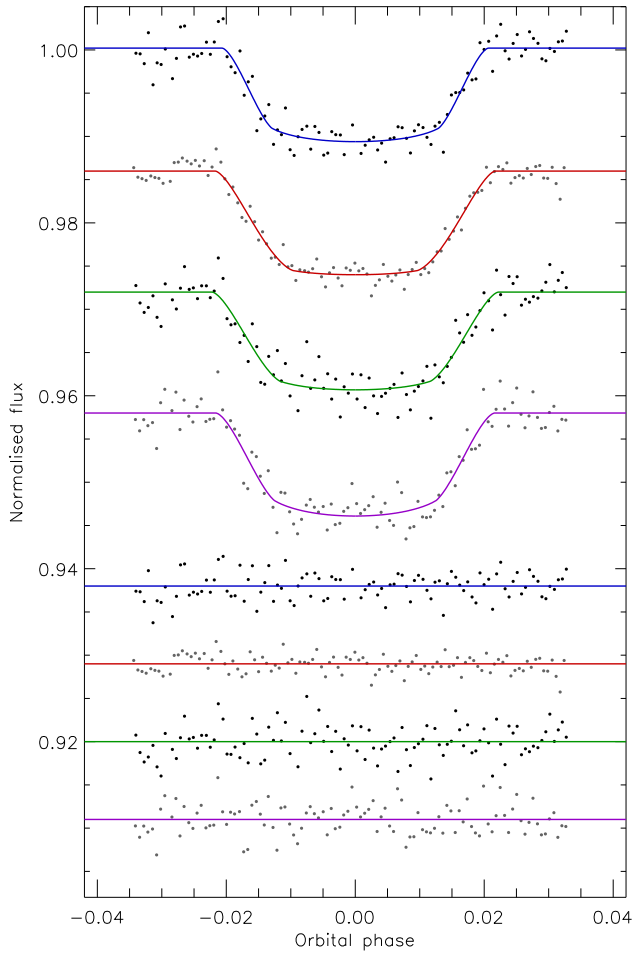}
\caption{\label{fig:oglel9:lc} Phased GROND light curves of OGLE-TR-L9 from
\citet{Snellen+09aa} compared to the best fit found using {\sc jktebop} and the
quadratic LD law. From top to bottom the light curves are $g$, $r$, $i$ and $z$.
The residuals are offset from zero to the base of the figure.} \end{figure}

OGLE-TR-L9 was discovered within the OGLE-II survey data \citep{Udalski++97aca} by \citet{Snellen+09aa}, and is a relatively massive planet orbiting a rapidly-rotating ($\Vsini = 39$\kms) F3\,V star. High-quality follow-up light curves were obtained by Snellen et al.\ using the newly-commissioned GROND instrument \citep{Greiner+08pasp} on the 2.2\,m telescope at ESO La Silla. GROND is a CCD imager which utilises dichroics to observe simultaneously in seven passbands (SDSS $griz$ and near-infrared $JHK$). In the case of OGLE-TR-L9 the $JHK$ data were too noisy to be useful, but the $griz$ data are of good quality (Fig.\,\ref{fig:oglel9:lc}).

The $griz$ observations have been analysed here (Tables A20 to A23). The $gri$ light curves are good enough to support LD fit/fix solutions but for the $z$ data both LD coefficients were held fixed. The parameters for the four light curves were combined to obtain the final photometric results (Table\,A24). Their agreement with those of \citet{Snellen+09aa} is not good -- $i$ and $r_{\rm A}$ are correlated and my solution corresponds to a significantly higher $i$ (3$\sigma$) and lower $r_{\rm A}$ and $r_{\rm b}$ (both 6$\sigma$).

The {\sc absdim} results for OGLE-TR-L9 (Table\,A25) are in reasonable agreement with those of \citet{Snellen+09aa}, which is surprising given the differences in the photometric parameters. One reason for this is that the {\sc absdim} solution is governed mainly by the observed \Teff\ (which has a relative uncertainty of 0.8\%) rather than by $r_{\rm A}$ (5\%). I find that OGLE-TR-L9\,b has one of the highest \Teq s ($2039 \pm 51$\,K) of any known TEP. Aside from its faintness, this planet is an excellent candidate for multicolour `transmission photometry' to detect variations in planetary radius with wavelength due to atmospheric opacity effects \citep{Fortney+08apj}. The system would benefit from additional spectroscopy to provide improved measurements of \FeH\ and $K_{\rm A}$. This subsection completes my analysis of the TEPs discovered from OGLE survey data.

%%%%%%%%%%%%%%%%%%%%%%%%%%%%%%%%%%%%%%%%%%%%%%%%%%%%%%%%%%%%%%%%%%%%%%%%%%%%%%%%%%%%%%%%%%%%%%%%%%%%%%%%%%%%%%%%%%%%%%%

\subsection{TrES-2}                                                                              \label{sec:teps:tres2}

\begin{figure} \includegraphics[width=\columnwidth,angle=0]{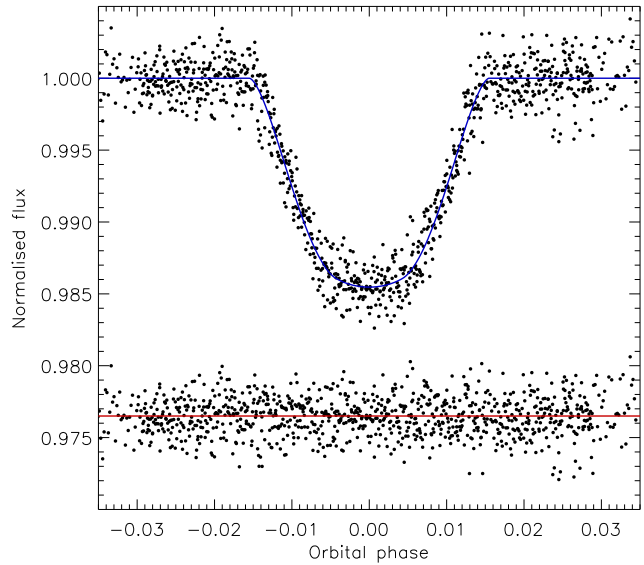}
\caption{\label{fig:tres2:lc} Phased light curve of the transits of TrES-2
from \citet{Holman+07apj} compared to the best fit found using {\sc jktebop}
and the quadratic LD law with the linear LD coefficient included as a fitted
parameter. The residuals of the fit are plotted at the bottom of the figure,
offset from zero.} \end{figure}

TrES-2 was the second TEP discovered by the Trans-Atlantic Exoplanet Survey \citep{Odonovan+06apj} and is both larger and more massive than Jupiter. Excellent ground-based light curves of three transits were obtained and studied by \citet{Holman+07apj}, in close collaboration with \citet{Sozzetti+07apj}. The relatively low orbital inclination of this system ($i = 83.8^\circ$) help $r_{\rm A}$ and $r_{\rm b}$ to be determined to a high accuracy.

The Holman $z$-band light curve of TrES-2 was studied in Paper\,I, but is revisited here because \citet{Daemgen+09aa} have since found a fainter star separated from TrES-2 by $1.089 \pm 0.008$ arcsec and with a magnitude difference of $\Delta z = 3.429 \pm 0.010$ (see Sect.\,\ref{sec:lc:l3}).

TrES-2 is of additional interest because \citet{MislisSchmitt09aa} found a possible decrease in the transit duration between their own and Holman's observations. This would most likely indicate that the orbital inclination is getting lower, which in turn points to the presence of a low-mass third body in the TrES-2 system. \citet{Scuderi+09xxx} obtained new data which did not confirm this hypothesis, but \citet{Mislis+09xxx} have since \reff{resurrected} the changing $i$. In addition, \citet{Rabus+09aa} presented a transit timing study of TrES-2 which found a small sinusoidal perturbation with a 0.2 day period, but with only moderate statistical significance. TrES-2 is in the field of view of the NASA {\it Kepler} satellite \citep{Koch+10xxx}, so a light curve of remarkable quality should become available for future investigations of these possibilities.

Here I reanalyse the $z$-band light curve of \citet{Holman+07apj}, this time with the incorporation of a third light value of $L_3 (z) = 0.0408 \pm 0.0004$. From the adopted \Teff\ of TrES-2\,A (Table\,\ref{tab:teps:synth}) and the magnitude differences in $i$ and $z$ \citep{Daemgen+09aa} I find that the fainter companion star has $\Teff = 4390 \pm 70$\,K (consistent with the $\sim$K5 spectral type found by \citealt{Daemgen+09aa}), and is substantially further away than TrES-2 so is not physically bound to the system.

The results of the {\sc jktebop} analysis are given in Table\,A26 and I adopt the LD fit/fix results. Correlated noise is not important. Table\,A27 shows a comparison with the results from Paper\,I, in which the analysis did not account for the faint companion star: $k$ has decreased by $1\sigma$ whilst $r_{\rm A}+r_{\rm b}$ and $r_{\rm A}$ become smaller by less than $0.5\sigma$. The best fit is shown in Fig.\,\ref{fig:tres2:lc}.

The physical properties of the TrES-2 system are summarised in Table\,A28. My results agree with literature studies within the errors, although $M_{\rm A}$ and $M_{\rm b}$ are larger by roughly 1$\sigma$. This can be traced back to the slightly smaller $r_{\rm A}$ found above. More precise spectroscopic \Teff\ and \FeH\ values would allow improved system parameters to be obtained, as would better photometry. This is currently being obtained by the {\it Kepler} satellite, and a stunning light curve of TrES-2 can already be inspected in \citet{Gilliland+10}.

%%%%%%%%%%%%%%%%%%%%%%%%%%%%%%%%%%%%%%%%%%%%%%%%%%%%%%%%%%%%%%%%%%%%%%%%%%%%%%%%%%%%%%%%%%%%%%%%%%%%%%%%%%%%%%%%%%%%%%%

\subsection{TrES-3}                                                                              \label{sec:teps:tres3}

\begin{figure} \includegraphics[width=\columnwidth,angle=0]{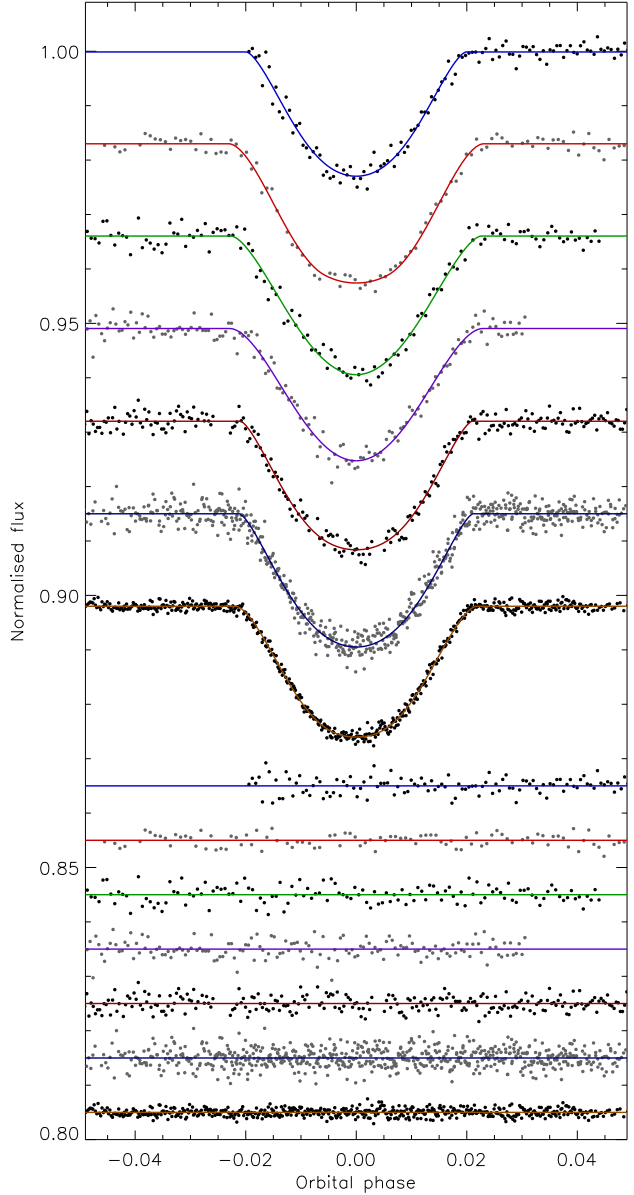}
\caption{\label{fig:tres3:lc} Phased light curves of the transits of TrES-3
compared to the best fits found using {\sc jktebop} and the quadratic LD law.
Successive datasets and residuals are offset in flux for display purposes. From
top to bottom are the $B$ and $z$ data \citep{Odonovan+07apj}, $Vgri$ observations
\citep{Sozzetti+09apj} and LT $V$$+$$R$ measurements \citep{Gibson+09apj}.} \end{figure}

TrES-3 is one of the more massive TEPs ($M_{\rm b} = 1.91$\Mjup) and orbits a rather cool star ($\Teff = 5650$\,K) with an orbital period of only 1.3\,d. It was identified as a TEP by \citet{Odonovan+07apj} and a discovery-quality light curve has also been obtained by the SuperWASP survey \citep{Cameron+07mn2}. Follow-up transit photometry has been presented by \citet{Sozzetti+09apj} and \citet{Gibson+09apj}, and occultation (secondary eclipse) observations have been secured by \citet{Winn+08aj}. A space-based light curve was obtained by the EPOXI mission \citep{Ballard+09iaus} but has not yet been published.

The \citet{Odonovan+07apj} $B$- and $z$-band photometry of TrES-3 is of good quality, and the transit observations of the follow-up papers \citep{Sozzetti+09apj,Gibson+09apj} are excellent. This, plus the fact that TrES-3 has a relatively low orbital inclination ($82^\circ$), means that the light curve parameters can be obtained to \reff{an unusually} high precision. In this work I analyse the $B$- and $z$-band data from \citet{Odonovan+07apj}, the $Vgri$-band observations from \citet{Sozzetti+09apj}, and the Liverpool Telescope (LT) RISE measurements from \citet{Gibson+09apj}. The last of the datasets, which has a passband of approximately $V$$+$$R$, was sorted in phase and then binned down by a factor of 20 (from 11\,350 to 568 datapoints) to ease the computational burden.

For the solutions of the seven light curves I adopt those with both LD coefficients fixed for $V$, $r$ and $i$, and the LD fit/fix solutions for the remainder (Fig.\,\ref{fig:tres3:lc}). Correlated noise is important for the $V$, $g$ and LT datasets. For most of the datasets I find that the residual-permutation algorithm returns errorbars which are significantly asymmetric, with larger upper errorbars than lower errorbars for both $r_{\rm A}$ and $r_{\rm b}$, but that the Monte Carlo errorbars are close to symmetric. This implies that the red noise in the light curves is biasing the results towards larger component radii, and would not have come to light if correlated noise was accounted for simply by rescaling the observational errors.

I accordingly end up with asymmetric errorbars for each light curve solution. The $V$ and $g$ solutions are quite uncertain and discrepant with the other results, so were rejected. To calculate the final photometric result I combined the solutions of the remaining five light curves by multiplying their probability density functions. The result is given in Table\,A36 and is found to be in good agreement with literature results.

The physical properties of the TrES-3 system are summarised in Table\,A37, and agree well with published values. For all five stellar model grids the best solution was found for zero age, and there is a possibility that edge effects will cause the uncertainties to be slightly underestimated for this object (for an example of this phenomenon see HD\,189733 in Paper\,I).

%%%%%%%%%%%%%%%%%%%%%%%%%%%%%%%%%%%%%%%%%%%%%%%%%%%%%%%%%%%%%%%%%%%%%%%%%%%%%%%%%%%%%%%%%%%%%%%%%%%%%%%%%%%%%%%%%%%%%%%

\subsection{TrES-4}                                                                              \label{sec:teps:tres4}

\begin{figure} \includegraphics[width=\columnwidth,angle=0]{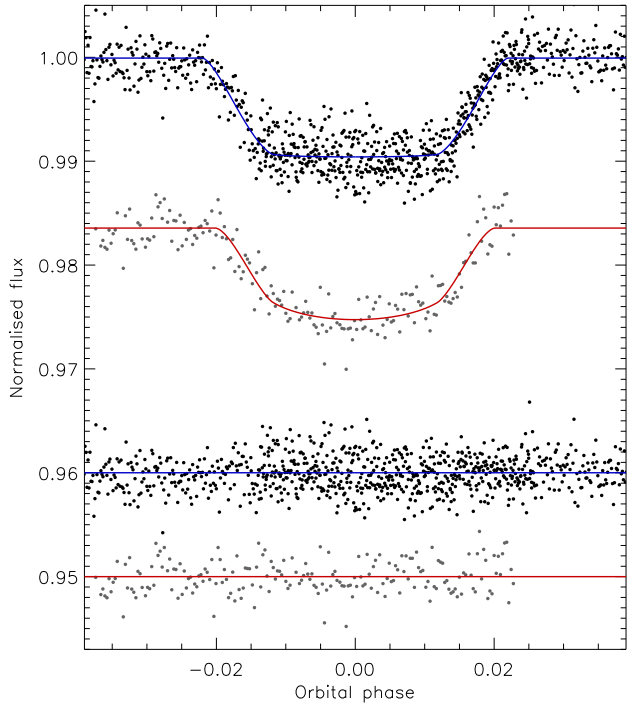}
\caption{\label{fig:tres4:lc} Phased light curves of the transits of TrES-4
compared to the best fits found using {\sc jktebop} and the quadratic LD law.
The upper data are in the $z$ band and the lower data are in the $B$ band.
The fits and residuals are offset in flux for display purposes.} \end{figure}

The planetary system TrES-4 was discovered by \citet{Mandushev+07apj}, and is noteworthy for having a very \reff{hot ($\Teq = 1861$\,K) and low-density (0.15\pjup) planet}. Revised physical properties of the system have been presented by TWH08 and \citet{Sozzetti+09apj}, and near-IR observations of the secondary eclipse show no evidence for orbital eccentricity \citep{Knutson+09apj}.

Like TrES-2 and WASP-2, high-resolution imaging observations by \citet{Daemgen+09aa} have detected a faint companion to TrES-4. It resides at an angular separation of $1.555 \pm 0.005$ arcsec and is fainter by $\Delta i = 4.560 \pm 0.017$ and $\Delta z = 4.232 \pm 0.025$ than TrES-4\,A. These result in third light contributions of $L_3(i) = 0.0150 \pm 0.0002$ and $L3(z) = 0.0199 \pm 0.0005$, which are taken into account in the light curve analysis as described in Sect.\,\ref{sec:lc:l3}.

In this work I study the high-precision $B$ and $z$ light curves presented in the discovery paper \citep{Mandushev+07apj}, and provide the first results to fully account for the third light contribution. Using {\sc atlas9} model spectra I have propagated the third light to the $B$-band, finding $L_3(B) = 0.0040 \pm 0.0003$ and a $\Teff = 4206 \pm 78$\,K for the companion star. The fainter star is more than twice as distant as TrES-4, so is not bound to the planetary system.

The $B$ data are rather sparse and do not allow LD coefficients to be fitted for (Fig.\,\ref{fig:tres4:lc}). The $z$ observations cover two transits so were studied with \Porb\ left as a fitted parameter; the LD fit/fix solutions were adopted. Correlated errors are important for both datasets. The two light curve solutions agree reasonably well ($1.5\sigma$) within the rather large errorbars, and were combined by weighted mean to find the final parameter values. This solution corresponds to a smaller $i$, and larger star and planet than found previously (Table\,A40). \reff{From Fig.\,\ref{fig:lc:l3:zero} we would expect that accounting for third light would lead to a smaller $r_{\rm A}$ and larger $r_{\rm b}$ and $i$, which is only partially concordant with the current situation. The variation in parameter values must therefore be due to the different analysis methods used.}

The physical properties of TrES-4 are given in Table\,A41 and conform to a more evolved star than found by previous studies, as expected for the larger $r_{\rm A}$ found above. The properties of the planet agree well with literature values, but are quite uncertain. The {\sc absdim} error budget shows that an improved light curve and additional RV measurements would benefit this system. All four TEPs discovered by TrES have now been analysed in the current series of papers.

%%%%%%%%%%%%%%%%%%%%%%%%%%%%%%%%%%%%%%%%%%%%%%%%%%%%%%%%%%%%%%%%%%%%%%%%%%%%%%%%%%%%%%%%%%%%%%%%%%%%%%%%%%%%%%%%%%%%%%%

\subsection{WASP-3}                                                                              \label{sec:teps:wasp3}

\begin{figure} \includegraphics[width=\columnwidth,angle=0]{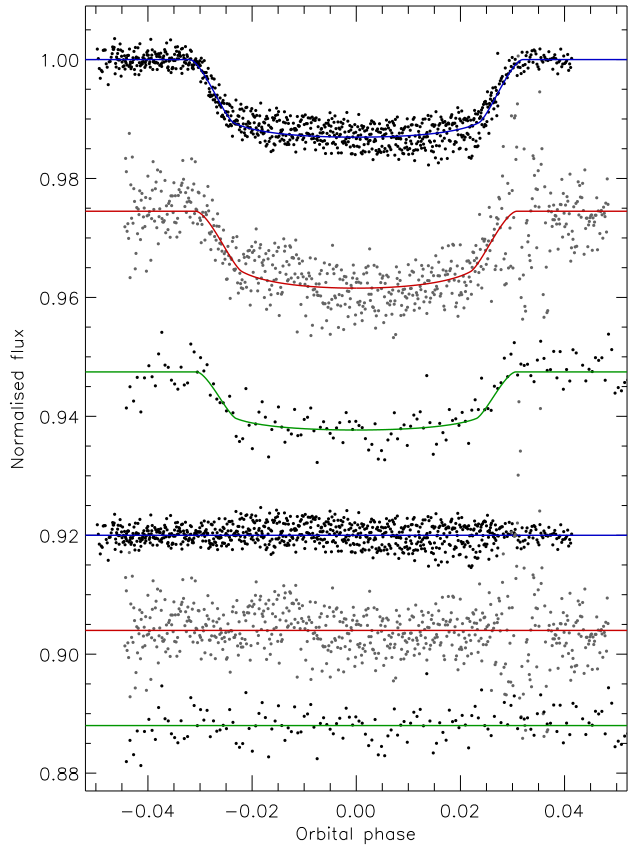}
\caption{\label{fig:wasp3:lc} Phased light curves of WASP-3 compared to the best
fits found using {\sc jktebop} and the quadratic LD law. From top to bottom the
light curves are LT $V$+$R$ \citep{Gibson+08aa}, which have been binned by a
factor of ten for analysis, Keele $R$ and IAC $I$ \citep{Pollacco+08mn}. The
residuals are offset from zero to the base of the figure.} \end{figure}

WASP-3 was identified as a possible TEP by \citet{Street+07mn} based on SuperWASP data \citep{Pollacco+06pasp}. Confirmation of its planetary nature was provided by \citet{Pollacco+08mn}, who presented four transit light curves obtained from various sources and found WASP-3\,b to be one of the most strongly irradiated TEPs ($\Teq = 2028$\,K). High-precision light curves from the LT have since been presented and analysed by \citet{Gibson+08aa}. Here I study these data, plus the Keele $R$-band and IAU 80\,cm $I$-band observations from \citet{Pollacco+08mn}. The other two light curves (IAC 80\,cm $V$ and SuperWASP) have either large systematics or a large scatter. I have binned consecutive sets of 10 datapoints of the LT light curve in order to limit CPU time; the sampling rate of the binned data is 30\,s.

For the LT data I adopt the LD fit/fix results and find significantly larger errorbars from the residual-permutation analysis; correlated noise is clearly visible in these data in Fig.\,\ref{fig:wasp3:lc}. The other two datasets also contain significant red noise, and the LD fixed solutions were adopted. The agreement between the three light curve solutions is excellent so they have been combined into a weighted mean (Table\,A45). Apart from a 2$\sigma$ larger $r_{\rm b}$, the final values agree well with the studies of \citet{Pollacco+08mn} and \citet{Gibson+08aa} but not with the preliminary results given by \citet{Damasso+09xxx}.

The physical properties of the WASP-3 system were originally calculated using $K_{\rm A} = 251.2 \pm 9.3$\kms\ \citep{Pollacco+08mn}. After this work was completed a revised $K_{\rm A}$ of $276.0 \pm 11.0$ was presented by \citet{Simpson+09}, who also found that the angle between the planetary orbit and the stellar spin was $\lambda = \er{15}{10}{9}\,^\circ$. Shortly before the current work was submitted a further study of WASP-3 was produced \citep{Tripathi+10}, containing new results including $K_{\rm A} = \er{290.5}{9.8}{9.2}$\kms\ and $\lambda = \er{3.3}{2.5}{4.4}\,^\circ$.

The third and most recent $K_{\rm A}$ has been used to obtain the physical properties of the WASP-3 system (Table\,A46). I find that WASP-3\,b has a rather larger mass and radius than most literature studies, except for that of \citet{Tripathi+10}. More precise measurements of \Teff\ and \FeH\ for WASP-3 would be useful to improve our understanding of its physical properties. More extensive spectroscopy would also be useful to pin down $K_{\rm A}$, for which a variety of measurements are currently available.

%%%%%%%%%%%%%%%%%%%%%%%%%%%%%%%%%%%%%%%%%%%%%%%%%%%%%%%%%%%%%%%%%%%%%%%%%%%%%%%%%%%%%%%%%%%%%%%%%%%%%%%%%%%%%%%%%%%%%%%

\subsection{WASP-10}                                                                            \label{sec:teps:wasp10}

\begin{figure} \includegraphics[width=\columnwidth,angle=0]{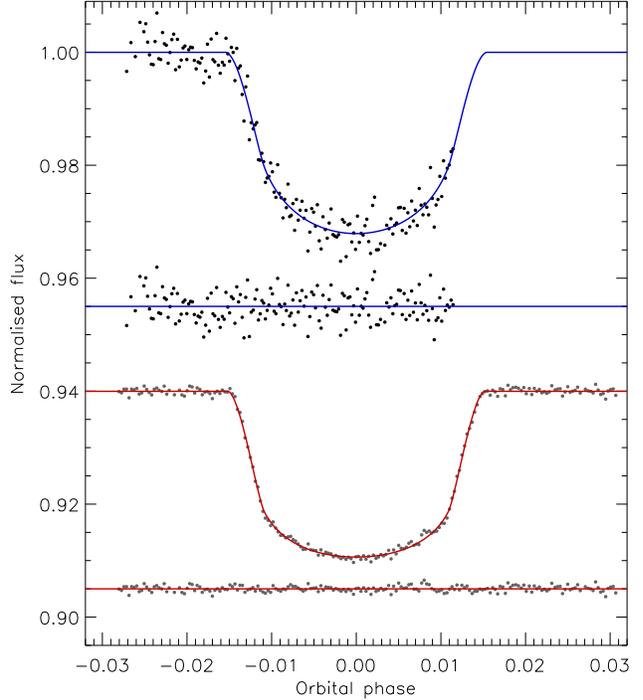}
\caption{\label{fig:wasp10:lc} Phased light curves of WASP-10 compared to the
best fit found using {\sc jktebop} and the quadratic LD law. The upper curve
and residuals represent the Mercator data \citep{Christian+09mn} and the lower
curve and residuals are the OPTIC data \citep{Johnson+09apj}. All offsets are
additive in flux.} \end{figure}

WASP-10 was found to be a TEP system by \citet{Christian+09mn}, and is notable for containing a fairly massive ($3.2$\Mjup) planet transiting a small ($0.70$\Rsun) and low-mass ($0.75$\Msun) star. The transit events are a generous 3.0\% deep, so photometric follow-up of this system is comparatively easy. \citet{Christian+09mn} obtained data from the 0.8\,m Tenagra and 1.2\,m Mercator telescopes, but unfortunately none of the datasets cover a full transit event. High-precision follow-up photometry of one complete transit of WASP-10 was obtained by \citet{Johnson+09apj}, using a novel orthogonal frame transfer CCD (OPTIC) to shape the point spread function and thus obtain a scatter of only 0.5\,mmag per datapoint at a reasonable sampling rate. I analyse the OPTIC observations to obtain the photometric parameters, and the Mercator data to provide a consistency check.

One complication for WASP-10 is its eccentric orbit. This is handled in the way described in Sect.\,\ref{sec:lc:ecc}, using the constraints $e \cos \omega = -0.045 \pm 0.02$ and $e \sin \omega = 0.023 \pm 0.04$ \citep{Johnson+09apj}. For both light curves I find that correlated noise is unimportant and that the LD fit/fix solutions are best (Fig.\,\ref{fig:wasp10:lc}). The two sets of results agree well (Table\,A49) and I adopt the OPTIC ones as final. The values agree well with those of \citet{Johnson+09apj} but my errorbars are rather larger, due in part to the inclusion of several different LD laws rather than the reliance on only one. The agreement with \citet{Christian+09mn} is less good but still acceptable.

The {\sc absdim} analysis returns results (Table\,A50) which are again in good agreement with those of \citet{Johnson+09apj} but with larger errorbars. The agreement between different model sets is unusually poor for WASP-10, resulting in systematic errorbars which are a significant fraction of the random errorbars and as large as the total errorbars given by \citeauthor{Johnson+09apj} Table\,A50 provides the first measurement of \safronov\ for WASP-10 and also corrects a calculation error in the \Teq\ listed in one of the published studies of this system. An improved photometric study of WASP-10 would settle the existing disagreement on its $k$ value, and the system would also be favoured by more precise \Teff\ and \FeH\ measurements.

After the above study of WASP-10 was completed, new datasets on this star were presented by \citet{Dittmann+10xxx} and \citet{Krejcova++10xxx} and a disagreement over the system properties became manifest. The two new studies both prefer system properties similar to those of \citet{Christian+09mn} but discrepant with the results -- based on a much better light curve -- of \citet{Johnson+09apj}. This will be revisited in the future, once the newer data and perhaps further observations become available.

%%%%%%%%%%%%%%%%%%%%%%%%%%%%%%%%%%%%%%%%%%%%%%%%%%%%%%%%%%%%%%%%%%%%%%%%%%%%%%%%%%%%%%%%%%%%%%%%%%%%%%%%%%%%%%%%%%%%%%%

\subsection{XO-2}                                                                                  \label{sec:teps:xo2}

\begin{figure} \includegraphics[width=\columnwidth,angle=0]{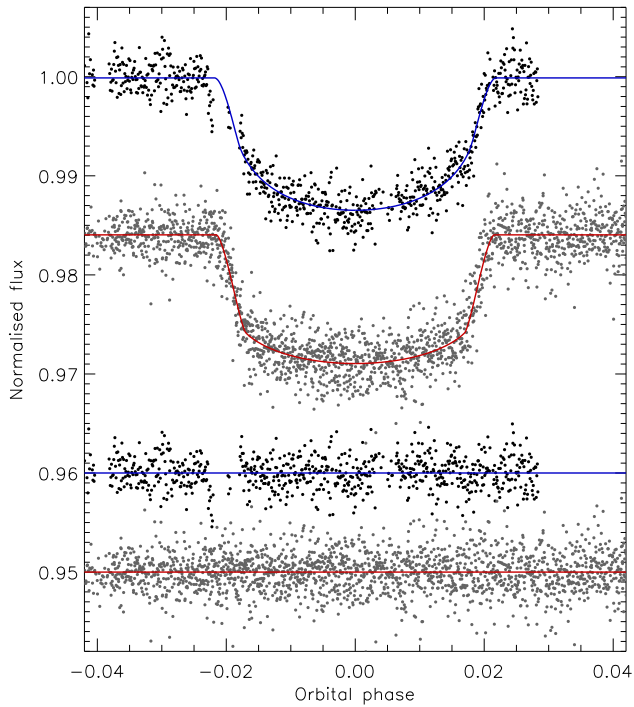}
\caption{\label{fig:xo2:lc} Phased light curves of XO-2 compared to the best
fits found using {\sc jktebop} and the quadratic LD law. The upper light curve
is $R$-band from the Perkins telescope \citep{Burke+07apj} and the lower
one is $z$-band from the FLWO 1.2\,m telescope \citep{Fernandez+09aj}. The
residuals are offset from zero to the base of the figure.} \end{figure}

The second planet discovered by the XO project \citep{Mccullough+06apj} is noteworthy for having a parent star which is very metal-rich ($\FeH = 0.45$) and a member of a common proper motion binary system \citep{Burke+07apj}. Good transit light curves from the Perkins telescope were published in the discovery paper \citep{Mccullough+06apj}, and from the FLWO 1.2\,m by the Transit Light Curve Project \citep{Fernandez+09aj}. The system has also been observed as part of the NASA EPOXI mission \citep{Ballard+09iaus}.

In this work I analyse the Perkins and FLWO observations (Fig.\,\ref{fig:xo2:lc}). In the former case correlated noise is important and in the latter it is not. I adopt the LD fit/fix results for both datasets and combine their probability density functions to find the final results. The solutions have $i \sim 90^\circ$ and thus asymmetric errorbars. The agreement between the two light curves and versus published values is good (Table\,A53).

The {\sc absdim} analysis is complicated by the high \FeH\ (Table\,A54). The results using the five different theoretical models are scattered, giving systematic errors which are larger than the statistical ones for the two most-affected quantities, $M_{\rm A}$ and $a$. However, the agreement with other studies is good. A more precise $K_{\rm A}$ value would be profitable.

%%%%%%%%%%%%%%%%%%%%%%%%%%%%%%%%%%%%%%%%%%%%%%%%%%%%%%%%%%%%%%%%%%%%%%%%%%%%%%%%%%%%%%%%%%%%%%%%%%%%%%%%%%%%%%%%%%%%%%%

\subsection{XO-3}                                                                                  \label{sec:teps:xo3}

\begin{figure} \includegraphics[width=\columnwidth,angle=0]{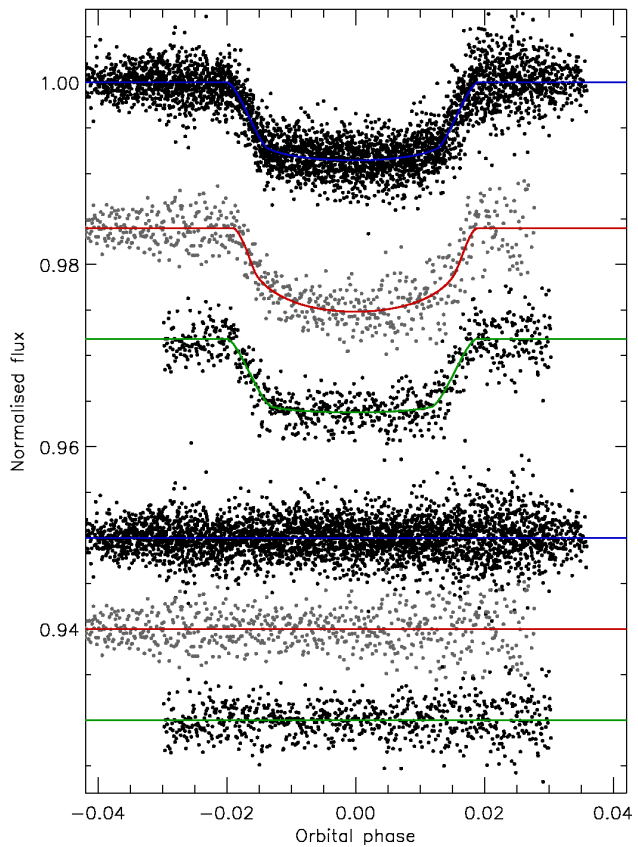}
\caption{\label{fig:xo3:lc} Phased light curves of XO-3 compared to the
best fits found using {\sc jktebop} and the quadratic LD law. From top to
bottom the light curves are KeplerCam $z$-band \citep{Winn+08apj2}, and
KeplerCam $r$-band and FLWO $I$-band \citep{Winn+09apj}. The residuals are
offset from zero.} \end{figure}

XO-3 was discovered by \citet{Johnskrull+08apj} to be a TEP which is so massive it is near the 13\Mjup\ value which represents the minimum mass of a brown dwarf. \citet{Johnskrull+08apj} presented two alternative sets of physical properties for the system, the first of which put XO-3\,b at $13.25 \pm 0.64$\Mjup\ but yielded a relatively poor fit to the observed transit light curve. The second set ignored the spectroscopic measurement of $\log g_{\rm A}$, \reff{yielding} $M_2 = 12.03 \pm 0.46$\Mjup\ and a much better fit to the light curve. The latter solution is preferable because light curve shapes are more reliable than spectroscopically-derived surface gravities for late-type dwarf stars.

A follow-up study of XO-3 was presented by \citet{Winn+08apj2}, based on high-quality new light curves and also ignoring the spectroscopic $\log g_{\rm A}$. Additional observations have been presented by \citet{Winn+09apj}. In this work I analyse the $z$-band photometry obtained by \citet{Winn+08apj2} using the FLWO 1.2\,m, and the $r$-band FLWO 1.2\,m and $I$-band Nickel datasets presented by \citet{Winn+09apj}. The photometric observations presented by \citet{Johnskrull+08apj} are not included because they comprise many datasets of only moderate quality.

An important property of XO-3 is its substantial orbital eccentricity ($e = 0.28$), which is a common feature of the more massive planets \citep{Me+09apj} and might indicate that they are a different population of objects to their less massive cousins. XO-3 is also known to exhibit a large spin-orbit misalignment \citep{Hebrard+09aa,Winn+09apj} suggestive of dynamical evolution through gravitational interactions \citep[e.g.][]{FabryckyWinn09apj}. In the following analysis I adopt the constraints $e = 0.2884 \pm 0.0035$ and $\omega = 346.3 \pm 1.3$, taken from \citet{Winn+09apj}. Compared to a solution assuming a circular orbit, $r_{\rm A}$ and $r_{\rm b}$ decrease by 0.0033 and 0.00029 respectively, which is less than 1$\sigma$ in both cases.

% with e/w:
% Fractional primary radius:                 0.1438919897
% Fractional secondary radius:               0.0131597161
% Chi-squared from errorbars:             3742.3216552776
%
% with e=w=0:
% Fractional primary radius:                 0.1471902423
% Fractional secondary radius:               0.0134518756
% Chi-squared from errorbars:             3742.3195207863

Correlated errors are unimportant for all three light curves, and in each case the best solutions are LD fit/fix. The $z$ and $I$ data agree well but the $r$ results have a higher $i$ and a 2.5$\sigma$ lower $r_{\rm A}$ (Table\,A58). The most extensive dataset is $z$, so I have combined the results from this and $I$ and rejected the $r$ results as discrepant. The best fits are plotted in Fig\,\ref{fig:xo3:lc}. My results agree well with those of \citet{Winn+09apj} and with the second of the two alternative solutions given by \citet{Johnskrull+08apj}.

When determining the physical properties of the XO-3 system I adopted increased uncertainties of 75\,K in \Teff\ and 0.05\,dex in \FeH, to allow for the possibility of systematic errors in these values for low-mass stars \reff{such as XO-3\,A}, and to account for the low spectroscopic gravity value ($3.95 \pm 0.06$ versus $4.23 \pm 0.04$) in the discovery paper \citep{Johnskrull+08apj}. Like the {\sc jktebop} outcome, the results of the {\sc absdim} analysis agree well with those of \citet{Winn+09apj} but not with the preferred solution of \citet{Johnskrull+08apj}. The mass of the planet is $M_{\rm b} = 11.8 \pm 0.5$\Mjup, which is close to but below the 13\Mjup\ dividing line between planets and brown dwarfs. The {\it VRSS} model results disagree strongly with those of the other models, so were not included in the final analysis. The \Teq\ of XO-3\,b is high at $1729 \pm 34$\,K, making it an interesting object for the study of planetary atmospheres. Aside from the {\it VRSS} models, XO-3 is one of the best-measured TEPs. An improved spectroscopic study, incorporating the best $\log g_{\rm A}$ value given in Table\,A59, would be the best way of improving this understanding even further.

%%%%%%%%%%%%%%%%%%%%%%%%%%%%%%%%%%%%%%%%%%%%%%%%%%%%%%%%%%%%%%%%%%%%%%%%%%%%%%%%%%%%%%%%%%%%%%%%%%%%%%%%%%%%%%%%%%%%%%%

\subsection{XO-4}                                                                                  \label{sec:teps:xo4}

\begin{figure} \includegraphics[width=\columnwidth,angle=0]{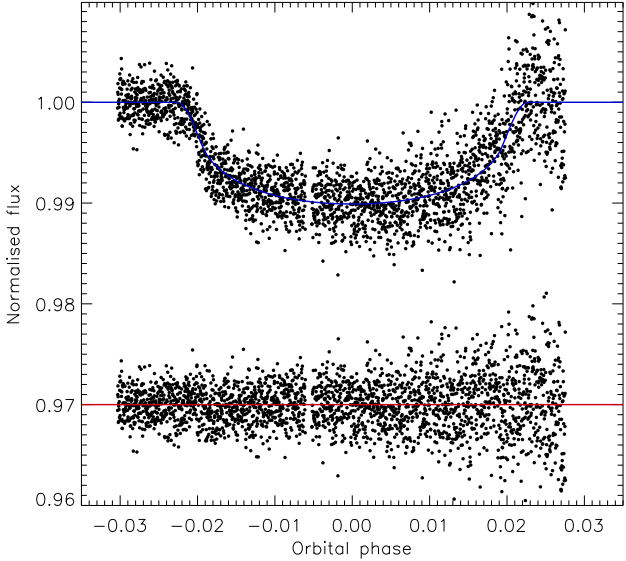}
\caption{\label{fig:xo4:lc} Phased light curve of XO-4 from the Perkins
telescope \citep{Mccullough+08xxx}. The blue line shows the best fit
from {\sc jktebop} using the quadratic LD law. The residuals are offset
from zero.} \end{figure}

XO-4 was discovered to be a TEP by \citet{Mccullough+08xxx}; the parent star is one of the hottest of the known planetary hosts. Here I analyse the $R$-band Perkins telescope light curve obtained by \citet{Mccullough+08xxx}. Correlated errors are unimportant and the best solutions are LD fit/fix. The inclination is near 90$^\circ$, resulting in asymmetric errorbars. My photometric solution (Fig.\,\ref{fig:xo4:lc}) is in excellent agreement with that of \citeauthor{Mccullough+08xxx} (Table\,A61).

The results of the {\sc absdim} analysis are given in Table\,A62, and include the first reported measurements of the \Teq, Safronov number, $g_{\rm b}$, $\rho_{\rm A}$ and $\rho_{\rm b}$ of the XO-4 system. The other output parameters agree well with those of \citet{Mccullough+08xxx}; $M_{\rm b}$ is 0.2\Mjup\ smaller in my solution but this is within the errorbars. Improved photometric and RV observations of XO-4 would be beneficial.

%%%%%%%%%%%%%%%%%%%%%%%%%%%%%%%%%%%%%%%%%%%%%%%%%%%%%%%%%%%%%%%%%%%%%%%%%%%%%%%%%%%%%%%%%%%%%%%%%%%%%%%%%%%%%%%%%%%%%%%

\subsection{XO-5}                                                                                  \label{sec:teps:xo5}

\begin{figure} \includegraphics[width=\columnwidth,angle=0]{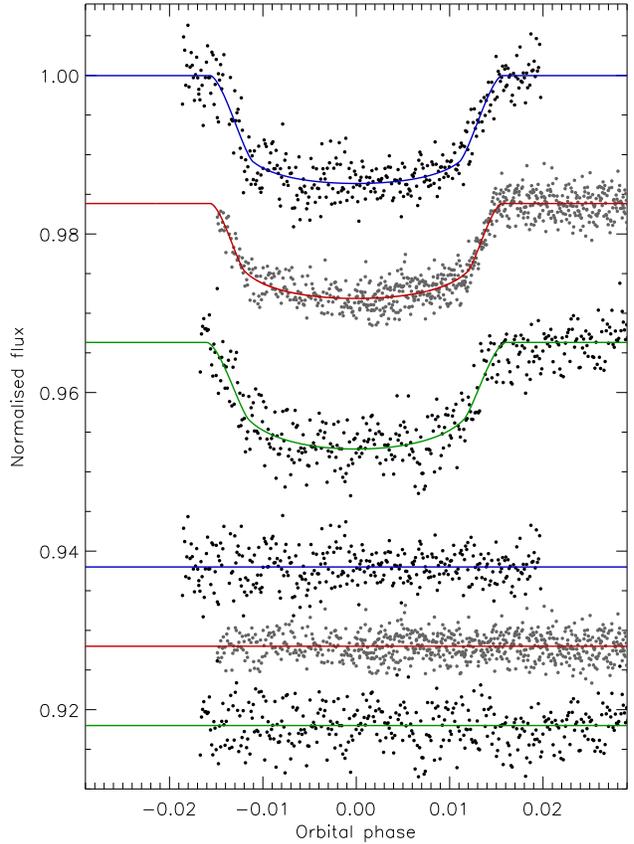}
\caption{\label{fig:xo5:lc} Phased light curves of XO-5 compared to the
best fits found using {\sc jktebop} and the quadratic LD law. From top to
bottom the light curves are Perkins $R$-band \citep{Burke+08apj}, and
KeplerCam $i$-band and $z$-band \citep{Pal+09apj}. The residuals are
offset from zero.} \end{figure}

The discovery that XO-5 is a TEP system was presented by \citet{Burke+08apj} and extensive follow-up observations in the context of the HAT consortium were presented by \citet{Pal+09apj}. I adopt the $K_{\rm A}$ value from \citet{Pal+09apj}, which agrees with but is much more precise than that given by \citet{Burke+08apj}. A comparison of the \Teff\ and \FeH\ given by the two studies shows a concerning disagreement, which may be due to the treatment of $\log g_{\rm A}$. \citet{Pal+09apj} fix $\log g_{\rm A}$ at the transit-derived value during their spectral synthesis analysis, and also have spectra with a higher signal to noise ratio, so I have preferred their spectroscopic results.

Three high-quality light curves of XO-5 are available: $R$-band coverage of one full transit using the Perkins telescope \citep{Burke+08apj}, and $i$- and $z$-band observations (both covering one full and one partial transit) obtained with the FLWO 1.2\,m telescope \citep{Pal+09apj}. I have analysed all three datasets, and the best fits are shown in Fig.\,\ref{fig:xo5:lc}. In the case of the $R$ and $z$ data correlated noise is important, and in all cases the best solutions are LD fit/fix. The resulting photometric parameters are in good agreement except for $k$, for which there is a large scatter of 5$\sigma$ (Table\,A66). The three parameter sets have therefore been combined and the errorbar in $k$ increased to account for this discrepancy.

The physical properties of XO-5 are given in Table\,A67, and are in good agreement with established values. This system is a good candidate for improved photometric observations, which would allow to sort out the discrepancy in $k$ and improve the precision of the system properties.

%%%%%%%%%%%%%%%%%%%%%%%%%%%%%%%%%%%%%%%%%%%%%%%%%%%%%%%%%%%%%%%%%%%%%%%%%%%%%%%%%%%%%%%%%%%%%%%%%%%%%%%%%%%%%%%%%%%%%%%

\subsection{TEPs studied in previous papers}                                                      \label{sec:teps:done}

The preceding subsections have presented full studies of fifteen TEPs. In this subsection I apply my improved {\sc absdim} analysis to the other fifteen systems for which photometric results have already been calculated (Paper\,I; \citealt{Me+09mn,Me+09mn2,Me+09apj,Me+10}). Their physical properties are collected and compared to literature results in Tables A68 to A81, and relevant points are discussed below.

{\it GJ\,436.} Attempts to obtain solutions for $\Teff = 3350$\,K failed. Inspection of Fig.\,\ref{fig:model:mrt} explains this: for a 0.5\Msun\ star we expect \Teff\ values in the range 3500--4000\,K (except for the {\it Claret} models which prefer 3100\,K). I have therefore adopted the higher \Teff\ of 3500\,K from \citet{Bean++06apj}, with an errorbar doubled to 100\,K. The results with the {\it Claret} models are discrepant so were not incorporated into the final solution. This does not mean that the {\it Claret} models are wrong -- they are in fact closer to the measured properties of eclipsing binaries (primarily CM\,Dra) in this mass regime -- but that they depart from the consensus established by the other model sets. The lesson here is that we require an improved understanding of low-mass stars to better measure the physical properties of the important GJ\,436 system. The revised results for GJ\,436 (Table\,A68) are slightly smaller than those found in Paper\,II, and are in good agreement with literature studies.

{\it HAT-P-1.} \ I used a new $K_{\rm A}$ value from \citet{Johnson+08apj}. My {\sc absdim} solutions prefer the \Teff\ value used in Paper\,II \citep{Bakos+07apj} to the higher one found by \citet{Ammler+09aa}.

{\it HD\,189733.} \ A new $K_{\rm A}$ value is available from \citet{Boisse+09aa}, supported by the value given by \citet{Triaud+09aa}. The {\sc absdim} solutions with different models converged on either a very young ($\sim$1\,Gyr) or old (9--13\,Gyr) age, resulting in large systematic errorbars for most output parameters. HD\,189733 is an active star \citep{Bouchy+05aa} with starspots \citep{Pont+07aa2}, a short rotation period of 12\,d \citep{Winn+07aj}, RV jitter \citep{Boisse+09aa}, and Ca\,H\,\&\,K emission modulated on its rotation period \citep{Moutou+07aa}. These facts imply a young age \citep{Skumanich72apj,MamajekHillenbrand08apj} so the search for the best solution was restricted to ages below 5\,Gyr. This resulted in much more consistent solutions, with ages of 0.1--2.9\,Gyr, which were accepted as the final results (Table\,A70).

{\it HD\,209458.} \ I used the revised value of $K_{\rm A} = 84.67 \pm 0.70$ given by TWH08. The revised results are in good agreement with those from Paper\,II. $R_{\rm A}$ has been measured directly by interferometric means \citep{BelleBraun09apj} to be $2\sigma$ larger than found here and in all other studies of this object given in Table\,A71.

{\it OGLE-TR-10.} \ The light curve solution used in Paper\,II was unintentionally a preliminary rather than the final one from Paper\,I. This has been corrected here, resulting in a less dense planet. The properties of OGLE-TR-10 are rather uncertain and the system is badly in need of improved spectroscopy and photometry.

{\it OGLE-TR-132.} \ I use the more precise $K_{\rm A}$ value of $167 \pm 18$\kms\ given by \citet{Moutou+04aa} which was overlooked in Papers I and II. OGLE-TR-132\,b has a high \Teq\ of $2017 \pm 97$\,K (Table\,75).

{\it WASP-2.} \ The results for this system are reproduced from the dedicated study of \citet{Me+10}.

{\it WASP-4, WASP-5, WASP-18.} \ \reff{I use the new and improved $K_{\rm A}$ values for these three TEPs from \citet{Triaud+10aa}.}

%%%%%%%%%%%%%%%%%%%%%%%%%%%%%%%%%%%%%%%%%%%%%%%%%%%%%%%%%%%%%%%%%%%%%%%%%%%%%%%%%%%%%%%%%%%%%%%%%%%%%%%%%%%%%%%%%%%%%%%

\section{Tracking the systematic errors in the properties of transiting systems}                     \label{sec:syserr}

\begin{figure*} \includegraphics[width=\textwidth,angle=0]{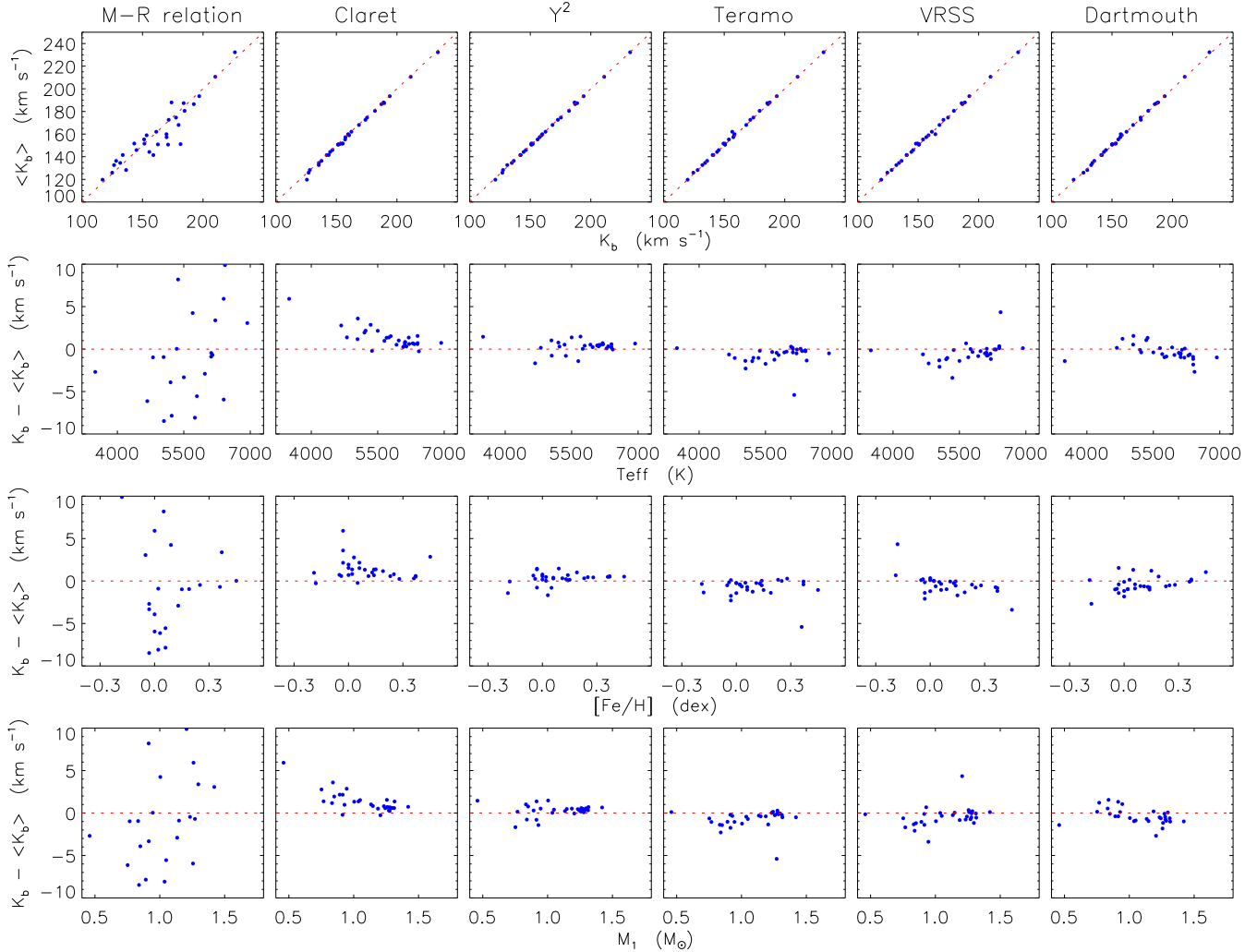}
\caption{\label{fig:syserr} Comparisons between the $K_{\rm b}$ values obtained
using specific sets of stellar evolutionary models and the unweighted mean value,
$\langle$$K_{\rm b}$$\rangle$, for each TEP. From left to right the panels show
results for the mass--radius relation and then the five stellar model sets. The
top panels compare $K_{\rm b}$ to $\langle$$K_{\rm b}$$\rangle$ for each model
set, with parity indicated by a dotted line. Lower panels show the difference
($K_{\rm b} - \langle$$K_{\rm b}$$\rangle$) as functions of effective temperature,
metal abundance and stellar mass. Errorbars have been ignored for clarity; their
median values are $\pm3$\kms\ (statistical) and $\pm1.4$\kms\ (systematic).}
\end{figure*}

A major result of the current work is a detailed understanding of where model-dependent systematic errors surface in the analysis of TEPs, and the importance of these systematics for the various physical properties which can be calculated. The approach used in this work means that all of the model dependence is combined into $K_{\rm b}$ (the velocity amplitude of the planet), making it an excellent tracer of systematic errors. For each TEP a separate value of $K_{\rm b}$ is found using each of the five sets of stellar models, as well as for the empirical mass--radius relation (Paper\,II). I have converted these into `consensus values', $\langle$$K_{\rm b}$$\rangle$, using the same algorithm as for the other measured physical properties: the unweighted mean of the values from the five different stellar model sets. Remember that $\langle$$K_{\rm b}$$\rangle$ should not be taken as an indicator of {\em correctness}, only of concordance.

From Eqs.\ 4 to 12 in Paper\,II it can be seen that the component masses are most sensitive to systematic errors ($M_{\rm A} \propto K_{\rm b}^{\ 3}$ and $M_{\rm b} \propto K_{\rm b}^{\ 2}$) and that their radii, gravities and densities are less so (all directly proportional to $K_{\rm b}$). The semimajor axis is also rather model-dependent as the other quantities required to calculate it all have much smaller uncertainties than $K_{\rm b}$ does.

A detailed exploration of model-dependence is presented in Fig.\,\ref{fig:syserr} as a function of \Teff, \FeH\ and $M_{\rm A}$. The top panels in this Figure highlight the generally good agreement between different model sets. The mass--radius relation is in much poorer agreement and is biased to high values as it does not account for the effects of evolution through the main-sequence phase (particularly apparent for HAT-P-2, TrES-4 and XO-3). This bias to large $K_{\rm b}$ comes from trying to reproduce the low density of an evolved star, and pushes the mass and radius of the planet and star to high and incorrect values.

Turning to the stellar model sets in Fig.\,\ref{fig:syserr}, clear trends with respect to \Teff, \FeH\ and $M_{\rm A}$ can be seen in many cases. The {\it Claret} models yielded larger $K_{\rm b}$ values on average, particularly for low values of \Teff\ and \FeH. This predilection for high values is balanced by other model sets: the {\it Teramo} and {\it VRSS} models both tend to produce lower $K_{\rm b}$ values and the {\it DSEP} models trend to low $K_{\rm b}$ for cooler and more metal-poor stars. Within this m\'elange, the {\it Y$^2$} models are the closest to the consensus value and do not exhibit significant trends with the stellar properties. The {\it Y$^2$} models are thus the best choice to obtain quick results, although of course other model sets are required for the assessment of systematic errors.

\begin{figure} \includegraphics[width=\columnwidth,angle=0]{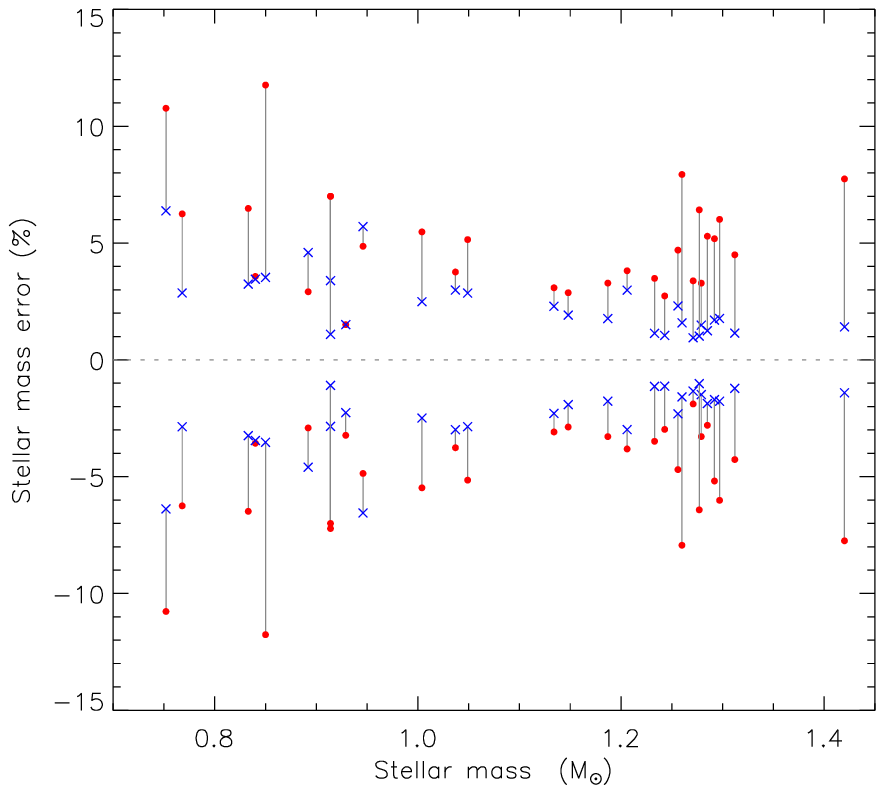}
\caption{\label{fig:syserr:m1sys} Plot of the sizes of the statistical
(red filled circles) and systematic (blue crosses) errorbars for the
TEPs studied in this work, versus the stellar masses. The errors
are plotted as fractions of $M_{\rm A}$ and the statistical and systematic
errors for each TEP are connected by grey lines.} \end{figure}

\begin{figure} \includegraphics[width=\columnwidth,angle=0]{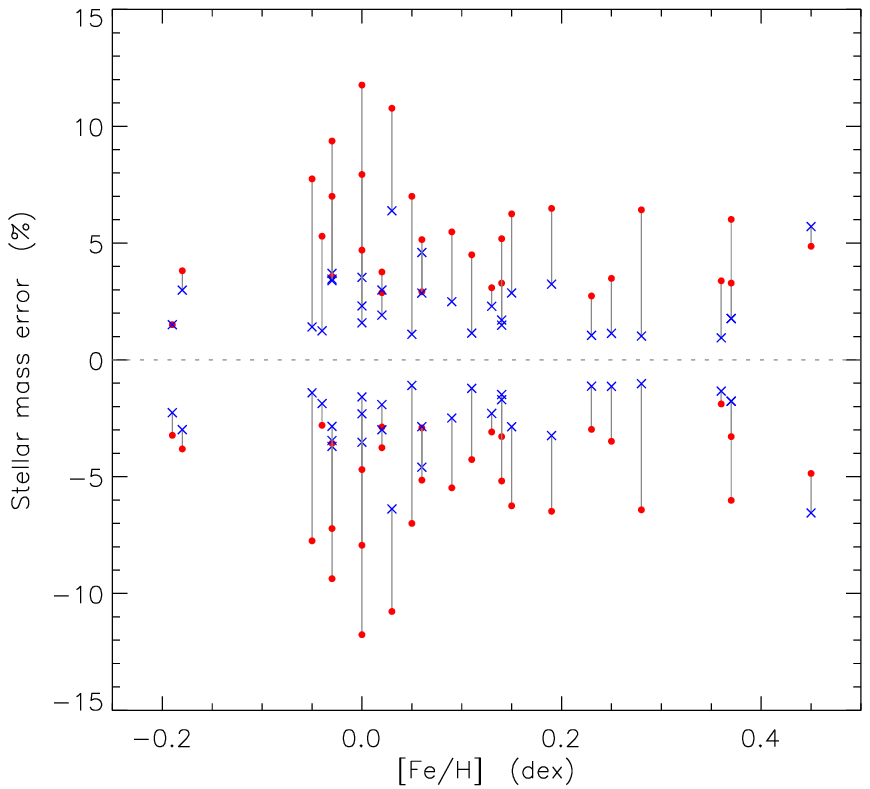}
\caption{\label{fig:syserr:fehsys} Same as Fig.\,\ref{fig:syserr:m1sys},
except that the error sizes are plotted versus stellar \FeH\ rather
than mass.} \end{figure}

Figs. \ref{fig:syserr:m1sys} and \ref{fig:syserr:fehsys} compare the sizes of the random and systematic errors in stellar mass, as a function of $M_{\rm A}$ and of \FeH, respectively. $M_{\rm A}$ was chosen for this comparison because it is one of the parameters more affected by systematics; it was also expected that these systematic effects would be minimised in the region of 1\Msun\ and $\FeH = 0$ as all of the stellar model sets are calibrated on the Sun. Surprisingly, this does not turn out to be the case. There is a hard lower limit of 1\% on the errors in $M_{\rm A}$ arising from model-dependent systematics, but this lower limit is reached at a wide range of masses and \FeH\ values. The systematic errors are clearly larger for lower-mass stars ($<$0.9\Msun) and for a high metal abundance ($\FeH > 0.4$). The hard lower limit is unavoidable until stellar model sets are in much better agreement, and the mass--radius \reff{relation} results hint that the real systematic errors are probably somewhat higher than this. The random errors in $M_{\rm A}$ tend to decrease towards higher masses (temporarily ignoring the much fainter OGLE systems), as expected because more massive stars are intrinsically brighter and therefore easier to obtain good data for.

\subsection{An external test: WASP-11 versus HAT-P-10}

One of the best ways to investigate the presence of systematic errors is to compare two independent studies of the same object. Whilst multiple discovery and characterisation publications have been presented for several TEPs (e.g.\ XO-1 and XO-5), successive papers on these objects have in each case been informed by the initial discovery papers. There are two exceptions: HD\,80606 and WASP-11\,/\,HAT-P-10. In the case of HD\,80606 three groups discovered its transiting nature essentially simultaneously, but their analyses were heavily dependent on the same published spectroscopic observations. WASP-11\,/\,HAT-P-10 is therefore the only TEP which was discovered and fully characterised by two groups working without knowledge of each others' analyses. The discovery papers were submitted within a week of each other, but the agreed name of the system is WASP-11\,/\,HAT-P-10 because the WASP group submitted their paper first.

\begin{table} \caption{\label{tab:syserr:wasp11} Physical properties
of the WASP-11\,/\,HAT-P-14 system determined by the two discovery
papers. Aside from the orbital period, quantities without uncertainties
were calculated from the results given in the respective papers.}
\begin{tabular}{l r@{\,$\pm$\,}l r@{\,$\pm$\,}l} \hline \hline
Parameter               & \mc{\citet{West+09aa}}      & \mc{\citet{Bakos+09apj}}    \\
\hline
\Porb\ (d)              & \mc{3.722465}               & \mc{3.7224747}              \\
\Teff\ (K)              & 4800 & 100                  & 4980 & 60                   \\
\Vsini\ (\kms)          & \mc{$<$ 6}                  & 0.5 & 0.2                   \\
\MoH,\FeH\ (dex)        & 0.0 & 0.2                   & 0.13 & 0.08                 \\[2pt]
$K_{\rm A}$ (\ms)       & 82.1 & 7.4                  & 74.5 & 1.8                  \\
$e$                     & \mc{0 fixed}                & \mc{0 fixed}                \\
$a$ (AU)                & 0.043 & 0.002               & 0.0435 & 0.0006             \\[2pt]
$r_{\rm A}$             & \mc{0.0801}                 & 0.0842 & 0.0019             \\
$r_{\rm b}$             & \mc{0.01011}                & \mc{0.01104}                \\
$k$                     & \erc{0.1273}{0.0011}{0.0008}& 0.1313 & 0.0010             \\
$i$ ($^\circ$)          & \erc{89.8}{0.2}{0.8}        & \erc{88.6}{0.5}{0.4}        \\[2pt]
$M_{\rm A}$ (\Msun)     & \erc{0.77}{0.10}{0.08}      & 0.83 & 0.03                 \\
$R_{\rm A}$ (\Rsun)     & \erc{0.74}{0.04}{0.03}      & 0.79 & 0.02                 \\
$\logg_{\rm A}$ [cgs]   & 4.45 & 0.2                  & 4.56 & 0.02                 \\[2pt]
$M_{\rm b}$ (\Mjup)     & 0.53 & 0.07                 & 0.460 & 0.028               \\
$R_{\rm b}$ (\Rjup)     & \erc{0.91}{0.06}{0.03}      & \erc{1.005}{0.032}{0.027}   \\
$g_{\rm b}$ (\mss)      & \erc{14.5}{1.4}{1.6}        & 12.0 & 0.8                  \\
$\rho_{\rm b}$ (\pjup)  & \erc{0.69}{0.07}{0.11}      & 0.479 & 0.042               \\
\Teq\ (K)               & 960 & 70                    & 1020 & 17                   \\
\hline \hline \end{tabular} \end{table}

In Table\,\ref{tab:syserr:wasp11} I collect the properties of WASP-11 determined in the two discovery papers \citep{West+09aa,Bakos+09apj}. The values in general show a gratifying agreement, but two quantities stand out as being discrepant \reff{to some extent}. The \Teff\ values disagree by $1.6\sigma$ and the $k$ values by $2.7\sigma$. The divergent $k$ values have a large knock-on effect on the planetary properties ($R_{\rm b}$, $g_{\rm b}$ and $\rho_{\rm b}$).

The possibility of systematic errors in \Teff\ is a well-known phenomenon (see Sect.\,\ref{sec:model}), so the $1.6\sigma$ disagreement is unsurprising. Similarly, $k$ measurements primarily depend on the observed transit depth (see Sect.\,\ref{sec:lc}) and can be affected by imperfect normalisation as well as by astrophysical effects such as starspots. Discrepancies in $k$ have previously been found in the well-studied systems HD\,189733 and HD\,209458 (Paper\,I), WASP-4 \citep{Me+09mn2}, HAT-P-1 and WASP-10 (Sections \ref{sec:teps:hatp1} and \ref{sec:teps:wasp10}).

This implies that a high-quality study of a TEP cannot rely on photometric coverage of only one transit, irrespective of its quality, but must include observations of two or preferably more in order to pick up on well-camouflaged systematic errors affecting light curves. It is important to realise that a systematic over- or under-estimation of a transit depth has a big effect on measurements of the planet properties, but cannot be identified in any dataset covering only one transit.

%%%%%%%%%%%%%%%%%%%%%%%%%%%%%%%%%%%%%%%%%%%%%%%%%%%%%%%%%%%%%%%%%%%%%%%%%%%%%%%%%%%%%%%%%%%%%%%%%%%%%%%%%%%%%%%%%%%%%%%

\section{Physical properties of the transiting extrasolar planetary systems}                         \label{sec:absdim}

\begin{table*} \caption{\label{tab:absdim:stars} Physical properties of the stellar
components of the TEPs studied in this work. For each quantity the first uncertainty
is derived from a propagation of all observational errors and the second uncertainty
is an estimate of the systematic errors arising from the dependence on stellar theory.}
\setlength{\tabcolsep}{3pt}
\begin{tabular}{l l@{\,$\pm$\,}l@{\,$\pm$\,}l l@{\,$\pm$\,}l@{\,$\pm$\,}l l@{\,$\pm$\,}l@{\,$\pm$\,}l l@{\,$\pm$\,}l@{\,$\pm$\,}l l@{\,$\pm$\,}l@{\,$\pm$\,}l l@{\,$\pm$\,}l@{\,$\pm$\,}l}
\hline \hline
System & \mcc{Semimajor axis (AU)} & \mcc{Mass (\Msun)} & \mcc{Radius (\Rsun)} & \mcc{$\log g_{\rm A}$ [cm/s]} & \mcc{Density (\psun)} & \mcc{Age (Gyr)} \\
\hline
GJ\,436     & 0.02887   & 0.00089   & 0.00035     & 0.459     & 0.043     & 0.017       & 0.454     & 0.029     & 0.005       & 4.787     & 0.030     & 0.005       & \mcc{$4.92 \pm 0.55$}               & \mc{unconstrained}                      \\
HAT-P-1     & 0.05535   & 0.00057   & 0.00042     & 1.134     & 0.035     & 0.026       & 1.112     & 0.031     & 0.008       & 4.400     & 0.024     & 0.003       & \mcc{$0.824 \pm 0.066$}             & \ermcc{ 2.1}{ 1.4}{ 1.2}{ 0.5}{ 0.6}    \\
HAT-P-2     & 0.06740   & 0.00074   & 0.00034     & 1.279     & 0.042     & 0.019       & 1.68      & 0.15      & 0.01        & 4.092     & 0.074     & 0.002       & \mcc{$0.268 \pm 0.070$}             & \ermcc{ 2.6}{ 0.4}{ 0.7}{ 0.5}{ 0.3}    \\
HD\,149026  & \ermcc{0.04288}{0.00048}{0.00027}{0.00013}{0.00019} & \ermcc{1.271}{0.043}{0.024}{0.012}{0.017} & \ermcc{1.290}{0.120}{0.058}{0.004}{0.006}  & \ermcc{4.321}{0.042}{0.070}{0.001}{0.002}      & \ercc{0.592}{0.083}{0.129} & \ermcc{ 1.2}{ 1.1}{ 1.5}{ 0.3}{ 0.1} \\
HD\,189733  & 0.03142   & 0.00038   & 0.00036     & 0.840     & 0.030     & 0.029       & 0.752     & 0.023     & 0.009       & 4.610     & 0.026     & 0.005       & \mcc{$1.98 \pm 0.17$}               & \ermcc{ 1.4}{ 4.7}{ 1.4}{ 1.5}{ 1.3}    \\
HD\,209458  & 0.04747   & 0.00046   & 0.00031     & 1.148     & 0.033     & 0.022       & 1.162     & 0.012     & 0.008       & 4.368     & 0.005     & 0.003       & \mcc{$0.733 \pm 0.008$}             & \ermcc{ 2.3}{ 0.9}{ 0.7}{ 0.5}{ 0.4}    \\
OGLE-TR-10  & 0.04516   & 0.00099   & 0.00015     & 1.277     & 0.082     & 0.013       & 1.52      & 0.10      & 0.00        & 4.178     & 0.053     & 0.001       & \mcc{$0.361 \pm 0.063$}             & \ermcc{ 3.1}{ 3.3}{ 0.7}{ 0.3}{ 0.2}    \\
OGLE-TR-56  & 0.02386   & 0.00028   & 0.00009     & 1.233     & 0.043     & 0.014       & 1.26      & 0.14      & 0.00        & 4.331     & 0.094     & 0.002       & \mcc{$0.62 \pm 0.21$}               & \ermcc{ 1.7}{ 1.4}{ 2.0}{ 0.3}{ 0.2}    \\
OGLE-TR-111 & 0.04651   & 0.00099   & 0.00051     & 0.833     & 0.054     & 0.027       & 0.842     & 0.042     & 0.009       & 4.508     & 0.044     & 0.005       & \mcc{$1.40 \pm 0.19$}               & \mc{unconstrained}                      \\
OGLE-TR-113 & 0.02278   & 0.00047   & 0.00022     & 0.768     & 0.048     & 0.022       & 0.780     & 0.029     & 0.008       & 4.539     & 0.028     & 0.004       & \mcc{$1.62 \pm 0.13$}               & \mc{unconstrained}                      \\
OGLE-TR-132 & 0.03029   & 0.00062   & 0.00018     & 1.297     & 0.078     & 0.023       & 1.37      & 0.14      & 0.01        & 4.275     & 0.083     & 0.003       & \mcc{$0.50 \pm 0.15$}               & \ermcc{ 1.5}{ 4.2}{ 1.5}{ 0.3}{ 0.4}    \\
OGLE-TR-182 & 0.05205   & 0.00057   & 0.00031     & 1.187     & 0.039     & 0.021       & 1.53      & 0.17      & 0.01        & 4.142     & 0.089     & 0.003       & \mcc{$0.33 \pm 0.10$}               & \ermcc{ 4.3}{ 0.5}{ 1.9}{ 1.4}{ 1.3}    \\
OGLE-TR-211 & \ermcc{0.05105}{0.00076}{0.00073}{0.00020}{0.00021} & \ermcc{1.312}{0.059}{0.056}{0.015}{0.016} & \ermcc{1.56}{0.18}{0.10}{0.01}{0.01}       & \ermcc{4.170}{0.052}{0.085}{0.002}{0.002}      & \ercc{0.345}{0.068}{0.090} & \ermcc{ 2.6}{ 0.6}{ 0.7}{ 0.4}{ 0.3} \\
OGLE-TR-L9  & 0.0404    & 0.0011    & 0.0002      & 1.42      & 0.11      & 0.02        & 1.503     & 0.083     & 0.008       & 4.236     & 0.043     & 0.002       & \mcc{$0.418 \pm 0.061$}             & \ermcc{ 1.0}{ 0.6}{ 0.7}{ 0.3}{ 0.2}    \\
TrES-1      & 0.03946   & 0.00039   & 0.00060     & 0.892     & 0.026     & 0.041       & 0.818     & 0.017     & 0.013       & 4.563     & 0.019     & 0.007       & \mcc{$1.632 \pm 0.092$}             & \ermcc{ 3.4}{ 3.4}{ 3.0}{ 1.9}{ 2.9}    \\
% TrES-2    & 0.03635   & 0.00063   & 0.00035     & 1.049     & 0.054     & 0.030       & 1.002     & 0.029     & 0.010       & 4.457     & 0.027     & 0.004       & \mcc{$1.043 \pm 0.088$}             & \ermcc{ 2.5}{ 2.8}{ 2.5}{ 0.7}{ 0.8}    \\
TrES-2      & 0.03635   & 0.00063   & 0.00035     & 1.049     & 0.054     & 0.030       & 1.002     & 0.029     & 0.010       & 4.457     & 0.027     & 0.004       & \mcc{$1.043 \pm 0.088$}             & \ermcc{ 2.5}{ 2.8}{ 2.5}{ 0.7}{ 0.8}    \\
TrES-3      & \ermcc{0.02283}{0.00012}{0.00025}{0.00012}{0.00017} & \ermcc{0.929}{0.014}{0.030}{0.014}{0.021} & \ermcc{0.818}{0.011}{0.013}{0.004}{0.006}  & \ermcc{4.581}{0.007}{0.010}{0.002}{0.003}      & \ercc{1.700}{0.047}{0.051} & \ermcc{ 0.1}{ 0.7}{ 0.0}{ 0.0}{ 0.0} \\
% TrES-4    & 0.04994   & 0.00077   & 0.00045     & 1.315     & 0.061     & 0.036       & 1.94      & 0.10      & 0.02        & 3.983     & 0.040     & 0.004       & \mcc{$0.181 \pm 0.025$}             & \ermcc{ 3.3}{ 2.0}{ 0.6}{ 0.5}{ 0.8}    \\
TrES-4      & 0.04965   & 0.00087   & 0.00028     & 1.292     & 0.067     & 0.022       & 1.92      & 0.11      & 0.01        & 3.981     & 0.047     & 0.002       & \mcc{$0.182 \pm 0.030$}             & \ermcc{ 3.7}{ 1.6}{ 1.4}{ 0.2}{ 0.4}    \\
WASP-1      & \ermcc{0.03898}{0.00036}{0.00039}{0.00013}{0.00014} & \ermcc{1.243}{0.034}{0.037}{0.013}{0.014} & \ermcc{1.455}{0.052}{0.079}{0.005}{0.005}  & \ermcc{4.207}{0.045}{0.028}{0.002}{0.002}      & \ercc{0.403}{0.069}{0.037} & \ermcc{ 3.0}{ 0.7}{ 0.6}{ 0.3}{ 0.3} \\
% WASP-2    & 0.03034   & 0.00060   & 0.00042     & 0.804     & 0.048     & 0.034       & 0.807     & 0.019     & 0.011       & 4.530     & 0.017     & 0.006       & \mcc{$1.531 \pm 0.067$}             & \ermcc{11.8}{ 8.1}{ 4.4}{ 3.3}{ 2.5}    \\
WASP-2      & 0.03033   & 0.00060   & 0.00043     & 0.803     & 0.049     & 0.034       & 0.807     & 0.019     & 0.011       & 4.529     & 0.017     & 0.006       & \mcc{$1.527 \pm 0.067$}             & \ermcc{11.9}{ 8.1}{ 4.3}{ 3.3}{ 2.5}    \\
WASP-3      & 0.03187   & 0.00086   & 0.00020     & 1.26      & 0.10      & 0.02        & 1.377     & 0.085     & 0.009       & 4.262     & 0.044     & 0.003       & \mcc{$0.484 \pm 0.073$}             & \ermcc{ 2.1}{ 1.5}{ 1.2}{ 0.4}{ 0.3}    \\
% WASP-4    & \ermcc{0.02308}{0.00053}{0.00057}{0.00027}{0.00021} & \ermcc{0.915}{0.064}{0.069}{0.032}{0.025} & \ermcc{0.905}{0.021}{0.023}{0.010}{0.0084} & \ermcc{4.486}{0.011}{0.012}{0.005}{0.004}      & \ercc{1.232}{0.020}{0.022} & \ermcc{ 6.9}{ 5.3}{ 4.5}{ 2.1}{ 1.8} \\
WASP-4      & \ermcc{0.02307}{0.00053}{0.00055}{0.00026}{0.00022} & \ermcc{0.914}{0.064}{0.066}{0.031}{0.026} & \ermcc{0.905}{0.021}{0.022}{0.010}{0.009}  & \ermcc{4.485}{0.011}{0.012}{0.005}{0.004}      & \ercc{1.233}{0.020}{0.022} & \ermcc{ 7.0}{ 5.2}{ 4.5}{ 2.1}{ 1.8} \\
% WASP-5    & 0.02714   & 0.00049   & 0.00022     & 1.004     & 0.055     & 0.025       & 1.077     & 0.042     & 0.009       & 4.375     & 0.030     & 0.004       & \mcc{$0.803 \pm 0.080$}             & \ermcc{ 7.0}{ 3.2}{ 3.0}{ 1.5}{ 1.5}    \\
WASP-5      & 0.02714   & 0.00049   & 0.00022     & 1.004     & 0.055     & 0.025       & 1.077     & 0.042     & 0.009       & 4.375     & 0.030     & 0.004       & \mcc{$0.803 \pm 0.080$}             & \ermcc{ 7.0}{ 3.2}{ 3.0}{ 1.5}{ 1.5}    \\
WASP-10     & 0.0378    & 0.0014    & 0.0008      & 0.752     & 0.081     & 0.048       & 0.703     & 0.036     & 0.015       & 4.620     & 0.049     & 0.009       & \mcc{$2.16 \pm 0.31$}               & \mc{unconstrained}                      \\
% WASP-18   & 0.02034   & 0.00032   & 0.00016     & 1.256     & 0.059     & 0.029       & 1.222     & 0.042     & 0.010       & 4.363     & 0.027     & 0.003       & \mcc{$0.689 \pm 0.062$}             & \ermcc{ 0.5}{ 1.2}{ 0.9}{ 0.6}{ 0.4}    \\
WASP-18     & 0.02034   & 0.00032   & 0.00016     & 1.256     & 0.059     & 0.029       & 1.222     & 0.042     & 0.010       & 4.363     & 0.027     & 0.003       & \mcc{$0.689 \pm 0.062$}             & \ermcc{ 0.5}{ 1.2}{ 0.9}{ 0.6}{ 0.4}    \\
XO-1        & 0.04944   & 0.00062   & 0.00050     & 1.037     & 0.039     & 0.031       & 0.942     & 0.022     & 0.009       & 4.506     & 0.021     & 0.004       & \mcc{$1.242 \pm 0.080$}             & \ermcc{ 0.9}{ 2.4}{ 0.9}{ 0.8}{ 0.8}    \\
XO-2        & \ermcc{0.03647}{0.00059}{0.00058}{0.00069}{0.00081} & \ermcc{0.946}{0.046}{0.046}{0.054}{0.062} & \ermcc{0.970}{0.028}{0.035}{0.018}{0.022}  & \ermcc{4.440}{0.037}{0.021}{0.008}{0.010}      & \ercc{1.037}{0.128}{0.058} & \ermcc{ 1.9}{ 4.5}{ 1.9}{ 4.7}{ 1.9} \\
XO-3        & 0.04529   & 0.00057   & 0.00045     & 1.206     & 0.046     & 0.036       & 1.409     & 0.054     & 0.014       & 4.222     & 0.027     & 0.004       & \mcc{$0.431 \pm 0.041$}             & \ermcc{ 3.0}{ 0.9}{ 0.6}{ 0.5}{ 0.4}    \\
XO-4        & \ermcc{0.05475}{0.00094}{0.00051}{0.00022}{0.00035} & \ermcc{1.285}{0.068}{0.036}{0.016}{0.024} & \ermcc{1.530}{0.362}{0.069}{0.006}{0.010}  & \ermcc{4.178}{0.034}{0.169}{0.002}{0.002}      & \ercc{0.359}{0.046}{0.160} & \ermcc{ 2.7}{ 0.6}{ 0.5}{ 0.2}{ 0.3} \\
XO-5        & 0.0494    & 0.0011    & 0.0002      & 0.914     & 0.064     & 0.010       & 1.065     & 0.064     & 0.004       & 4.344     & 0.043     & 0.002       & \mcc{$0.76 \pm 0.11$}               & \mc{unconstrained}                      \\
\hline \hline \end{tabular} \end{table*}

\begin{table*} \caption{\label{tab:absdim:planets} Physical properties of the planetary
components of the TEPs studied in this work. For each quantity the first uncertainty is
derived from a propagation of all observational errors and the second uncertainty is an
estimate of the systematic errors arising from the dependence on stellar theory.}
\setlength{\tabcolsep}{4pt}
\begin{tabular}{l l@{\,$\pm$\,}l@{\,$\pm$\,}l l@{\,$\pm$\,}l@{\,$\pm$\,}l l@{\,$\pm$\,}l@{\,$\pm$\,}l l@{\,$\pm$\,}l@{\,$\pm$\,}l l@{\,$\pm$\,}l@{\,$\pm$\,}l l@{\,$\pm$\,}l@{\,$\pm$\,}l}
\hline \hline
System & \mcc{Mass (\Mjup)} & \mcc{Radius (\Rjup)} & \mcc{$g_{\rm b}$ (\mss)} & \mcc{Density (\pjup)} & \mcc{\Teq\ (K)} & \mcc{\safronov} \\
\hline
GJ\,436     & 0.0737    & 0.0051    & 0.0018      & 0.365     & 0.018     & 0.004       & \mcc{$13.7 \pm  1.1$}               & 1.51      & 0.19      & 0.02        & \mcc{$669 \pm  22$}                 & 0.0253    & 0.0015    & 0.0003       \\
HAT-P-1     & 0.524     & 0.016     & 0.008       & 1.217     & 0.038     & 0.009       & \mcc{$8.77 \pm 0.56$}               & 0.290     & 0.027     & 0.002       & \mcc{$1291 \pm   20$}               & 0.0419    & 0.0016    & 0.0003       \\
HAT-P-2     & 8.74      & 0.25      & 0.09        & 1.19      & 0.12      & 0.01        & \mcc{$152 \pm  30$}                 & 5.1       & 1.5       & 0.0         & \mcc{$1516 \pm   66$}               & 0.771     & 0.077     & 0.004        \\
HD\,149026  & \ermcc{0.356}{0.013}{0.011}{0.002}{0.003} & \ermcc{0.610}{0.099}{0.072}{0.002}{0.003}  & \ercc{23.7}{ 6.8}{ 6.2}     & \ermcc{1.57}{0.72}{0.57}{0.01}{0.01}      & \ercc{1626}{  69}{  37} & \ermcc{0.0393}{0.0054}{0.0056}{0.0002}{0.0001} \\
HD\,189733  & 1.150     & 0.028     & 0.027       & 1.151     & 0.036     & 0.013       & \mcc{$21.5 \pm  1.2$}               & 0.755     & 0.066     & 0.009       & \mcc{$1191 \pm   20$}               & 0.0747    & 0.0024    & 0.0009       \\
HD\,209458  & 0.714     & 0.014     & 0.009       & 1.380     & 0.015     & 0.009       & \mcc{$9.30 \pm 0.08$}               & 0.272     & 0.004     & 0.002       & \mcc{$1459 \pm   12$}               & 0.0427    & 0.0005    & 0.0003       \\
OGLE-TR-10  & 0.68      & 0.15      & 0.00        & 1.72      & 0.11      & 0.01        & \mcc{$5.7 \pm 1.4$}                 & 0.134     & 0.038     & 0.000       & \mcc{$1702 \pm   54$}               & 0.0279    & 0.0062    & 0.0001       \\
OGLE-TR-56  & 1.300     & 0.080     & 0.010       & 1.20      & 0.17      & 0.00        & \mcc{$22.3 \pm  6.7$}               & 0.75      & 0.34      & 0.34        & \mcc{$2140 \pm  120$}               & 0.0418    & 0.0065    & 0.0002       \\
OGLE-TR-111 & 0.54      & 0.10      & 0.01        & 1.077     & 0.072     & 0.012       & \mcc{$11.5 \pm  2.5$}               & 0.43      & 0.11      & 0.00        & \mcc{$1034 \pm   28$}               & 0.056     & 0.011     & 0.001        \\
OGLE-TR-113 & 1.24      & 0.17      & 0.02        & 1.111     & 0.049     & 0.011       & \mcc{$25.0 \pm  3.7$}               & 0.91      & 0.16      & 0.01        & \mcc{$1355 \pm   35$}               & 0.0664    & 0.0090    & 0.0007       \\
OGLE-TR-132 & 1.17      & 0.14      & 0.01        & 1.25      & 0.16      & 0.01        & \mcc{$18.5 \pm  5.0$}               & 0.59      & 0.24      & 0.00        & \mcc{$2017 \pm   97$}               & 0.0436    & 0.0072    & 0.0003       \\
OGLE-TR-182 & 1.06      & 0.15      & 0.01        & 1.47      & 0.14      & 0.01        & \mcc{$12.1 \pm  2.9$}               & 0.332     & 0.111     & 0.002       & \mcc{$1550 \pm   81$}               & 0.0628    & 0.0109    & 0.0004       \\
OGLE-TR-211 & \ermcc{0.75}{0.15}{0.15}{0.01}{0.01}      & \ermcc{1.262}{0.158}{0.091}{0.005}{0.005} & \ercc{11.6}{ 2.9}{ 3.3}    & \ermcc{0.372}{0.117}{0.132}{0.002}{0.001} & \ercc{1686}{  90}{  55} & \ermcc{0.0460}{0.0097}{0.0104}{0.0002}{0.0002} \\
OGLE-TR-L9  & 4.34      & 1.47      & 0.05        & 1.614     & 0.083     & 0.009       & \mcc{$41 \pm 14$}                   & 1.03      & 0.37      & 0.01        & \mcc{$2039 \pm   51$}               & 0.153     & 0.052     & 0.001        \\
TrES-1      & 0.761     & 0.045     & 0.023       & 1.099     & 0.031     & 0.017       & \mcc{$15.6 \pm  1.2$}               & 0.573     & 0.056     & 0.009       & \mcc{$1147 \pm   15$}               & 0.0612    & 0.0037    & 0.0009       \\
% TrES-2    & 1.253     & 0.047     & 0.024       & 1.260     & 0.039     & 0.012       & \mcc{$19.6 \pm  1.1$}               & 0.626     & 0.051     & 0.006       & \mcc{$1467 \pm   27$}               & 0.0688    & 0.0024    & 0.0007       \\
TrES-2      & 1.253     & 0.047     & 0.024       & 1.261     & 0.039     & 0.012       & \mcc{$19.5 \pm  1.1$}               & 0.625     & 0.051     & 0.006       & \mcc{$1467 \pm   27$}               & 0.0688    & 0.0024    & 0.0007       \\
TrES-3      & \ermcc{1.910}{0.060}{0.070}{0.020}{0.029} & \ermcc{1.305}{0.027}{0.025}{0.007}{0.010} & \ercc{27.8}{ 1.2}{ 1.4}    & \ermcc{0.860}{0.050}{0.057}{0.007}{0.004} & \ercc{1630}{  23}{  22} & \ermcc{0.0719}{0.0026}{0.0026}{0.0006}{0.0004} \\
% TrES-4    & 0.888     & 0.072     & 0.016       & 1.82      & 0.13      & 0.02        & \mcc{$6.7 \pm 1.1$}                 & 0.148     & 0.033     & 0.001       & \mcc{$1861 \pm   48$}               & 0.0370    & 0.0038    & 0.0003       \\
TrES-4      & 0.877     & 0.072     & 0.010       & 1.81      & 0.15      & 0.01        & \mcc{$6.7 \pm 1.2$}                 & 0.148     & 0.039     & 0.001       & \mcc{$1861 \pm   54$}               & 0.0373    & 0.0042    & 0.0002       \\
WASP-1      & \ermcc{0.860}{0.072}{0.072}{0.006}{0.006} & \ermcc{1.484}{0.059}{0.091}{0.005}{0.006} & \ercc{9.7}{1.5}{1.1}       & \ermcc{0.263}{0.058}{0.036}{0.001}{0.001} & \ercc{1800}{  32}{  49} & \ermcc{0.0363}{0.0038}{0.0033}{0.0001}{0.0001} \\
% WASP-2    & 0.847     & 0.038     & 0.024       & 1.044     & 0.029     & 0.015       & \mcc{$19.30 \pm  0.82$}             & 0.745     & 0.048     & 0.010       & \mcc{$1280 \pm   21$}               & 0.0612    & 0.0021    & 0.0009       \\
WASP-2      & 0.847     & 0.038     & 0.024       & 1.043     & 0.029     & 0.015       & \mcc{$19.32 \pm  0.80$}             & 0.747     & 0.047     & 0.010       & \mcc{$1281 \pm   21$}               & 0.0613    & 0.0021    & 0.0009       \\
WASP-3      & 2.06      & 0.13      & 0.03        & 1.454     & 0.083     & 0.009       & \mcc{$24.2 \pm  2.6$}               & 0.67      & 0.11      & 0.00        & \mcc{$2028 \pm   59$}               & 0.0715    & 0.0048    & 0.0004       \\
% WASP-4    & \ermcc{1.266}{0.091}{0.072}{0.029}{0.023} & \ermcc{1.358}{0.033}{0.034}{0.016}{0.013} & \ercc{17.03}{ 0.97}{ 0.54} & \ermcc{0.506}{0.032}{0.022}{0.005}{0.006} & \ercc{1661}{  30}{  30} & \ermcc{0.0470}{0.0029}{0.0017}{0.0004}{0.0005} \\
WASP-4      & \ermcc{1.237}{0.059}{0.062}{0.028}{0.023} & \ermcc{1.357}{0.033}{0.033}{0.015}{0.013} & \ercc{16.65}{ 0.26}{ 0.33} & \ermcc{0.495}{0.015}{0.018}{0.005}{0.006} & \ercc{1661}{  30}{  30} & \ermcc{0.0460}{0.0012}{0.0013}{0.0004}{0.0005} \\
% WASP-5    & 1.619     & 0.076     & 0.026       & 1.164     & 0.056     & 0.009       & \mcc{$29.6 \pm  2.8$}               & 1.03      & 0.14      & 0.01        & \mcc{$1732 \pm   41$}               & 0.0751    & 0.0042    & 0.0006       \\
WASP-5      & 1.565     & 0.058     & 0.026       & 1.164     & 0.056     & 0.009       & \mcc{$28.7 \pm  2.6$}               & 0.99      & 0.14      & 0.01        & \mcc{$1732 \pm   41$}               & 0.0726    & 0.0035    & 0.0006       \\
WASP-10     & 3.16      & 0.23      & 0.13        & 1.067     & 0.064     & 0.022       & \mcc{$68.9 \pm  6.7$}               & 2.60      & 0.39      & 0.05        & \mcc{$972 \pm  31$}                 & 0.298     & 0.019     & 0.006        \\
% WASP-18   & 10.30     &  0.33     &  0.16       & 1.158     & 0.054     & 0.009       & \mcc{$190 \pm  16$}                 & 6.64      & 0.90      & 0.05        & \mcc{$2392 \pm   51$}               & 0.288     & 0.014     & 0.002        \\
WASP-18     & 10.29     &  0.33     &  0.16       & 1.158     & 0.054     & 0.009       & \mcc{$190 \pm  16$}                 & 6.64      & 0.90      & 0.05        & \mcc{$2392 \pm   51$}               & 0.288     & 0.014     & 0.002        \\
XO-1        & 0.924     & 0.075     & 0.019       & 1.206     & 0.039     & 0.012       & \mcc{$15.8 \pm  1.5$}               & 0.526     & 0.063     & 0.005       & \mcc{$1210 \pm   16$}               & 0.0730    & 0.0062    & 0.0007       \\
XO-2        & \ermcc{0.555}{0.062}{0.057}{0.021}{0.025} & \ermcc{0.992}{0.034}{0.057}{0.019}{0.022} & \ercc{14.0}{ 2.1}{ 1.5}    & \ermcc{0.569}{0.119}{0.072}{0.013}{0.011} & \ercc{1328}{  17}{  28} & \ermcc{0.0432}{0.0049}{0.0045}{0.0010}{0.0008} \\
XO-3        & 11.83     &  0.31     &  0.23       & 1.248     & 0.047     & 0.012       & \mcc{$188 \pm  13$}                 & 6.08      & 0.67      & 0.06        & \mcc{$1729 \pm   34$}               & 0.711     & 0.027     & 0.007        \\
XO-4        & \ermcc{1.521}{0.160}{0.153}{0.012}{0.019} & \ermcc{1.29}{0.38}{0.06}{0.01}{0.01}      & \ercc{22.8}{ 3.2}{ 9.5}    & \ermcc{0.71}{0.13}{0.40}{0.01}{0.00}      & \ercc{1630}{ 169}{  36} & \ermcc{0.1006}{0.0112}{0.0253}{0.0006}{0.0004} \\
XO-5        & 1.084     & 0.055     & 0.008       & 1.089     & 0.082     & 0.004       & \mcc{$22.7 \pm  3.2$}               & 0.84      & 0.18      & 0.00        & \mcc{$1203 \pm   33$}               & 0.1075    & 0.0082    & 0.0004       \\
\hline \hline \end{tabular} \end{table*}

The main result of this work is the determination of the basic physical properties of thirty TEPs using homogeneous methods. Detailed results for each TEP are given in the many Tables in the Appendix, and the final results for all systems have been summarised in Tables \ref{tab:absdim:stars} and \ref{tab:absdim:planets}. The homogeneous nature of these numbers means they are well suited for comparing different TEPs, for planning follow-up observations, and for performing detailed statistical studies.

\begin{figure} \includegraphics[width=0.48\textwidth,angle=0]{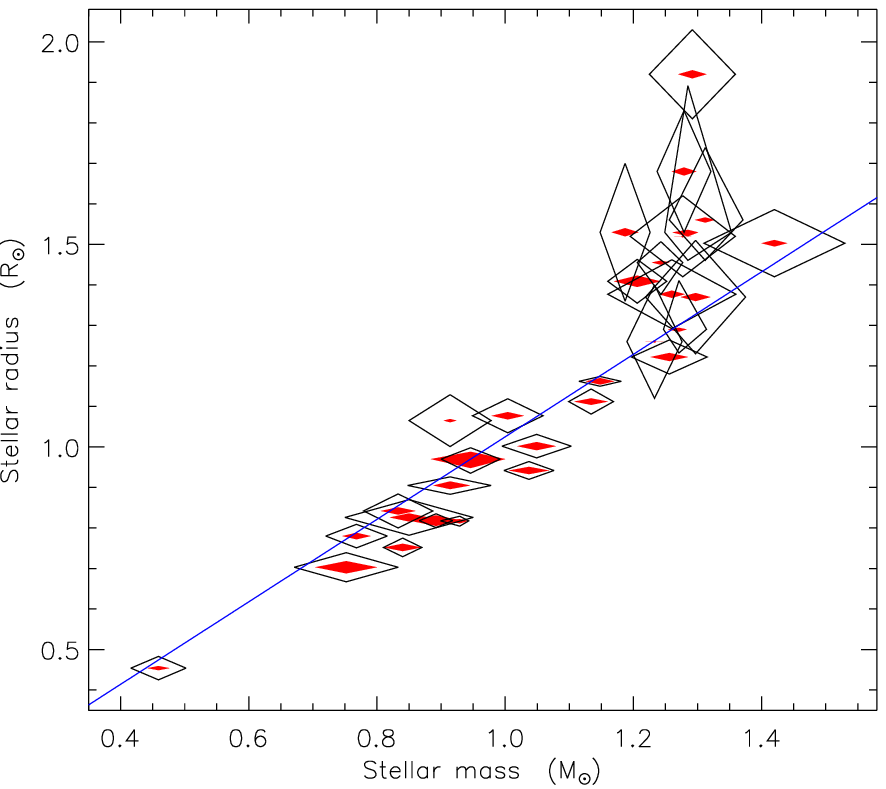}
\caption{\label{fig:absdim:M1R1} Plot of the masses versus the radii of the
stars in the thirty TEPs studied in this work. The statistical uncertainties
are shown by black open diamonds and the systematic uncertainties by red
filled diamonds. The empirical mass--radius relation from Paper\,I is shown
with a blue line.} \end{figure}

The masses and radii of the stars considered in this work are plotted in Fig.\,\ref{fig:absdim:M1R1} with both their random and systematic errorbars, and with the empirical mass--radius relation (Paper\,II) indicated. A striking aspect of Fig.\,\ref{fig:absdim:M1R1} is that for $M_{\rm A} < 1.2$\Msun\ the stars generally stick close to the main sequence, but beyond this point their distribution turns almost vertically upwards. The two exceptions are XO-5, which has a more evolved lower-mass star, and OGLE-TR-L9, which contains a more massive star (1.4\Msun).

That low-mass stars stick closely to the main sequence is to be expected due to their long evolutionary timescales. The shift to larger $R_{\rm A}$ at 1.2\Msun\ is reasonable because a larger stellar radius makes it more likely that a given planet is transiting (see Paper\,II). The avoidance of more massive stars is likely due to difficulty of measuring RVs for these objects, making them lower-priority and more observationally expensive targets for TEP search consortia at the follow-up stage. There may be a large and presently neglected population of TEPs around more massive stars\footnote{Whilst RV surveys mostly concentrate on F, G and K stars, they have not been neglecting ones more massive than this; see \citet{Galland+05aa}, \citet{Johnson+07apj} \citet{Lagrange+09aa} and \citet{Bowler+10apj}.}.

Another notable feature of Fig.\,\ref{fig:absdim:M1R1} is that the systematic errors (red filled diamonds in the Figure) are not negligible compared to the random errors, and in some cases (HD\,189733, TrES-1, TrES-3, XO-2) are of a similar size or even larger.

\begin{figure} \includegraphics[width=0.48\textwidth,angle=0]{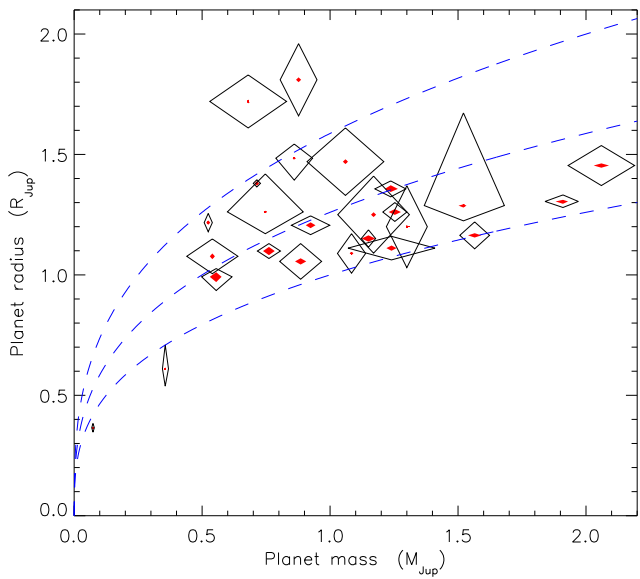}
\caption{\label{fig:absdim:M2R2} Plot of the masses versus the radii of
the planets in the thirty TEPs studied in this work. The statistical
uncertainties are shown by black open diamonds and the systematic
uncertainties by red filled diamonds. Blue dotted lines show where
density is 1.0, 0.5 and 0.25 \pjup.} \end{figure}

A similar diagram for the planets (Fig.\,\ref{fig:absdim:M2R2}) shows a much larger scatter in the planet properties, but also that systematic errors are much less important for these objects. The two outliers with very low densities are OGLE-TR-10 and TrES-4.

\subsection{Physical properties of all known TEPs}

\begin{figure} \includegraphics[width=\columnwidth,angle=0]{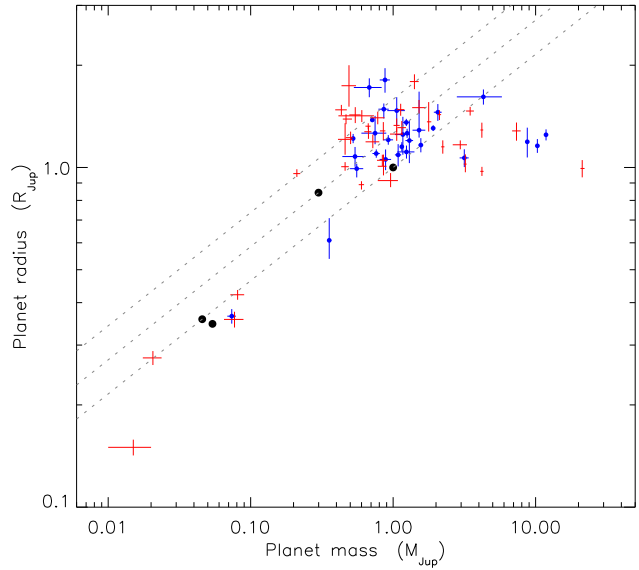}
\caption{\label{fig:absdim:m2r2} Mass--radius plot for the known transiting
extrasolar planets. Those objects studied in this work are shown with (blue)
filled circles and numbers taken from the literature with (red) crosses. The
four gas giant planets in our Solar System are denoted with (black) filled circles.
Dotted lines show loci where density equals 1.0, 0.5 and 0.25 \pjup.} \end{figure}

The physical properties of the thirty TEPs studied in this work have been augmented with the literature values for the other \reff{41 known systems (as of 31/05/2010)} to yield a larger but inhomogeneous sample. Fig.\,\ref{fig:absdim:m2r2} shows their masses and radii on logarithmic axes, alongside those of the four Solar system gas giants. The dominant population of known TEPs continues to reside in a cloud of points centred roughly on 1.0\Mjup\ and 1.2\Rjup. There is a clear tail of high-mass planets, culminating in the brown dwarf CoRoT-3, as well as an increasing number of systems of much smaller mass and radius, typified by GJ\,1214 and the first possible rocky exoplanet, CoRoT-7.

\begin{figure} \includegraphics[width=\columnwidth,angle=0]{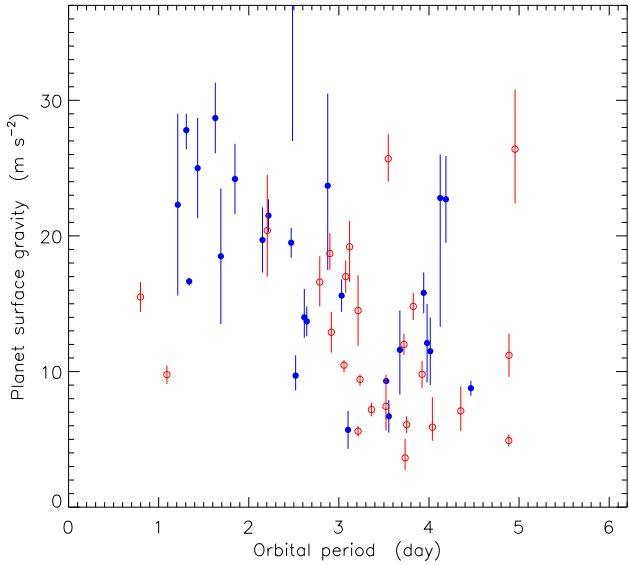}
\caption{\label{fig:absdim:pg2} Plot of the orbital periods versus the
surface gravities of known TEPs. The objects studied in this work are
shown with (blue) filled circles and other objects with (red) open
circles.} \end{figure}

\begin{figure} \includegraphics[width=\columnwidth,angle=0]{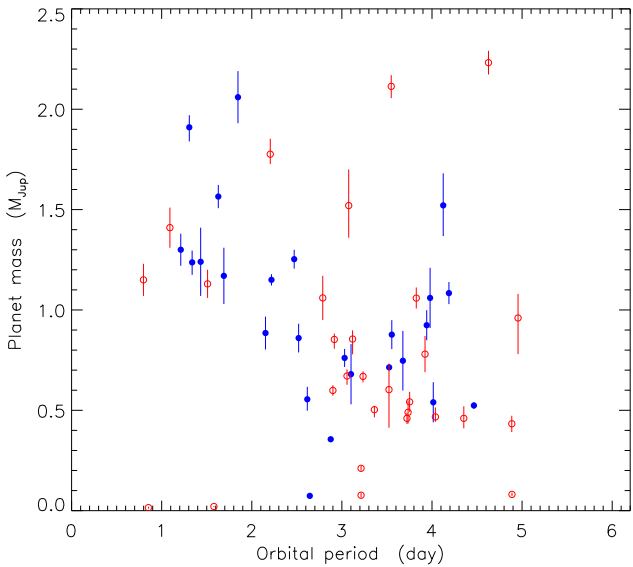}
\caption{\label{fig:absdim:pm2} Same as Fig.\,\ref{fig:absdim:pg2}, but
for planetary mass instead of $g_{\rm b}$.} \end{figure}

Correlations have previously been noticed between \Porb\ and $g_{\rm b}$ \citep{Me++07mn} and \Porb\ and $M_{\rm b}$ \citep{Mazeh++05mn}. The relevant plots are shown in Figs.\ \ref{fig:absdim:pg2} and \ref{fig:absdim:pm2}. In both cases there are ten planets whose properties put them outside the range of \Porb\ shown in these Figures. Neglecting this population of massive planets (which may be a fundamentally different population to the lower mass ones; \citealt{FabryckyWinn09apj}; \citealt{Me+09apj}), the rank correlation test of \citet{Spearman1904} returns a probability of \reff{99.80}\% ($3.1\sigma$) that the $\Porb$--$g_{\rm b}$ correlation is real and \reff{97.6}\% ($2.3\sigma$) that the $\Porb$--$M_{\rm b}$ correlation is real. The corresponding figures from Paper\,II, based on 44 TEPs, are 98.3\% and 97.5\%. Two concerns with these correlations are apparent. Firstly, they are insignificant if the longer-period planets are not rejected\footnote{Merrill's theorem should be borne in mind: ``When one throws away discrepant results, the remainder are found to agree well.''}. Secondly, \reff{the period--mass correlation has not increased} in significance with the addition of 21 new TEPs since Paper\,II.

\begin{figure} \includegraphics[width=\columnwidth,angle=0]{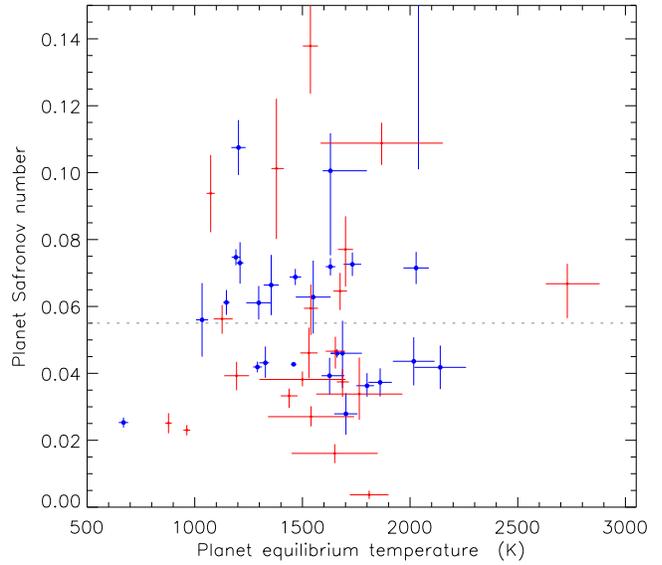}
\caption{\label{fig:analysis:teqsaf} Plot of equilibrium temperature versus
Safronov number for the full sample of planets. Objects shown with (blue)
circles were studied in this work; those which are just (red) errorbars
were not. The dotted line shows the separation between Class A and Class
B proposed by \citet{HansenBarman07apj}.} \end{figure}

\citet{HansenBarman07apj} divided up eighteen of the twenty TEPs then known into two classes based on their position in a diagram of \safronov\ versus $T_{\rm eq}$. An updated version of the diagram can be seen in Fig.\,\ref{fig:analysis:teqsaf}, and agrees with previous conclusions (Paper\,II) that the division between the classes was blurred into insignificance. A dotted line at $\Theta = 0.055$ has been drawn to show the expected boundaries between Class\,I ($\Theta \approx 0.07 \pm 0.01$) and Class\,II ($\Theta \approx 0.04 \pm 0.01$). There are several other equally possible lines which could be drawn through this diagram, which does not inspire confidence in their reality. It is probable that the addition of new TEPs in the future will fill out this diagram with objects drawn from a single distribution.
%I have also drawn two dashed revised lines corresponding to $\Theta = 0.027 + \frac{\Teq}{60000\,{\rm K}}$ and $\Theta = 0.060 + \frac{\Teq}{60000\,{\rm K}}$ which do a reasonable job at separating the two existing classes and one new one without actually inspiring any confidence in their reality. It is probable that the addition of new TEPs in the future will fill out this diagram with a single distribution of objects.

%%%%%%%%%%%%%%%%%%%%%%%%%%%%%%%%%%%%%%%%%%%%%%%%%%%%%%%%%%%%%%%%%%%%%%%%%%%%%%%%%%%%%%%%%%%%%%%%%%%%%%%%%%%%%%%%%%%%%%%

\section{Summary}                                                                              \label{sec:teps:summary}

Measurements of the physical properties of many transiting extrasolar planets are available in the literature, but are not directly comparable as they have been derived by different researchers using a variety of methods. In this series of papers I aim to derive the physical properties of the known TEPs using a rigorous and homogeneous approach, resulting in quantities which are suitable for statistical analysis. Each TEP is tackled in two steps, the first being analysis of its transit light curves and the second being the determination of its physical properties.

The transit light curves are studied using the {\sc jktebop} code, which models binary systems using biaxial spheroids. I pay special attention to the limb darkening of the parent star; solutions are obtained for every light curve using each of five different limb darkening laws. Orbital eccentricity is accounted for by including $e$ and $\omega$ as fitted parameters constrained by values and uncertainties measured from radial velocity observations of the parent star. Parameter errors are assessed using Monte Carlo simulations, a residual-permutation algorithm, and the inter-agreement between solutions with different limb darkening laws.

\begin{table} \centering \caption{\label{tab:absdim:obs} Summary
of which types of additional observations would be useful for the
thirty TEPs studied in this series of papers. $\star$ denotes where
additional data would be useful, and $\star\star$ indicates where
it would be useful but difficult to either obtain or interpret.}
\setlength{\tabcolsep}{8pt}
\begin{tabular}{lccc}
\hline \hline
System      & Photometric  & Radial       & Spectral     \\
            & observations & velocities   & synthesis    \\
\hline
GJ\,436     &              &              & $\star\star$ \\
HAT-P-1     & $\star$      &              &              \\
HAT-P-2     & $\star\star$ &              &              \\
HD\,149026  & $\star$      &              &              \\
HD\,189733  &              &              &              \\
HD\,209458  &              &              &              \\
OGLE-TR-10  & $\star\star$ & $\star\star$ &              \\
OGLE-TR-56  & $\star\star$ & $\star\star$ &              \\
OGLE-TR-111 &              & $\star\star$ &              \\
OGLE-TR-113 & $\star\star$ & $\star\star$ & $\star$      \\
OGLE-TR-132 & $\star\star$ & $\star\star$ &              \\
OGLE-TR-182 & $\star\star$ & $\star\star$ &              \\
OGLE-TR-211 & $\star\star$ & $\star\star$ &              \\
OGLE-TR-L9  & $\star$      & $\star\star$ &              \\
TrES-1      & $\star$      & $\star$      &              \\
TrES-2      &              &              &              \\
TrES-3      &              &              &              \\
TrES-4      & $\star$      & $\star$      &              \\
WASP-1      & $\star$      & $\star$      &              \\
WASP-2      &              &              &              \\
WASP-3      &              & $\star$      &              \\
WASP-4      &              &              &              \\
WASP-5      & $\star$      &              &              \\
WASP-10     & $\star$      &              & $\star$      \\
WASP-18     & $\star$      &              &              \\
XO-1        & $\star$      & $\star$      &              \\
XO-2        &              & $\star$      &              \\
XO-3        &              &              & $\star$      \\
XO-4        & $\star$      & $\star$      &              \\
XO-5        & $\star$      &              & $\star$      \\
\hline \hline \end{tabular} \end{table}

In this paper I present analyses of published transit light curves of fifteen TEPs (Table\,\ref{tab:teps:lcpar}). Three of these (TrES-2, TrES-4 and WASP-2) have faint companion stars within 1.6\as\ on the sky \citep{Daemgen+09aa}. If neglected, this `third light' dilutes the transit depth and causes systematic errors in parameters measured from the light curves. As a guideline, $L_3 = 5$\% causes the radius of a representative planet to be underestimated by 2\%. I show that it is not possible to detect third light in synthetic light curves representative of good ground-based observations. A third light value must instead be measured and used to constrain the light curve solutions.

In the second stage of the analysis of each TEP, I augment the light curve results with measurements of the velocity amplitude, \Teff\ and \FeH\ for its parent star. I then interpolate within tabulated predictions from theoretical stellar models to find the overall best-fitting physical properties of the star and the planet. This model dependence causes systematic errors in the resulting quantities, which I assess by comparing the individual results found using five independent sets of theoretical model predictions. This leads also to `consensus values' for the physical properties; the Yonsei-Yale models are overall the closest to this consensus.

Measurements of three \reff{physical properties} are exempt from this model dependence: the planetary surface gravity can be determined from only observed quantities; the density of the star has only a negligible dependence on stellar models; and the planet's equilibrium temperature does not depend on stellar models but does rest on the effective temperature scale of low-mass \reff{(F, G and K)} stars. The model dependence specifies a minimum level of uncertainty for other physical properties, and this ranges from 1\% for $M_{\rm A}$ to 0.6\% for $M_{\rm b}$ and 0.3\% for $R_{\rm A}$, $R_{\rm b}$ and $a$. Reducing these minimum levels will require improvements in our understanding of low-mass stars. An external test of systematic errors was obtained by comparing the discovery papers of the WASP-11\,/\,HAT-P-10 system. The agreement between the two studies is good for most parameters, but less so for the transit depth and the stellar \Teff.

Tables \ref{tab:absdim:stars} and \ref{tab:absdim:planets} summarise the physical properties of thirty TEPs: the fifteen with light curve analyses in the current work plus another fifteen with photometric analyses in Paper\,I and later works. The statistical errors in these quantities are calculated using a perturbation analysis which returns complete error budgets for each output parameter. These error budgets indicate which type of follow-up observations are the best way of improving the parameter measurements for each TEP. In most the quality of the light curve is the primary limitation on measurements of the properties of the planet; many systems would also benefit from additional spectroscopic observations. This is summarised in Table\,\ref{tab:absdim:obs} for the benefit of follow-up programmes.

The measured properties of the 30 TEPs were then reinforced with those of the other 35 known objects and plotted in several parameter diagrams. I find that the correlation between orbital period and planetary surface gravity is significant at the \reff{3.1}$\sigma$ level once clear outliers are thrown out -- this significance has \reff{increased} with the addition of new systems since Paper\,II. \reff{The period vs.\ planetary mass correlation has a significance of only \reff{2.3}$\sigma$, and this \reff{ {\em not} increased} since Paper\,II.} Similarly, the distinction between Class\,I and Class\,II planets in the \Teq\ vs.\ \safronov\ diagram is blurring into irrelevance.

In this work I have now treated 30 TEPs, including all currently known from the OGLE, TrES and XO surveys. The discovery rate of transiting planets is expected to continue rising, as the ground-based surveys HAT and WASP mature and new areas of parameter space are probed by the space missions CoRoT and {\it Kepler}. A homogeneous study of the atmospheric parameters of all TEP host stars would be very complementary to this work. Whilst the light curves obtained by CoRoT and {\it Kepler} are of extremely high quality, it should not be forgotten that extensive velocity observations are also needed to study transiting systems in detail. Precise physical properties of transiting planets are a fundamental requirement of constraining the atmospheric characteristics, formation and evolution of planetary systems.

%%%%%%%%%%%%%%%%%%%%%%%%%%%%%%%%%%%%%%%%%%%%%%%%%%%%%%%%%%%%%%%%%%%%%%%%%%%%%%%%%%%%%%%%%%%%%%%%%%%%%%%%%%%%%%%%%%%%%%%

\section*{Acknowledgments}

I acknowledge financial support from STFC in the form of a postdoctoral research assistant position, and a timely and useful report from the anonymous referee. I am grateful to Barry Smalley, Andrew Collier Cameron, Pierre Maxted and Aaron Dotter for discussions, and to Adriano Pietrinferni, Emma Nasi and Antonio Claret for calculating stellar model sets for me. I thank Ignas Snellen, G\'asp\'ar Bakos, Neale Gibson and Peter McCullough for sending me their data and to the NSTeD and CDS websites for making many other datasets available. The following internet-based resources were used in research for this paper: the ESO Digitized Sky Survey; the NASA Astrophysics Data System; the SIMBAD database operated at CDS, Strasbourg, France; the ar$\chi$iv scientific paper preprint service operated by Cornell University; and the BaSTI web tools.

%%%%%%%%%%%%%%%%%%%%%%%%%%%%%%%%%%%%%%%%%%%%%%%%%%%%%%%%%%%%%%%%%%%%%%%%%%%%%%%%%%%%%%%%%%%%%%%%%%%%%%%%%%%%%%

\bibliographystyle{mn_new}
% \bibliography{aamnem99,jkt}

\begin{thebibliography}{139}
\expandafter\ifx\csname natexlab\endcsname\relax\def\natexlab#1{#1}\fi

\bibitem[{{Alonso} et~al.(2004)}]{Alonso+04apj}
{Alonso}, R., et~al., 2004, ApJ, 613, L153

\bibitem[{{Ammler-von Eiff} et~al.(2009){Ammler-von Eiff}, {Santos}, {Sousa},
  {Fernandes}, {Guillot}, {Israelian}, {Mayor}, \& {Melo}}]{Ammler+09aa}
{Ammler-von Eiff}, M., {Santos}, N.~C., {Sousa}, S.~G., {Fernandes}, J.,
  {Guillot}, T., {Israelian}, G., {Mayor}, M., {Melo}, C., 2009, A\&A, 507, 523

\bibitem[{{Anderson} et~al.(2008)}]{Anderson+08mn}
{Anderson}, D.~R., et~al., 2008, MNRAS, 387, L4

\bibitem[{{Bakos} et~al.(2004){Bakos}, {Noyes}, {Kov{\'a}cs}, {Stanek},
  {Sasselov}, \& {Domsa}}]{Bakos+04pasp}
{Bakos}, G., {Noyes}, R.~W., {Kov{\'a}cs}, G., {Stanek}, K.~Z., {Sasselov},
  D.~D., {Domsa}, I., 2004, PASP, 116, 266

\bibitem[{{Bakos} et~al.(2002){Bakos}, {L{\'a}z{\'a}r}, {Papp}, {S{\'a}ri}, \&
  {Green}}]{Bakos+02pasp}
{Bakos}, G.~{\'A}., {L{\'a}z{\'a}r}, J., {Papp}, I., {S{\'a}ri}, P., {Green},
  E.~M., 2002, PASP, 114, 974

\bibitem[{{Bakos} et~al.(2007{\natexlab{a}})}]{Bakos+07apj}
{Bakos}, G.~{\'A}., et~al., 2007{\natexlab{a}}, ApJ, 656, 552

\bibitem[{{Bakos} et~al.(2007{\natexlab{b}})}]{Bakos+07apj3}
{Bakos}, G.~{\'A}., et~al., 2007{\natexlab{b}}, ApJ, 670, 826

\bibitem[{{Bakos} et~al.(2009)}]{Bakos+09apj}
{Bakos}, G.~{\'A}., et~al., 2009, apJ, 696, 1950

\bibitem[{{Ballard} et~al.(2009)}]{Ballard+09iaus}
{Ballard}, S., et~al., 2009, in Transiting planets, vol. 253 of \emph{IAU
  Symposium}, p. 470

\bibitem[{{Bean} et~al.(2006){Bean}, {Benedict}, \& {Endl}}]{Bean++06apj}
{Bean}, J.~L., {Benedict}, G.~F., {Endl}, M., 2006, ApJ, 653, L65

\bibitem[{{Boisse} et~al.(2009)}]{Boisse+09aa}
{Boisse}, I., et~al., 2009, A\&A, 495, 959

\bibitem[{{Bonfils} et~al.(2005){Bonfils}, {Delfosse}, {Udry}, {Santos},
  {Forveille}, \& {S{\'e}gransan}}]{Bonfils+05aa}
{Bonfils}, X., {Delfosse}, X., {Udry}, S., {Santos}, N.~C., {Forveille}, T.,
  {S{\'e}gransan}, D., 2005, A\&A, 442, 635

\bibitem[{{Bouchy} et~al.(2004){Bouchy}, {Pont}, {Santos}, {Melo}, {Mayor},
  {Queloz}, \& {Udry}}]{Bouchy+04aa}
{Bouchy}, F., {Pont}, F., {Santos}, N.~C., {Melo}, C., {Mayor}, M., {Queloz},
  D., {Udry}, S., 2004, A\&A, 421, L13

\bibitem[{{Bouchy} et~al.(2005{\natexlab{a}}){Bouchy}, {Pont}, {Melo},
  {Santos}, {Mayor}, {Queloz}, \& {Udry}}]{Bouchy+05aa2}
{Bouchy}, F., {Pont}, F., {Melo}, C., {Santos}, N.~C., {Mayor}, M., {Queloz},
  D., {Udry}, S., 2005{\natexlab{a}}, A\&A, 431, 1105

\bibitem[{{Bouchy} et~al.(2005{\natexlab{b}})}]{Bouchy+05aa}
{Bouchy}, F., et~al., 2005{\natexlab{b}}, A\&A, 444, L15

\bibitem[{{Bowler} et~al.(2010)}]{Bowler+10apj}
{Bowler}, B.~P., et~al., 2010, ApJ, 709, 396

\bibitem[{{Bruntt} et~al.(2006){Bruntt}, {Southworth}, {Torres}, {Penny},
  {Clausen}, \& {Buzasi}}]{Bruntt+06aa}
{Bruntt}, H., {Southworth}, J., {Torres}, G., {Penny}, A.~J., {Clausen}, J.~V.,
  {Buzasi}, D.~L., 2006, A\&A, 456, 651

\bibitem[{{Burke} et~al.(2007)}]{Burke+07apj}
{Burke}, C.~J., et~al., 2007, ApJ, 671, 2115

\bibitem[{{Burke} et~al.(2008)}]{Burke+08apj}
{Burke}, C.~J., et~al., 2008, ApJ, 686, 1331

\bibitem[{{Christian} et~al.(2009)}]{Christian+09mn}
{Christian}, D.~J., et~al., 2009, MNRAS, 392, 1585

\bibitem[{{Claret}(2000)}]{Claret00aa}
{Claret}, A., 2000, A\&A, 363, 1081

\bibitem[{{Claret}(2004{\natexlab{a}})}]{Claret04aa2}
{Claret}, A., 2004{\natexlab{a}}, A\&A, 428, 1001

\bibitem[{{Claret}(2004{\natexlab{b}})}]{Claret04aa}
{Claret}, A., 2004{\natexlab{b}}, A\&A, 424, 919

\bibitem[{{Claret}(2005)}]{Claret05aa}
{Claret}, A., 2005, A\&A, 440, 647

\bibitem[{{Claret}(2006)}]{Claret06aa2}
{Claret}, A., 2006, A\&A, 453, 769

\bibitem[{{Claret}(2007)}]{Claret07aa2}
{Claret}, A., 2007, A\&A, 467, 1389

\bibitem[{{Claret} \& {Hauschildt}(2003)}]{ClaretHauschildt03aa}
{Claret}, A., {Hauschildt}, P.~H., 2003, A\&A, 412, 241

\bibitem[{{Collier Cameron} et~al.(2007)}]{Cameron+07mn2}
{Collier Cameron}, A., et~al., 2007, MNRAS, 380, 1230

\bibitem[{{Daemgen} et~al.(2009){Daemgen}, {Hormuth}, {Brandner}, {Bergfors},
  {Janson}, {Hippler}, \& {Henning}}]{Daemgen+09aa}
{Daemgen}, S., {Hormuth}, F., {Brandner}, W., {Bergfors}, C., {Janson}, M.,
  {Hippler}, S., {Henning}, T., 2009, A\&A, 498, 567

\bibitem[{{Damasso} et~al.(2009){Damasso}, {Calcidese}, {Bernagozzi},
  {Bertolini}, {Giacobbe}, {Lattanzi}, {Smart}, \& {Sozzetti}}]{Damasso+09xxx}
{Damasso}, M., {Calcidese}, P., {Bernagozzi}, A., {Bertolini}, E., {Giacobbe},
  P., {Lattanzi}, M.~G., {Smart}, R., {Sozzetti}, A., 2009, in ``Pathways
  towards habitable planets'' conference, in press, {\tt arXiv:0911.3587}

\bibitem[{{Demarque} et~al.(2004){Demarque}, {Woo}, {Kim}, \&
  {Yi}}]{Demarque+04apjs}
{Demarque}, P., {Woo}, J.-H., {Kim}, Y.-C., {Yi}, S.~K., 2004, ApJS, 155, 667

\bibitem[{{D{\'{\i}}az} et~al.(2007)}]{Diaz+07apj}
{D{\'{\i}}az}, R.~F., et~al., 2007, ApJ, 660, 850

\bibitem[{{Dittmann} et~al.(2010){Dittmann}, {Close}, {Scuderi}, \&
  {Morris}}]{Dittmann+10xxx}
{Dittmann}, J.~A., {Close}, L.~M., {Scuderi}, L.~J., {Morris}, M.~D., 2010,
  ApJ, submitted, {\tt arXiv:1003.1762}

\bibitem[{{Dotter} et~al.(2008){Dotter}, {Chaboyer}, {Jevremovi{\'c}},
  {Kostov}, {Baron}, \& {Ferguson}}]{Dotter+08apjs}
{Dotter}, A., {Chaboyer}, B., {Jevremovi{\'c}}, D., {Kostov}, V., {Baron}, E.,
  {Ferguson}, J.~W., 2008, ApJS, 178, 89

\bibitem[{{Etzel}(1981)}]{Etzel81conf}
{Etzel}, P.~B., 1981, in {Carling}, E.~B., {Kopal}, Z., eds., Photometric and
  Spectroscopic Binary Systems, NATO ASI Ser. C., 69, Kluwer, Dordrecht, p. 111

\bibitem[{{Fabrycky} \& {Winn}(2009)}]{FabryckyWinn09apj}
{Fabrycky}, D.~C., {Winn}, J.~N., 2009, ApJ, 696, 1230

\bibitem[{{Fernandez} et~al.(2009){Fernandez}, {Holman}, {Winn}, {Torres},
  {Shporer}, {Mazeh}, {Esquerdo}, \& {Everett}}]{Fernandez+09aj}
{Fernandez}, J.~M., {Holman}, M.~J., {Winn}, J.~N., {Torres}, G., {Shporer},
  A., {Mazeh}, T., {Esquerdo}, G.~A., {Everett}, M.~E., 2009, AJ, 137, 4911

\bibitem[{{Fortney} et~al.(2008){Fortney}, {Lodders}, {Marley}, \&
  {Freedman}}]{Fortney+08apj}
{Fortney}, J.~J., {Lodders}, K., {Marley}, M.~S., {Freedman}, R.~S., 2008, ApJ,
  678, 1419

\bibitem[{{Galland} et~al.(2005){Galland}, {Lagrange}, {Udry}, {Chelli},
  {Pepe}, {Queloz}, {Beuzit}, \& {Mayor}}]{Galland+05aa}
{Galland}, F., {Lagrange}, A., {Udry}, S., {Chelli}, A., {Pepe}, F., {Queloz},
  D., {Beuzit}, J., {Mayor}, M., 2005, A\&A, 443, 337

\bibitem[{{Ghezzi} et~al.(2010){Ghezzi}, {Cunha}, {de Ara{\'u}jo}, {Smith}, {de
  la Reza}, \& {Schuler}}]{Ghezzi+10iaus}
{Ghezzi}, L., {Cunha}, K., {de Ara{\'u}jo}, F.~X., {Smith}, V.~V., {de la
  Reza}, R., {Schuler}, S., 2010, in {K.~Cunha, M.~Spite, \& B.~Barbuy}, ed.,
  IAU Symposium, vol. 265, p. 432

\bibitem[{{Gibson} et~al.(2008)}]{Gibson+08aa}
{Gibson}, N.~P., et~al., 2008, A\&A, 492, 603

\bibitem[{{Gibson} et~al.(2009)}]{Gibson+09apj}
{Gibson}, N.~P., et~al., 2009, ApJ, 700, 1078

\bibitem[{{Gilliland} et~al.(2010)}]{Gilliland+10}
{Gilliland}, R.~L., et~al., 2010, ApJ Letters, in press, {\tt arXiv:1001.0142}

\bibitem[{{Gillon} et~al.(2006){Gillon}, {Pont}, {Moutou}, {Bouchy}, {Courbin},
  {Sohy}, \& {Magain}}]{Gillon+06aa}
{Gillon}, M., {Pont}, F., {Moutou}, C., {Bouchy}, F., {Courbin}, F., {Sohy},
  S., {Magain}, P., 2006, A\&A, 459, 249

\bibitem[{{Gillon} et~al.(2007)}]{Gillon+07aa3}
{Gillon}, M., et~al., 2007, A\&A, 466, 743

\bibitem[{{Gillon} et~al.(2009)}]{Gillon+09aa}
{Gillon}, M., et~al., 2009, A\&A, 496, 259

\bibitem[{{Greiner} et~al.(2008)}]{Greiner+08pasp}
{Greiner}, J., et~al., 2008, PASP, 120, 405

\bibitem[{{Hansen} \& {Barman}(2007)}]{HansenBarman07apj}
{Hansen}, B.~M.~S., {Barman}, T., 2007, ApJ, 671, 861

\bibitem[{{H{\'e}brard} et~al.(2008)}]{Hebrard+09aa}
{H{\'e}brard}, G., et~al., 2008, A\&A, 488, 763

\bibitem[{{Hellier} et~al.(2009)}]{Hellier+09natur}
{Hellier}, C., et~al., 2009, Nature, 460, 1098

\bibitem[{{Hilditch}(2001)}]{Hilditch01book}
{Hilditch}, R.~W., 2001, {An Introduction to Close Binary Stars}, Cambridge
  University Press, Cambridge, UK

\bibitem[{{Holman} et~al.(2007)}]{Holman+07apj}
{Holman}, M.~J., et~al., 2007, ApJ, 664, 1185

\bibitem[{{Jenkins} et~al.(2002){Jenkins}, {Caldwell}, \&
  {Borucki}}]{Jenkins++02apj}
{Jenkins}, J.~M., {Caldwell}, D.~A., {Borucki}, W.~J., 2002, ApJ, 564, 495

\bibitem[{{Johns-Krull} et~al.(2008)}]{Johnskrull+08apj}
{Johns-Krull}, C.~M., et~al., 2008, ApJ, 677, 657

\bibitem[{{Johnson} et~al.(2009{\natexlab{a}}){Johnson}, {Winn}, {Albrecht},
  {Howard}, {Marcy}, \& {Gazak}}]{Johnson+09pasp}
{Johnson}, J.~A., {Winn}, J.~N., {Albrecht}, S., {Howard}, A.~W., {Marcy},
  G.~W., {Gazak}, J.~Z., 2009{\natexlab{a}}, PASP, 121, 1104

\bibitem[{{Johnson} et~al.(2009{\natexlab{b}}){Johnson}, {Winn}, {Cabrera}, \&
  {Carter}}]{Johnson+09apj}
{Johnson}, J.~A., {Winn}, J.~N., {Cabrera}, N.~E., {Carter}, J.~A.,
  2009{\natexlab{b}}, ApJ, 692, L100

\bibitem[{{Johnson} et~al.(2007)}]{Johnson+07apj}
{Johnson}, J.~A., et~al., 2007, ApJ, 665, 785

\bibitem[{{Johnson} et~al.(2008)}]{Johnson+08apj}
{Johnson}, J.~A., et~al., 2008, ApJ, 686, 649

\bibitem[{{Kipping}(2008)}]{Kipping08mn}
{Kipping}, D.~M., 2008, MNRAS, 389, 1383

\bibitem[{{Knutson} et~al.(2009){Knutson}, {Charbonneau}, {Burrows},
  {O'Donovan}, \& {Mandushev}}]{Knutson+09apj}
{Knutson}, H.~A., {Charbonneau}, D., {Burrows}, A., {O'Donovan}, F.~T.,
  {Mandushev}, G., 2009, ApJ, 691, 866

\bibitem[{{Koch} et~al.(2010)}]{Koch+10xxx}
{Koch}, D.~G., et~al., 2010, ApJ, in press, {\tt arXiv:1001.0268}

\bibitem[{{Konacki} et~al.(2005){Konacki}, {Torres}, {Sasselov}, \&
  {Jha}}]{Konacki+05apj}
{Konacki}, M., {Torres}, G., {Sasselov}, D.~D., {Jha}, S., 2005, ApJ, 624, 372

\bibitem[{{Konacki} et~al.(2004)}]{Konacki+04apj}
{Konacki}, M., et~al., 2004, ApJ, 609, L37

\bibitem[{{Krejcova} et~al.(2010){Krejcova}, {Budaj}, \&
  {Krushevska}}]{Krejcova++10xxx}
{Krejcova}, T., {Budaj}, J., {Krushevska}, V., 2010, Communications in
  Asteroseismology, submitted, {\tt arXiv:1003.1301}

\bibitem[{{Kurucz}(1993)}]{Kurucz93}
{Kurucz}, R., 1993, ATLAS9 stellar atmosphere programs and 2 km/s grid.~Kurucz
  CD-ROM No.~13

\bibitem[{{Kurucz}(1979)}]{Kurucz79apjs}
{Kurucz}, R.~L., 1979, ApJS, 40, 1

\bibitem[{{Lagrange} et~al.(2009){Lagrange}, {Desort}, {Galland}, {Udry}, \&
  {Mayor}}]{Lagrange+09aa}
{Lagrange}, A., {Desort}, M., {Galland}, F., {Udry}, S., {Mayor}, M., 2009,
  A\&A, 495, 335

\bibitem[{{Loeillet} et~al.(2008)}]{Loeillet+08aa}
{Loeillet}, B., et~al., 2008, A\&A, 481, 529

\bibitem[{{Mamajek} \& {Hillenbrand}(2008)}]{MamajekHillenbrand08apj}
{Mamajek}, E.~E., {Hillenbrand}, L.~A., 2008, ApJ, 687, 1264

\bibitem[{{Mandushev} et~al.(2007)}]{Mandushev+07apj}
{Mandushev}, G., et~al., 2007, ApJ, 667, L195

\bibitem[{{Maness} et~al.(2007){Maness}, {Marcy}, {Ford}, {Hauschildt},
  {Shreve}, {Basri}, {Butler}, \& {Vogt}}]{Maness+07pasp}
{Maness}, H.~L., {Marcy}, G.~W., {Ford}, E.~B., {Hauschildt}, P.~H., {Shreve},
  A.~T., {Basri}, G.~B., {Butler}, R.~P., {Vogt}, S.~S., 2007, PASP, 119, 90

\bibitem[{{Mayor} \& {Queloz}(1995)}]{MayorQueloz95nat}
{Mayor}, M., {Queloz}, D., 1995, Nature, 378, 355

\bibitem[{{Mazeh} et~al.(2005){Mazeh}, {Zucker}, \& {Pont}}]{Mazeh++05mn}
{Mazeh}, T., {Zucker}, S., {Pont}, F., 2005, MNRAS, 356, 955

\bibitem[{{McCullough} et~al.(2006)}]{Mccullough+06apj}
{McCullough}, P.~R., et~al., 2006, ApJ, 648, 1228

\bibitem[{{McCullough} et~al.(2008)}]{Mccullough+08xxx}
{McCullough}, P.~R., et~al., 2008, ApJ, submitted, {\tt arXiv:0805.2921}

\bibitem[{{Mislis} \& {Schmitt}(2009)}]{MislisSchmitt09aa}
{Mislis}, D., {Schmitt}, J.~H.~M.~M., 2009, A\&A, 500, L45

\bibitem[{{Mislis} et~al.(2009){Mislis}, {Schroter}, {Schmitt}, {Cordes}, \&
  {Reif}}]{Mislis+09xxx}
{Mislis}, D., {Schroter}, S., {Schmitt}, J.~H.~M.~M., {Cordes}, O., {Reif}, K.,
  2009, A\&A, in press, {\tt arXiv:0912.4428}

\bibitem[{{Moutou} et~al.(2004){Moutou}, {Pont}, {Bouchy}, \&
  {Mayor}}]{Moutou+04aa}
{Moutou}, C., {Pont}, F., {Bouchy}, F., {Mayor}, M., 2004, A\&A, 424, L31

\bibitem[{{Moutou} et~al.(2007)}]{Moutou+07aa}
{Moutou}, C., et~al., 2007, A\&A, 473, 651

\bibitem[{{Naef} et~al.(2004){Naef}, {Mayor}, {Beuzit}, {Perrier}, {Queloz},
  {Sivan}, \& {Udry}}]{Naef+04aa}
{Naef}, D., {Mayor}, M., {Beuzit}, J.~L., {Perrier}, C., {Queloz}, D., {Sivan},
  J.~P., {Udry}, S., 2004, A\&A, 414, 351

\bibitem[{{Nelson} \& {Davis}(1972)}]{NelsonDavis72apj}
{Nelson}, B., {Davis}, W.~D., 1972, ApJ, 174, 617

\bibitem[{{O'Donovan} et~al.(2006)}]{Odonovan+06apj}
{O'Donovan}, F.~T., et~al., 2006, ApJ, 651, L61

\bibitem[{{O'Donovan} et~al.(2007)}]{Odonovan+07apj}
{O'Donovan}, F.~T., et~al., 2007, ApJ, 663, L37

\bibitem[{{P{\'a}l} et~al.(2009)}]{Pal+09apj}
{P{\'a}l}, A., et~al., 2009, ApJ, 700, 783

\bibitem[{{P{\'a}l} et~al.(2010)}]{Pal+10mn}
{P{\'a}l}, A., et~al., 2010, MNRAS, 401, 2665

\bibitem[{{Pietrinferni} et~al.(2004){Pietrinferni}, {Cassisi}, {Salaris}, \&
  {Castelli}}]{Pietrinferni+04apj}
{Pietrinferni}, A., {Cassisi}, S., {Salaris}, M., {Castelli}, F., 2004, ApJ,
  612, 168

\bibitem[{{Pollacco} et~al.(2008)}]{Pollacco+08mn}
{Pollacco}, D., et~al., 2008, MNRAS, 385, 1576

\bibitem[{{Pollacco} et~al.(2006)}]{Pollacco+06pasp}
{Pollacco}, D.~L., et~al., 2006, PASP, 118, 1407

\bibitem[{{Pont} et~al.(2004){Pont}, {Bouchy}, {Queloz}, {Santos}, {Melo},
  {Mayor}, \& {Udry}}]{Pont+04aa}
{Pont}, F., {Bouchy}, F., {Queloz}, D., {Santos}, N.~C., {Melo}, C., {Mayor},
  M., {Udry}, S., 2004, A\&A, 426, L15

\bibitem[{{Pont} et~al.(2007)}]{Pont+07aa2}
{Pont}, F., et~al., 2007, A\&A, 476, 1347

\bibitem[{{Pont} et~al.(2008)}]{Pont+08aa}
{Pont}, F., et~al., 2008, A\&A, 487, 749

\bibitem[{{Popper} \& {Etzel}(1981)}]{PopperEtzel81aj}
{Popper}, D.~M., {Etzel}, P.~B., 1981, AJ, 86, 102

\bibitem[{{Rabus} et~al.(2009){Rabus}, {Deeg}, {Alonso}, {Belmonte}, \&
  {Almenara}}]{Rabus+09aa}
{Rabus}, M., {Deeg}, H.~J., {Alonso}, R., {Belmonte}, J.~A., {Almenara}, J.~M.,
  2009, A\&A, 508, 1011

\bibitem[{{Safronov}(1972)}]{Safronov72}
{Safronov}, V.~S., 1972, Evolution of the Protoplanetary Cloud and Formation of
  the Earth and Planets (Jerusalem: Israel Program for Scientific Translation)

\bibitem[{{Santos} et~al.(2004){Santos}, {Israelian}, \&
  {Mayor}}]{Santos++04aa}
{Santos}, N.~C., {Israelian}, G., {Mayor}, M., 2004, A\&A, 415, 1153

\bibitem[{{Santos} et~al.(2006)}]{Santos+06aa}
{Santos}, N.~C., et~al., 2006, A\&A, 458, 997

\bibitem[{{Sato} et~al.(2005)}]{Sato+05apj}
{Sato}, B., et~al., 2005, ApJ, 633, 465

\bibitem[{{Scuderi} et~al.(2009){Scuderi}, {Dittmann}, {Males}, {Green}, \&
  {Close}}]{Scuderi+09xxx}
{Scuderi}, L.~J., {Dittmann}, J.~A., {Males}, J.~R., {Green}, E.~M., {Close},
  L.~M., 2009, ApJ submitted, {\tt arXiv:0907.1685}

\bibitem[{{Seager} \& {Mall{\'e}n-Ornelas}(2003)}]{SeagerMallen03apj}
{Seager}, S., {Mall{\'e}n-Ornelas}, G., 2003, ApJ, 585, 1038

\bibitem[{{Simpson} et~al.(2009)}]{Simpson+09}
{Simpson}, E.~K., et~al., 2009, MNRAS, submitted, {\tt arXiv:0912.3643}

\bibitem[{{Skumanich}(1972)}]{Skumanich72apj}
{Skumanich}, A., 1972, ApJ, 171, 565

\bibitem[{{Snellen} \& {Covino}(2007)}]{SnellenCovino07mn}
{Snellen}, I.~A.~G., {Covino}, E., 2007, MNRAS, 375, 307

\bibitem[{{Snellen} et~al.(2009)}]{Snellen+09aa}
{Snellen}, I.~A.~G., et~al., 2009, A\&A, 497, 545

\bibitem[{{Southworth}(2008)}]{Me08mn}
{Southworth}, J., 2008, MNRAS, 386, 1644 \ (Paper\,I)

\bibitem[{{Southworth}(2009)}]{Me09mn}
{Southworth}, J., 2009, MNRAS, 394, 272 \ (Paper\,II)

\bibitem[{{Southworth} et~al.(2004{\natexlab{a}}){Southworth}, {Maxted}, \&
  {Smalley}}]{Me++04mn}
{Southworth}, J., {Maxted}, P.~F.~L., {Smalley}, B., 2004{\natexlab{a}}, MNRAS,
  349, 547

\bibitem[{{Southworth} et~al.(2004{\natexlab{b}}){Southworth}, {Maxted}, \&
  {Smalley}}]{Me++04mn2}
{Southworth}, J., {Maxted}, P.~F.~L., {Smalley}, B., 2004{\natexlab{b}}, MNRAS,
  351, 1277

\bibitem[{{Southworth} et~al.(2004{\natexlab{c}}){Southworth}, {Zucker},
  {Maxted}, \& {Smalley}}]{Me+04mn3}
{Southworth}, J., {Zucker}, S., {Maxted}, P.~F.~L., {Smalley}, B.,
  2004{\natexlab{c}}, MNRAS, 355, 986

\bibitem[{{Southworth} et~al.(2005){Southworth}, {Smalley}, {Maxted}, {Claret},
  \& {Etzel}}]{Me+05mn}
{Southworth}, J., {Smalley}, B., {Maxted}, P.~F.~L., {Claret}, A., {Etzel},
  P.~B., 2005, MNRAS, 363, 529

\bibitem[{{Southworth} et~al.(2007{\natexlab{a}}){Southworth}, {Bruntt}, \&
  {Buzasi}}]{Me++07aa}
{Southworth}, J., {Bruntt}, H., {Buzasi}, D.~L., 2007{\natexlab{a}}, A\&A, 467,
  1215

\bibitem[{{Southworth} et~al.(2007{\natexlab{b}}){Southworth}, {Wheatley}, \&
  {Sams}}]{Me++07mn}
{Southworth}, J., {Wheatley}, P.~J., {Sams}, G., 2007{\natexlab{b}}, MNRAS,
  379, L11

\bibitem[{{Southworth} et~al.(2009{\natexlab{a}})}]{Me+09mn}
{Southworth}, J., et~al., 2009{\natexlab{a}}, MNRAS, 396, 1023

\bibitem[{{Southworth} et~al.(2009{\natexlab{b}})}]{Me+09mn2}
{Southworth}, J., et~al., 2009{\natexlab{b}}, MNRAS, 399, 287

\bibitem[{{Southworth} et~al.(2009{\natexlab{c}})}]{Me+09apj}
{Southworth}, J., et~al., 2009{\natexlab{c}}, ApJ, 707, 167

\bibitem[{{Southworth} et~al.(2010)}]{Me+10}
{Southworth}, J., et~al., 2010, MNRAS submitted

\bibitem[{{Sozzetti} et~al.(2007){Sozzetti}, {Torres}, {Charbonneau}, {Latham},
  {Holman}, {Winn}, {Laird}, \& {O'Donovan}}]{Sozzetti+07apj}
{Sozzetti}, A., {Torres}, G., {Charbonneau}, D., {Latham}, D.~W., {Holman},
  M.~J., {Winn}, J.~N., {Laird}, J.~B., {O'Donovan}, F.~T., 2007, ApJ, 664,
  1190

\bibitem[{{Sozzetti} et~al.(2009)}]{Sozzetti+09apj}
{Sozzetti}, A., et~al., 2009, ApJ, 691, 1145

\bibitem[{{Spearman}(1904)}]{Spearman1904}
{Spearman}, C., 1904, American Journal of Psychology, 72

\bibitem[{{Stempels} et~al.(2007){Stempels}, {Collier Cameron}, {Hebb},
  {Smalley}, \& {Frandsen}}]{Stempels+07mn}
{Stempels}, H.~C., {Collier Cameron}, A., {Hebb}, L., {Smalley}, B.,
  {Frandsen}, S., 2007, MNRAS, 379, 773

\bibitem[{{Street} et~al.(2007)}]{Street+07mn}
{Street}, R.~A., et~al., 2007, MNRAS, 379, 816

\bibitem[{{Torres} et~al.(2008){Torres}, {Winn}, \& {Holman}}]{Torres++08apj}
{Torres}, G., {Winn}, J.~N., {Holman}, M.~J., 2008, ApJ, 677, 1324

\bibitem[{{Triaud} et~al.(2009)}]{Triaud+09aa}
{Triaud}, A.~H.~M.~J., et~al., 2009, A\&A, 506, 377

\bibitem[{{Triaud} et~al.(2010)}]{Triaud+10aa}
{Triaud}, A.~H.~M.~J., et~al., 2010, A\&A, submitted

\bibitem[{{Tripathi} et~al.(2010)}]{Tripathi+10}
{Tripathi}, A., et~al., 2010, ApJ, in press, {\tt arXiv:1004.0692}

\bibitem[{{Udalski} et~al.(1997){Udalski}, {Kubiak}, \&
  {Szyma{\'n}ski}}]{Udalski++97aca}
{Udalski}, A., {Kubiak}, M., {Szyma{\'n}ski}, M., 1997, AcA, 47, 319

\bibitem[{{Udalski} et~al.(2002){Udalski}, {Szewczyk}, {Zebrun}, {Pietrzynski},
  {Szymanski}, {Kubiak}, {Soszynski}, \& {Wyrzykowski}}]{Udalski+02aca2}
{Udalski}, A., {Szewczyk}, O., {Zebrun}, K., {Pietrzynski}, G., {Szymanski},
  M., {Kubiak}, M., {Soszynski}, I., {Wyrzykowski}, L., 2002, Acta Astronomica,
  52, 317

\bibitem[{{Udalski} et~al.(2008)}]{Udalski+08aa}
{Udalski}, A., et~al., 2008, A\&A, 482, 299

\bibitem[{{Valenti} \& {Fischer}(2005)}]{ValentiFischer05apjs}
{Valenti}, J.~A., {Fischer}, D.~A., 2005, ApJS, 159, 141

\bibitem[{{van Belle} \& {von Braun}(2009)}]{BelleBraun09apj}
{van Belle}, G.~T., {von Braun}, K., 2009, ApJ, 694, 1085

\bibitem[{{Van Hamme}(1993)}]{Vanhamme93aj}
{Van Hamme}, W., 1993, AJ, 106, 2096

\bibitem[{{VandenBerg} et~al.(2006){VandenBerg}, {Bergbusch}, \&
  {Dowler}}]{Vandenberg++06apjs}
{VandenBerg}, D.~A., {Bergbusch}, P.~A., {Dowler}, P.~D., 2006, ApJS, 162, 375

\bibitem[{{West} et~al.(2009)}]{West+09aa}
{West}, R.~G., et~al., 2009, A\&A, 502, 395

\bibitem[{{Wheatley} et~al.(2010)}]{Wheatley+10}
{Wheatley}, P.~J., et~al., 2010, ApJ, submitted, {\tt arXiv:1004.0836}

\bibitem[{{Winn} et~al.(2008{\natexlab{a}}){Winn}, {Holman}, {Shporer},
  {Fern{\'a}ndez}, {Mazeh}, {Latham}, {Charbonneau}, \& {Everett}}]{Winn+08aj}
{Winn}, J.~N., {Holman}, M.~J., {Shporer}, A., {Fern{\'a}ndez}, J., {Mazeh},
  T., {Latham}, D.~W., {Charbonneau}, D., {Everett}, M.~E., 2008{\natexlab{a}},
  AJ, 136, 267

\bibitem[{{Winn} et~al.(2007{\natexlab{a}})}]{Winn+07apj}
{Winn}, J.~N., et~al., 2007{\natexlab{a}}, ApJ, 665, L167

\bibitem[{{Winn} et~al.(2007{\natexlab{b}})}]{Winn+07aj}
{Winn}, J.~N., et~al., 2007{\natexlab{b}}, AJ, 133, 1828

\bibitem[{{Winn} et~al.(2007{\natexlab{c}})}]{Winn+07aj2}
{Winn}, J.~N., et~al., 2007{\natexlab{c}}, AJ, 134, 1707

\bibitem[{{Winn} et~al.(2008{\natexlab{b}})}]{Winn+08apj2}
{Winn}, J.~N., et~al., 2008{\natexlab{b}}, ApJ, 683, 1076

\bibitem[{{Winn} et~al.(2009)}]{Winn+09apj}
{Winn}, J.~N., et~al., 2009, ApJ, 700, 302

\end{thebibliography}
% \bsp

% Paper 1
% Paper 2

\label{lastpage}

%%%%%%%%%%%%%%%%%%%%%%%%%%%%%%%%%%%%%%%%%%%%%%%%%%%%%%%%%%%%%%%%%%%%%%%%%%%%%%%%%%%%%%%%%%%%%%%%%%%%%%%%%%%%%%
\end{document}